\documentclass[twocolumn]{aastex62}

\usepackage{longtable}
\usepackage{graphicx}
\usepackage{amsmath,amssymb}
\usepackage{color}
\usepackage{units}
\usepackage{epstopdf}
\usepackage{hyperref}
\usepackage{multirow}
\usepackage{url}
\usepackage{placeins}

\usepackage[figuresleft]{rotating}

\usepackage{subfigure}

\newcommand{\myRule}[3][black]{\textcolor{#1}{\rule{#2}{#3}}}

\newcommand{\beq}{\begin{equation}}
\newcommand{\eeq}{\end{equation}}
\newcommand{\bdm}{\begin{displaymath}}
\newcommand{\edm}{\end{displaymath}}

\definecolor{Gray}{gray}{0.9}
\definecolor{orange}{rgb}{0.9,0.5,0}

\newcommand{\beqn}{\begin{eqnarray}}
\newcommand{\eeqn}{\end{eqnarray}}

\graphicspath{{./plots/}}

\begin{document}

\title{Implications of the search for optical counterparts during the second part of the Advanced LIGO's and Advanced Virgo's third observing run: lessons learned for future follow-up observations}

\author[0000-0002-8262-2924]{Michael W. Coughlin}
\affiliation{School of Physics and Astronomy, University of Minnesota, Minneapolis, Minnesota 55455, USA}
\author[0000-0003-2374-307X]{Tim Dietrich}
\affiliation{Institut f\"{u}r Physik und Astronomie, Universit\"{a}t Potsdam, Haus 28, Karl-Liebknecht-Str. 24/25, 14476, Potsdam, Germany}
\author[0000-0002-7686-3334]{Sarah Antier}
\affiliation{Universit\'e de Paris, CNRS, Astroparticule et Cosmologie, F-75013 Paris, France}
\author[0000-0002-4694-7123]{Mouza Almualla}
\affiliation{American University of Sharjah, Physics Department, PO Box 26666, Sharjah, UAE}
\author[0000-0003-3768-7515]{Shreya Anand}
\affil{Division of Physics, Mathematics, and Astronomy, California Institute of Technology, Pasadena, CA 91125, USA}
\author[0000-0002-8255-5127]{Mattia Bulla}
\affiliation{Nordita, KTH Royal Institute of Technology and Stockholm University, Roslagstullsbacken 23, SE-106 91 Stockholm, Sweden}
\author[0000-0003-4617-4738]{Francois Foucart}
\affiliation{Department of Physics \& Astronomy, University of New Hampshire, 9 Library Way, Durham NH 03824, USA}
\author[0000-0003-1585-8205]{Nidhal Guessoum}
\affiliation{American University of Sharjah, Physics Department, PO Box 26666, Sharjah, UAE}
\author[0000-0002-2502-3730]{Kenta Hotokezaka}
\affiliation{Department of Astrophysical Sciences, Princeton University, Princeton, NJ 08544, USA}
\author[0000-0002-8359-9762]{Vishwesh Kumar}
\affiliation{American University of Sharjah, Physics Department, PO Box 26666, Sharjah, UAE}
\author[0000-0002-9397-786X]{Geert Raaijmakers}
\affiliation{GRAPPA, Anton Pannekoek Institute for Astronomy and Institute of High-Energy Physics,
University of Amsterdam, Science Park 904, 1098 XH Amsterdam, The Netherlands}
\author[0000-0001-6573-7773]{Samaya Nissanke}
\affiliation{GRAPPA, Anton Pannekoek Institute for Astronomy and Institute of High-Energy Physics,
University of Amsterdam, Science Park 904, 1098 XH Amsterdam, The Netherlands}
\affiliation{Nikhef, Science Park 105, 1098 XG Amsterdam, The Netherlands}

\begin{abstract}
Joint multi-messenger observations with gravitational waves and electromagnetic data offer new insights into the astrophysical studies of compact objects. The third Advanced LIGO and Advanced Virgo observing run began on April 1, 2019; during the eleven months of observation, there have been 14 compact binary systems candidates for which at least one component is potentially a neutron star. Although intensive follow-up campaigns involving tens of ground and space-based observatories searched for counterparts, no electromagnetic counterpart has been detected. Following on a previous study of the first six months of the campaign, we present in this paper the next five months of the campaign from October 2019 to March 2020. We highlight two neutron star - black hole candidates (S191205ah, S200105ae), two binary neutron star candidates (S191213g and S200213t) and a binary merger with a possible neutron star and a ``MassGap'' component, S200115j. Assuming that the gravitational-wave candidates are of astrophysical origin and their location was covered by optical telescopes, we derive possible constraints on the matter ejected during the events based on the non-detection of counterparts. We find that the follow-up observations during the second half of the third observing run did not meet the necessary sensitivity to constrain the source properties of the potential gravitational-wave candidate. Consequently, we suggest that different strategies have to be used to allow a better usage of the available telescope time. We examine different choices for follow-up surveys to optimize sky localization coverage vs.\ observational depth to understand the likelihood of counterpart detection. 
\end{abstract}

%\maketitle

%%%%%%%%%%%%%%%%%%%%%%%%%%%%%%%%%%%%%%%%%%%%%%%%%%%%%%%%%%%%%%%%%%%%%%%%%%%%%%%%%%%%%%%
%%%%%%%%%%%%%%%%%%%%%%%%%%%%%%%%%%%%%%%%%%%%%%%%%%%%%%%%%%%%%%%%%%%%%%%%%%%%%%%%%%%%%%%

\section{Introduction}

The observational campaigns of Advanced LIGO \citep{aLIGO} and Advanced Virgo \citep{adVirgo} revealed the existence of a diverse population of compact binary systems. Thanks to the continuous upgrades of the detectors from the first observing run (O1) over the second observing run (O2) up to the recent third observational campaign (O3), the gain in sensitivity leads to an increasing number of compact binary mergers candidates: 16 alerts of gravitational-wave (GW) candidates were sent to the astronomical community during O1 and O2, covering a total of 398 days \citep{AbEA2019}, compared to 80 alerts for O3a and O3b, covering a total of 330 days. Some of the candidates found during the online searches were retracted after further analysis, e.g., only 10 out of the 16 alerts were confirmed as candidates during the O1 and O2 runs \citep{AbEA2019,LIGOScientific:2018mvr}. Additional compact binary systems were found during the systematic offline analysis performed with re-calibrated data, e.g., \citealt{LIGOScientific:2018mvr}, resulting in 11 confirmed GW events. During O3a and O3b, 24 of 80 alerts have already been retracted due to data quality issues, e.g. \citealt{2019GCN.26413....1L,2020GCN.26665....1L}. 

GW detections improve our understanding of binary populations in the nearby Universe (distances less than $\sim$\,2 Gpc), and cover a large range of masses; these cover from $\sim$\,1--2.3 solar masses, e.g.~\citealt{La12,OzFr16,MaMe2017,ReMo2017,AbEA2019_GW190425}, for binary neutron stars (BNSs) to $\sim$\,100 solar masses for the most massive black hole remnants. They may also potentially constrain black hole spins \citep{2019ApJ...882L..24A}. For mergers including NSs, electromagnetic (EM) observations provide a complementary view, providing precise localizations of the event, required for redshift measurements which are important for cosmological constraints \citep{Sch1986}; these observations may last for years at wavelengths outside the optical spectrum; for instance, X-ray photons were detected almost 1000 days post-merger in the case of GW170817 \citep{troja2020thousand}.

The success of joint GW and EM observations to explore the compact binaries systems has been demonstrated by the success of GW170817, AT2017gfo, and GRB170817A, e.g.,~\citealt{AbEA2017b,GBM:2017lvd,ArHo2017,2017Sci...358.1556C,LiGo2017,MoNa2017,SaFe2017,SoHo2017,TaLe2013,TrPi2017,VaSa2017}.
GRB170817A, a short $\gamma$-ray burst (sGRB) \citep{1989Natur.340..126E, Paczynski1991, 1992ApJ...395L..83N, MocHer1993, LeeRR2007, Nakar2007}, and AT2017gfo, the associated kilonova \citep{SaPiNa1998,LiPa1998,MeMa2010,RoKa2011,KaMe2017}, 
were the EM counterparts of GW170817.
Overall, this multi-messenger event has been of interest for many reasons: 
to place constraints on the supranuclear equation of state describing the NS interior (e.g.~\citealt{Abbott:2018wiz,RaPe2018,RaDa2018,BaJu2017,MaMe2017,ReMo2017,CoDi2018,CoDi2018b,Capano:2019eae,DiCo2020}), to determine the expansion rate of the Universe~\citep{AbEA2017g,Hotokezaka:2018dfi,CoDi2019,DhBu2019,DiCo2020}, to provide tests for alternative theories of gravity~\citep{EzMa17,BaBe17,CrVe17,AbEA2019b}, to set bounds on the speed of GWs~\citep{GBM:2017lvd}, 
and to prove BNS mergers to be a production side for heavy elements, e.g.,~\citealt{2017Natur.551...67P,WaHa19}.

Numerical-relativity studies reveal that not all binary neutron star (BNS) and black hole- neutron star (BHNS) collisions will eject a sufficient amount of material to create bright EM signals, e.g.,~\citealt{Bauswein:2013jpa,HoKi13,DiUj2017,AbEA2017f,Koppel:2019pys,Agathos:2019sah,Kawaguchi:2016ana,Foucart:2018rjc,KrFo2020}. For example, there will be no bright EM signal if a black hole (BH) forms directly after merger of an almost equal-mass BNS, since the amount of ejected material and the mass of the potential debris disk are expected to be very small. Whether a merger remnant undergoes a prompt collapse depends mostly on its total mass but also seems to be sub-dominantly affected by the mass-ratio~\citep{Kiuchi:2019lls,Beea20}. EM bright signatures originating from BHNS systems depend on whether the NS gets tidally disrupted by the BH and thus ejects a large amount of material and forms a massive accretion disk. If the neutron star falls into the BH without disruption, EM signatures will not be produced. This outcome is mostly determined by the mass ratio of the binary, the spin of the black hole, and the compactness of the NS, with disruption being favored for low-mass, rapidly rotating BH and large NS radii~\citep{Etienne:2008re,Pannarale:2010vs,Foucart:2012nc,Kyutoku:2015gda,Kawaguchi:2016ana,Foucart:2018rjc}. In addition, beamed ejecta from the GRB can be weakened by the jet break \citep{2006ApJ...653..468B,2018ApJ...866L..16M} and may not escape from the ``cocoon'', which would change the luminosity evolution of the afterglow.

The observability and detectability of the EM signature depends on a variety of factors. First, and most practically, the event must be observable by telescopes, e.g., not too close to the Sun or majorly overlapping with the Galactic plane; 20\% of the O3 alerts were not observable by any of three major sites of astronomy; e.g. Palomar, the Cerro Tololo Inter-American Observatory, and Mauna Kea \citep{GRANDMA2020}. 

Secondly, the identification of counterparts depends on the duty cycle of instruments and the possibility to observe the skymap shortly after merger. For example, $\gamma$-ray observatories such as the {\it Fermi} Gamma-Ray Burst Monitor \citep{Connaughton12} can cover up to 70\% of the full sky, but due to their altitude and pointing restrictions, their field of view can be occluded by the Earth or when the satellite is passing through the South Atlantic Anomaly \citep{2019GCN.26342....1M,2020GCN.27361....1L}. The ability for telescopes to observe depends on the time of day of the event. For example, between 18\,hr -- 15\,hr UTC (although this level of coverage is available only portions of the year, and even then, it is twilight at the edges), both the Northern and Southern sky can, in theory, be covered thanks to observatories in South Africa, the Canary islands, Chile and North America; at other times, such as when night passes over the Pacific ocean or the Middle East, the dearth of observatories greatly reduces the chances of a ground detection.  

Third, counterpart searches are also affected by the viewing angle of the event with respect to the line of sight towards Earth. While the beamed jet of the burst can be viewed within a narrow cone, the kilonova signature is likely visible from all viewing angles; however, its color and luminosity evolution is likely to be viewing angle dependent \citep{RoKa2011,Bul2019,DaKa20,KaSh20,KoWo20}.
Finally, as the distance of the event changes, the number of instruments sensitive enough to perform an effective search changes. For example, compared to GW170817, detected at 40~Mpc, the O3 BNS candidates reported so far (with a BNS source probability of $>50\%$) have median estimated distances $\sim150-250$~Mpc.

\begin{figure}[t]
 \includegraphics[width=3.5in]{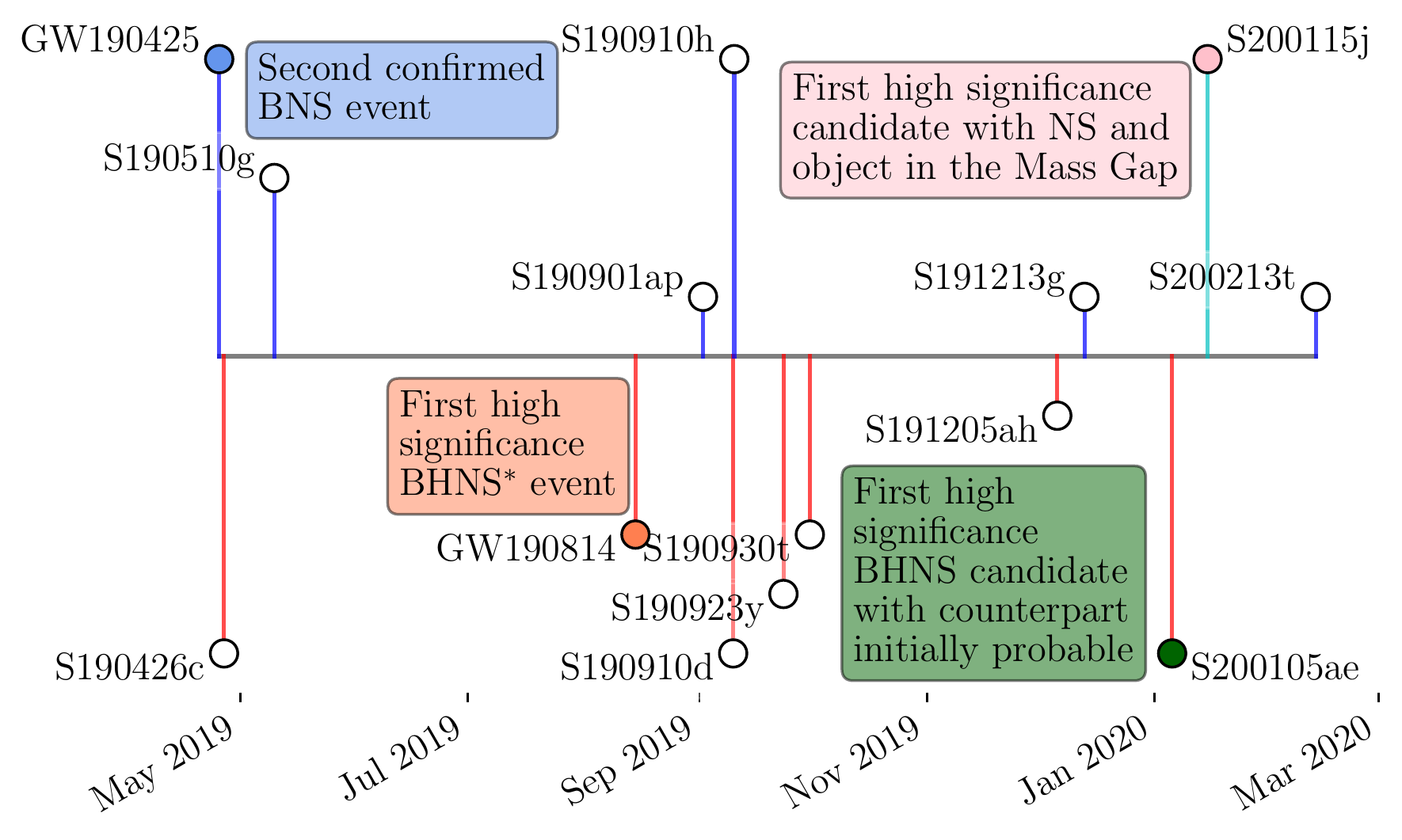}
  \caption{Timeline of O3 alerts with highest probability as being BNS, BHNS or MassGap, with highlights of some of the exceptional candidates released. The candidates, if astrophysical, on the top half of the plot are most likely BNSs (or a NS-MassGap candidate in the case of S200115j \citep{gcn26759}), while the candidates on the bottom half are most likely BHNSs. We highlight GW190425 \citep{AbEA2019_GW190425}, GW190814~\citep{AbEA2020}, S200105ae \citep{gcn26640,gcn26688}, and S200115j \citep{gcn26759}. We note that the initial estimate of p(remnant) for S200105ae was 12\% \citep{gcn26640}, but is now $<$\,1\% \citep{gcn26688}. It also has a significance likely greatly underestimated due to it being a single-instrument event, and a chirp-like structure in the spectrograms as mentioned in the public reports \citep{gcn26640, gcn26657}. \\
  $^*$We note that GW190814 contains either the highest mass neutron star or lowest mass black hole known~\citep{AbEA2020}.}
 \label{fig:timeline}
\end{figure} 

Despite those observational difficulties, the O3a and O3b observational campaigns were popular for searches of EM counterparts associated with the GW candidates (see Figure~\ref{fig:timeline} for a timeline for candidates with at least one NS component expected). They mobilized $\sim$\,100 groups covering multiple messengers, including neutrinos, cosmic rays, and the EM spectrum; about half of the participating groups are in the optical. In total, GW follow-up represented $\sim$\,50\% of the GCN service traffic (Gamma-ray Coordinates Network) with 1,558 circulars. The first half of the third observation run (O3a)  brought ten compact binary merger candidates that were expected to have low-mass components, including GW190425~\citep{SiEA2019a,SiEA2019b}, S190426c~\citep{ChEA2019a,ChEA2019b}, S190510g~\citep{gcn24489}, GW190814~\citep{AbEA2020}, S190901ap~\citep{gcn25606}, S190910h~\citep{gcn25707}, S190910d~\citep{gcn25695}, S190923y~\citep{gcn25814}, and S190930t~\citep{gcn25876}. The follow-up campaigns of these candidates have been extensive, with a myriad of instruments and teams scanning the sky localizations.\footnote{Amongst the wide field-of-view telescopes, ATLAS \citep{gcn24197,gcn24517,gcn25922,gcn25375}, ASAS-SN \citep{gcn24309}, CNEOST \citep{gcn24285,gcn24465,gcn24286}, Dabancheng/HMT \citep{gcn24476}, DESGW-DECam \citep{gcn25336}, DDOTI/OAN \citep{gcn24310,gcn25352,gcn25737}, GOTO \citep{gcn24224,gcn25654,gcn24291,gcn25337,GoCu2020}, GRANDMA \citep{GRANDMAO3A,GRANDMA2020}, GRAWITA-VST \citep{gcn24484,gcn25371}, GROWTH-DECAM \citep{AnGo2019,GoAn2019}, GROWTH-Gattini-IR \citep{gcn24187,gcn24284,gcn25358}, GROWTH-INDIA \citep{gcn24258}, HSC \citep{gcn24450}, J-GEM \citep{gcn24299}, KMTNet \citep{gcn24466,gcn25342}, MASTER-network \citep{gcn24167,gcn24436,gcn25609,gcn25712,gcn25712,gcn24236,gcn25322,gcn25694,gcn25812}, MeerLICHT \citep{gcn25340}, Pan-STARRS \citep{gcn24210,gcn24517,gcn24517}, SAGUARO \citep{2019arXiv190606345L}, SVOM-GWAC \citep{gcn25648}, Swope \citep{gcn25350}, Xinglong-Schmidt \citep{gcn24190,gcn24475}, and the Zwicky Transient  Facility \citep{gcn24191,gcn25616,gcn25722,gcn25899,gcn24283,gcn25343,gcn25706,CoAh2019b} participated.}

The follow-up of O3a yielded a number of interesting searches. For example, GW190425 \citep{AbEA2019_GW190425} brought stringent limits on potential counterparts from a number of teams, including GROWTH \citep{CoAh2019b} and MMT/SOAR \citep{HoCo2019}. GW190814, as a potential, well-localized BHNS candidate, also had extensive follow-up from a number of teams, including GROWTH \citep{Andreoni2020}, ENGRAVE \citep{AcAm2020}, GRANDMA \citep{GRANDMAO3A} and Magellan \citep{GoHo2019}. S190521g brought the first strong candidate counterpart to a BBH merger \cite{GrFo2020}.

The second half of the third observation run (O3b) has brought 23 new publicly announced compact binary merger candidates for which observational facilities performed follow-up searches, including two new BNS candidates, S191213g \citep{gcn26402} and S200213t \citep{gcn27042} and two new BHNS candidates, S191205ah \citep{gcn26350} and S200105ae \citep{gcn26640,gcn26688}. S200115j is special for having one NS component and one component object likely falling in the ``MassGap'' regime, indicating it is between 3--5~M$_\odot$\citep{gcn26759}.
After the second half of this intensive campaign, no significant counterpart (either GRB or kilonovae) was found. While this might be caused by the fact that the GW triggers have not been accompanied by bright EM counterparts, a likely reason for this lack of success in finding optical counterparts is the limited coordination of global EM follow-up surveys and the limited depth of the individual observations. 

In this article, we build on our summary of the O3a observations \citep{CoDi2019b} to explore constraints on potential counterparts based on the wide field-of-view telescope observations during O3b, and provide analyses summarizing how we may improve existing strategies with respect to the fourth observational run of advanced LIGO and advanced Virgo (O4).
In Sec.~\ref{sec:EM_follow_up_campaigns}, we review the optical follow-up campaigns for these sources.
In Sec.~\ref{sec:limits}, we summarize parameter constraints that are possible to achieve based on these follow-ups assuming that the candidate location was covered during the observations.
In Sec.~\ref{sec:strategies}, we use the results of these analyses and others to inform future observational strategies
trying to determine the optimal balance between coverage and exposure time. 
Finally, in Sec.~\ref{sec:summary}, we summarize our findings.

\section{EM follow-up campaigns}
\label{sec:EM_follow_up_campaigns}

\begin{table}
  \centering
  \caption{Current overview of non-retracted GW triggers
  with large probabilities of being BNS or BHNS systems.
  The individual columns refer to: 
  The name of the event, an estimate using the most up-to-date classification for the event to be 
  a BNS [p(BNS)], a BHNS [p(BHNS)], or terrestrial noise [p(terrestrial)] \citep{KaCa2019}, 
  and an indicator to estimate the probability of producing an EM signature assuming the candidate is of astrophysical origin [p(HasRemnant)] \citep{ChSh2019}, whose definition is in the \href{https://emfollow.docs.ligo.org/userguide/content.html}{LIGO-Virgo alert userguide}. Note that S200115j can also be classified as ``MassGap,'' completing the possible classifications. During O3b, a change in the template bank used led to a simplified version of the classification scheme where all of the astrophysical probabilities but one became 0, whereas during O3a, accounting for the mass uncertainty, more than one non-zero astrophysical class probability was generally obtained.
  } 
\begin{tabular}{l|cccc}
\hline
Name      & p(BNS)   & p(BHNS) & p(terr.) & p(HasRemn.) \\
\hline
\href{https://gracedb.ligo.org/superevents/S191205ah/view/}{S191205ah}  & $0\%$ & $93\%$    &   $7\%$      & $< 1\%$    \\  
\href{https://gracedb.ligo.org/superevents/S191213g/view/}{S191213g}  & $77\%$   & $ < 1\%$    &    $23\%$       & $> 99\%$   \\ 
\href{https://gracedb.ligo.org/superevents/S200105ae/view/}{S200105ae} & $0\%$   & $3\%$    &    $97\%$      & $< 1\%$    \\ 
\href{https://gracedb.ligo.org/superevents/S200115j/view/}{S200115j} & $< 1\%$ & $< 1\%$ & $< 1\%$ & $> 99\%$ \\
\href{https://gracedb.ligo.org/superevents/S200213t/view/}{S200213t} & $63\%$ & $< 1\%$ & $37\%$ & $> 99\%$
\end{tabular}
 \label{tab:allevents}
\end{table}

\begin{figure*}[t]
\centering
\textbf{S200115j} \hspace{3.0in} \textbf{S200213t} \\
 \includegraphics[width=3.5in]{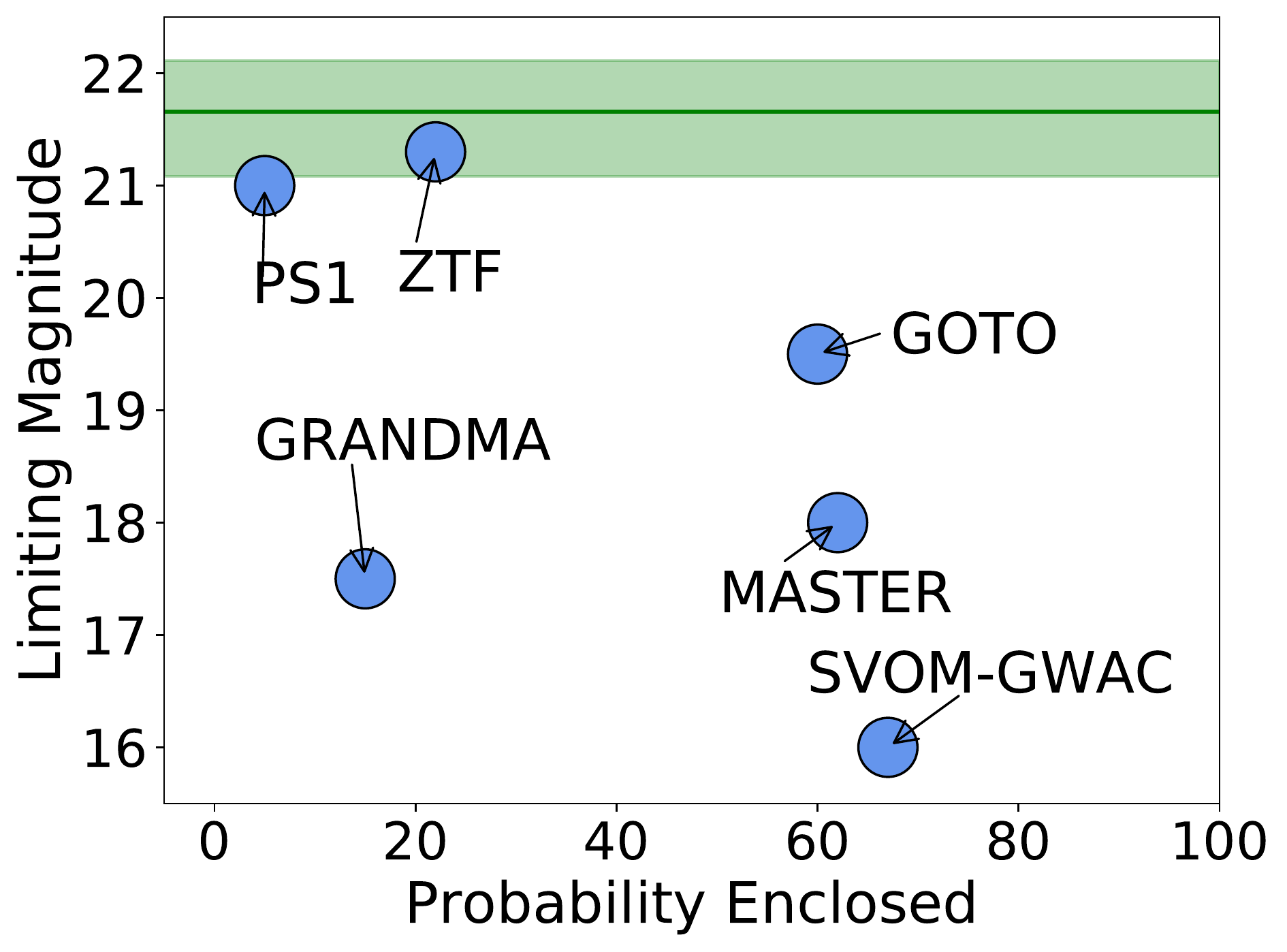}
 \includegraphics[width=3.5in]{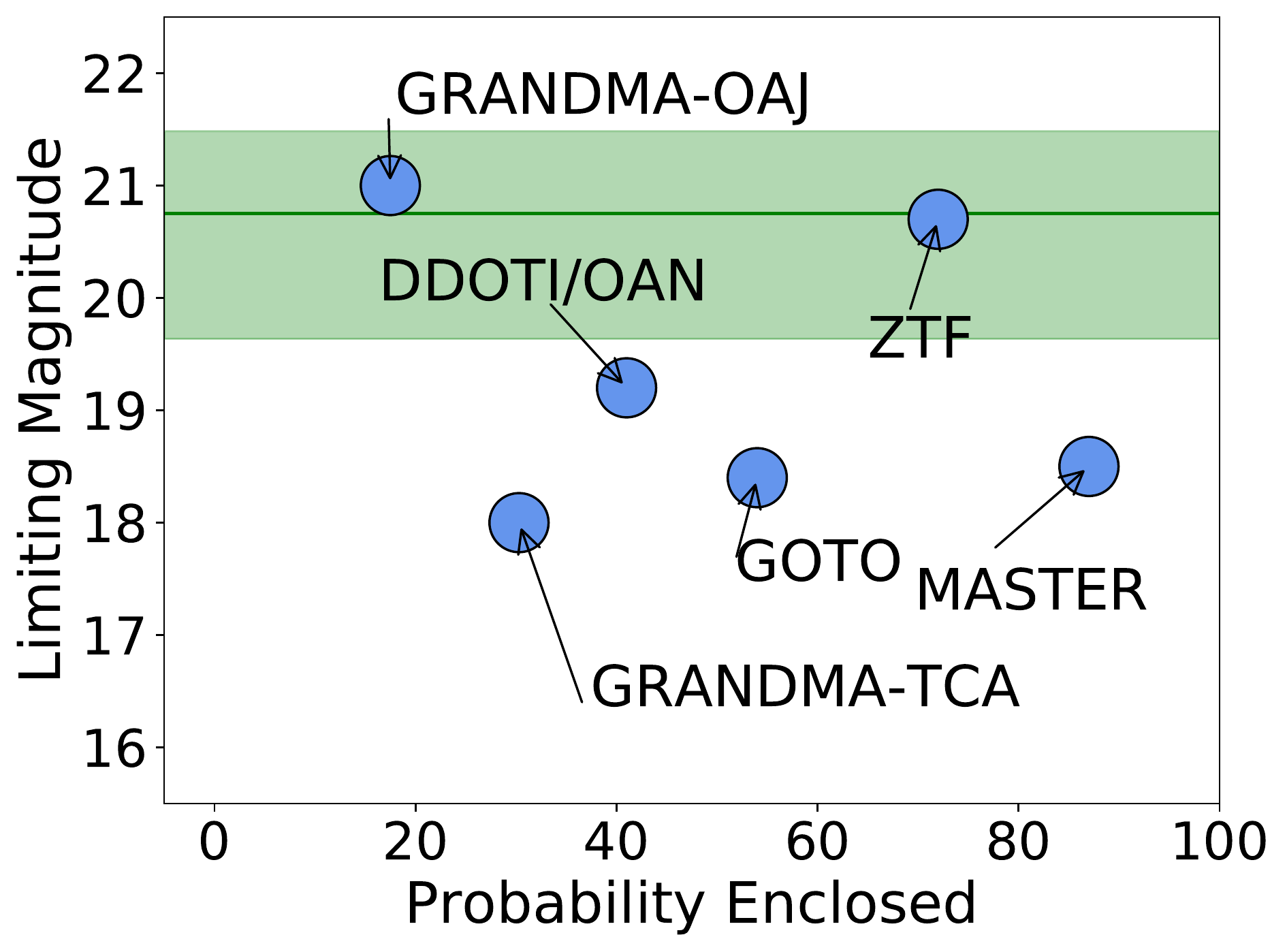}\\
 \caption{Comparison of limiting magnitudes and probabilities covered for S200115j (left) and S200213t (right). Observations span from immediately post merger up to a week after the GW trigger time. As a point of reference, we include as a green solid line, the apparent magnitude of an object with an absolute magnitude of $-16$, i.e., consistent with a signal similar to AT2017gfo. The green shaded region incorporated the $\pm$\,1\,$\sigma$ error bar of the distance that agrees with the two events.}
 \label{fig:efficiency_2}
\end{figure*}

We summarize the EM follow-up observations of the various teams that performed synoptic coverage of the sky localization area and circulated their findings in publicly available circulars during the second half of Advanced LIGO and Advanced Virgo's third observing run. The LIGO-Scientific and Virgo collaborations used the same near-time alert system during O3b as during O3a, releasing alerts within 2--6 minutes in general (with an important exception, S200105ae, discussed below). For a summary of the second observing run, please see \cite{AbEA2019}, and for the first six months of the third observing run, see \cite{CoDi2019b} and references therein. In addition to the classifications for the event in categories BNS, BHNS, ``MassGap,'' or terrestrial noise \citep{KaCa2019} and an indicator to estimate the probability of producing an EM signature assuming the candidate is of astrophysical origin, p(HasRemnant) \citep{ChSh2019}, skymaps using BAYESTAR \citep{SiPr2014} are also released. At later times, updated LALInference \citep{VeRa2015} skymaps are also sent to the community.

\begin{figure*}[t]
\centering
\textbf{S200115j} \hspace{3.0in} \textbf{S200213t} \\
\vspace{0.1in}
\textbf{Within 12 hours of merger} \\
 \includegraphics[width=3.5in]{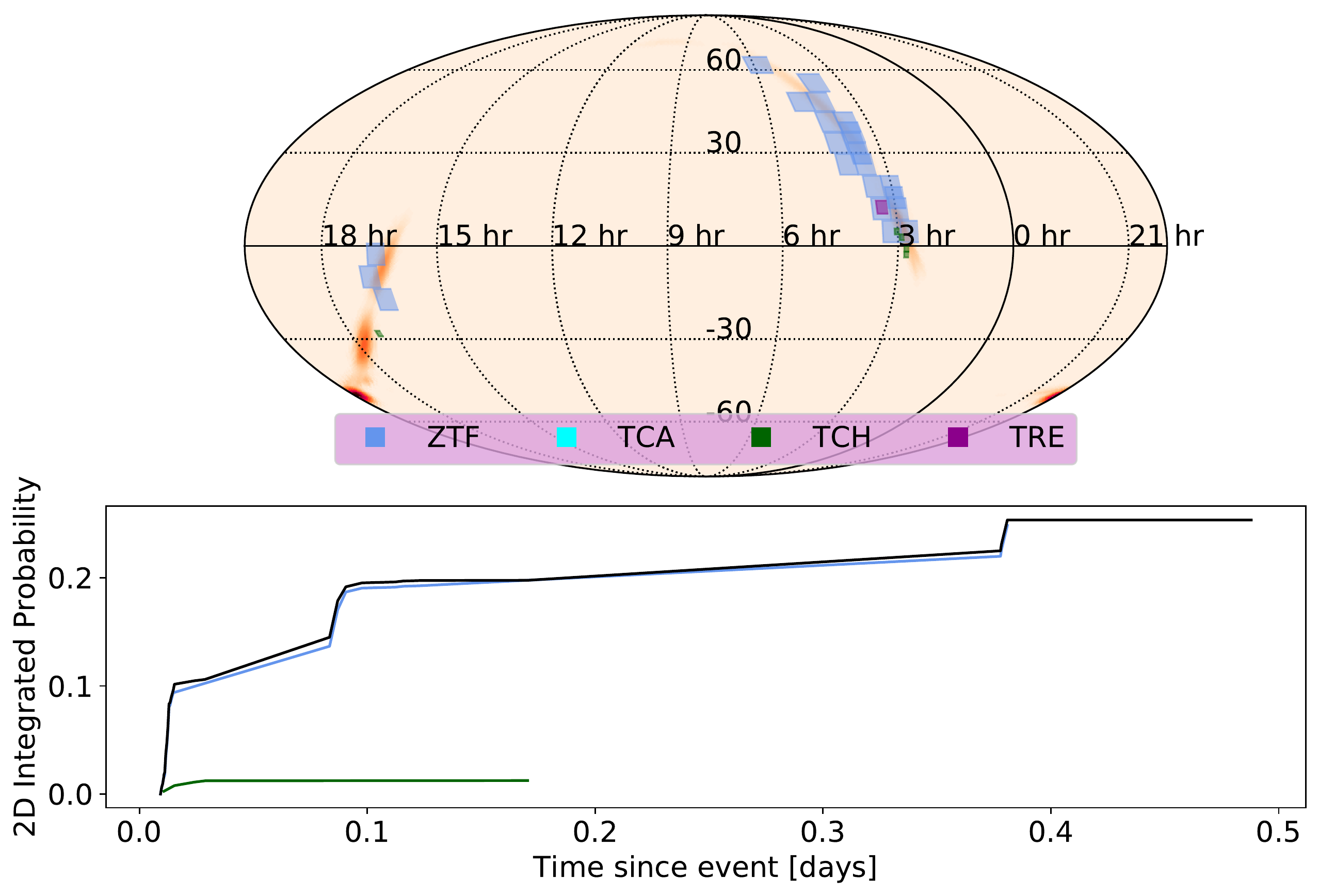}
 \includegraphics[width=3.5in]{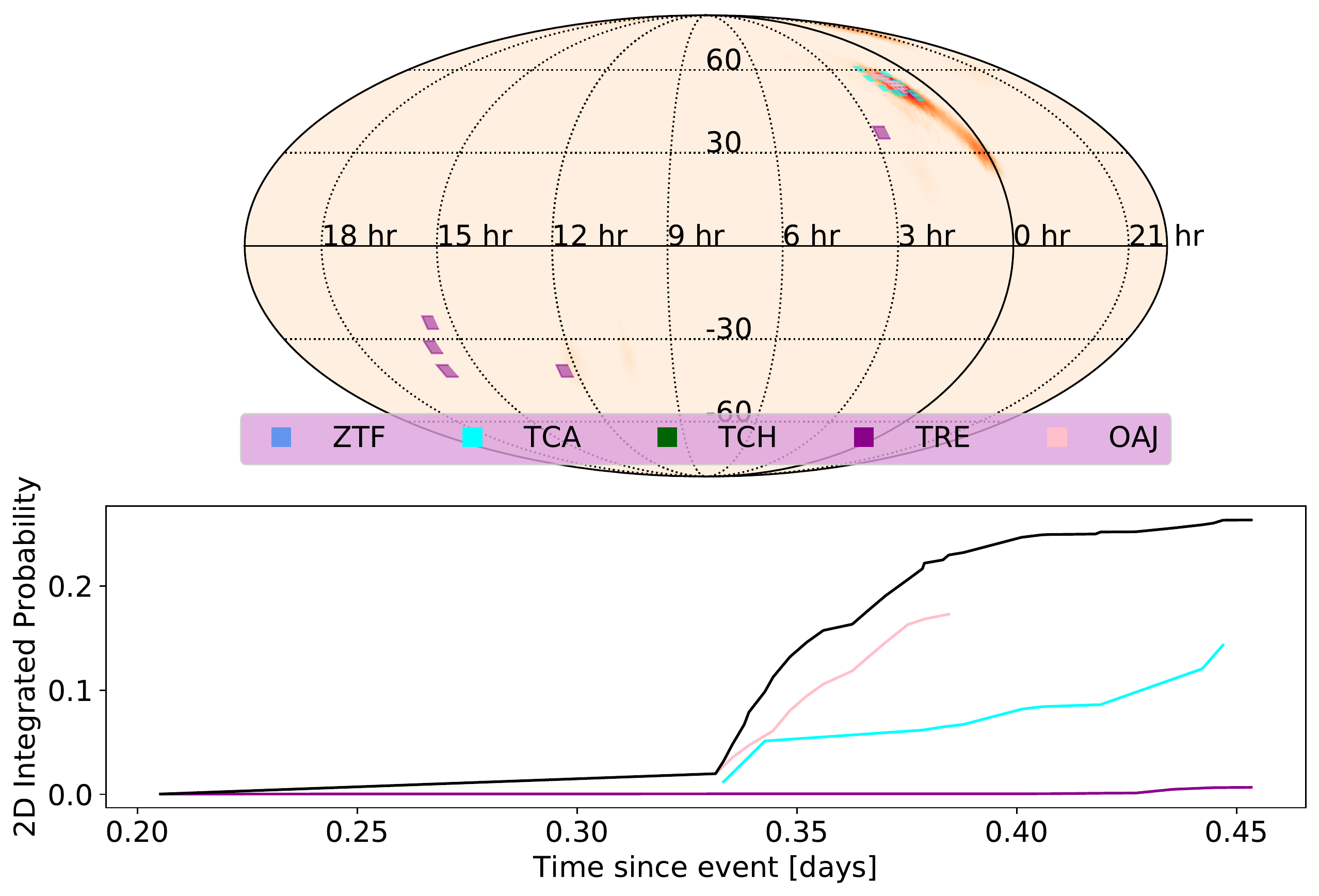} \\
\textbf{Within 24 hours of merger} \\
 \includegraphics[width=3.5in]{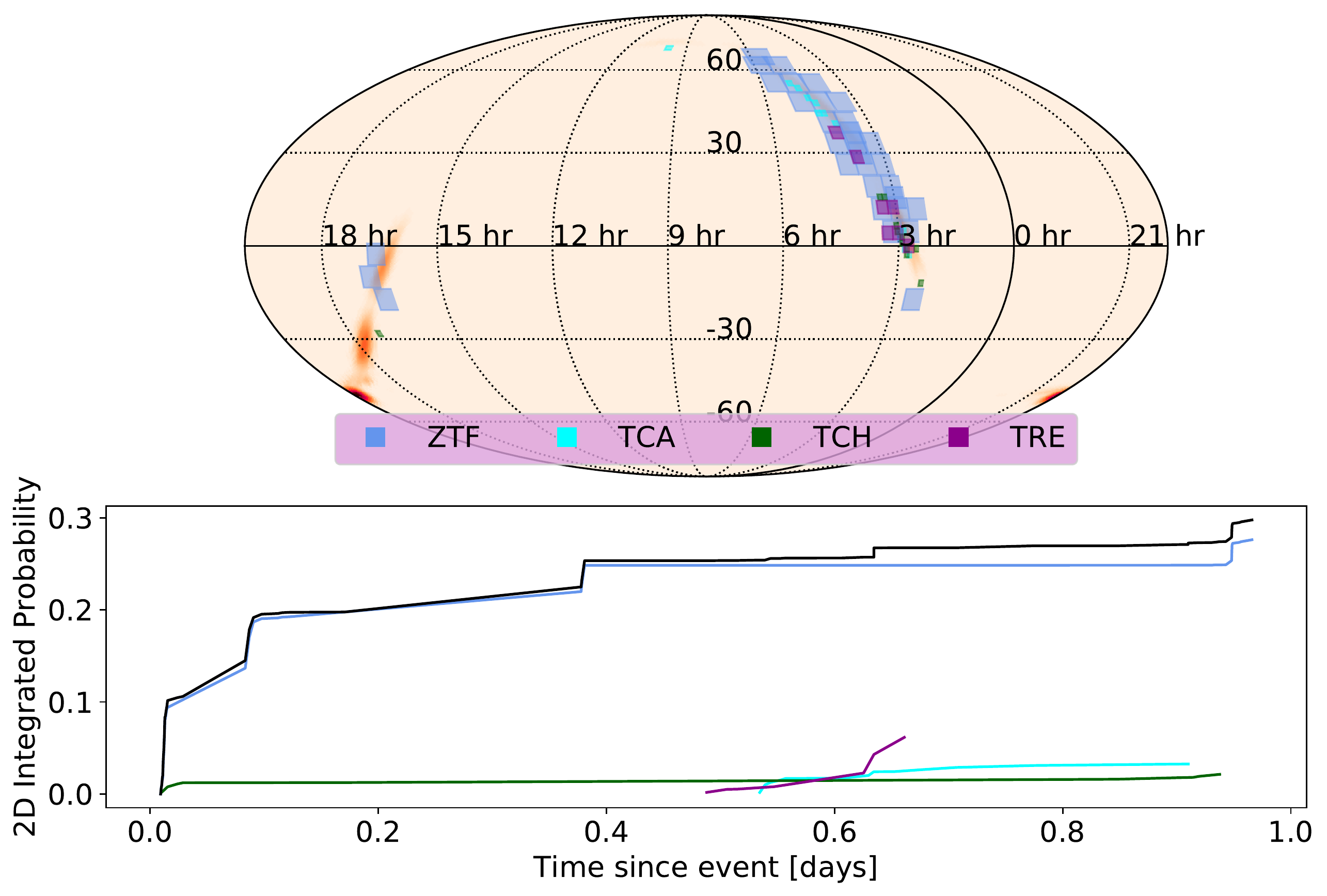} 
 \includegraphics[width=3.5in]{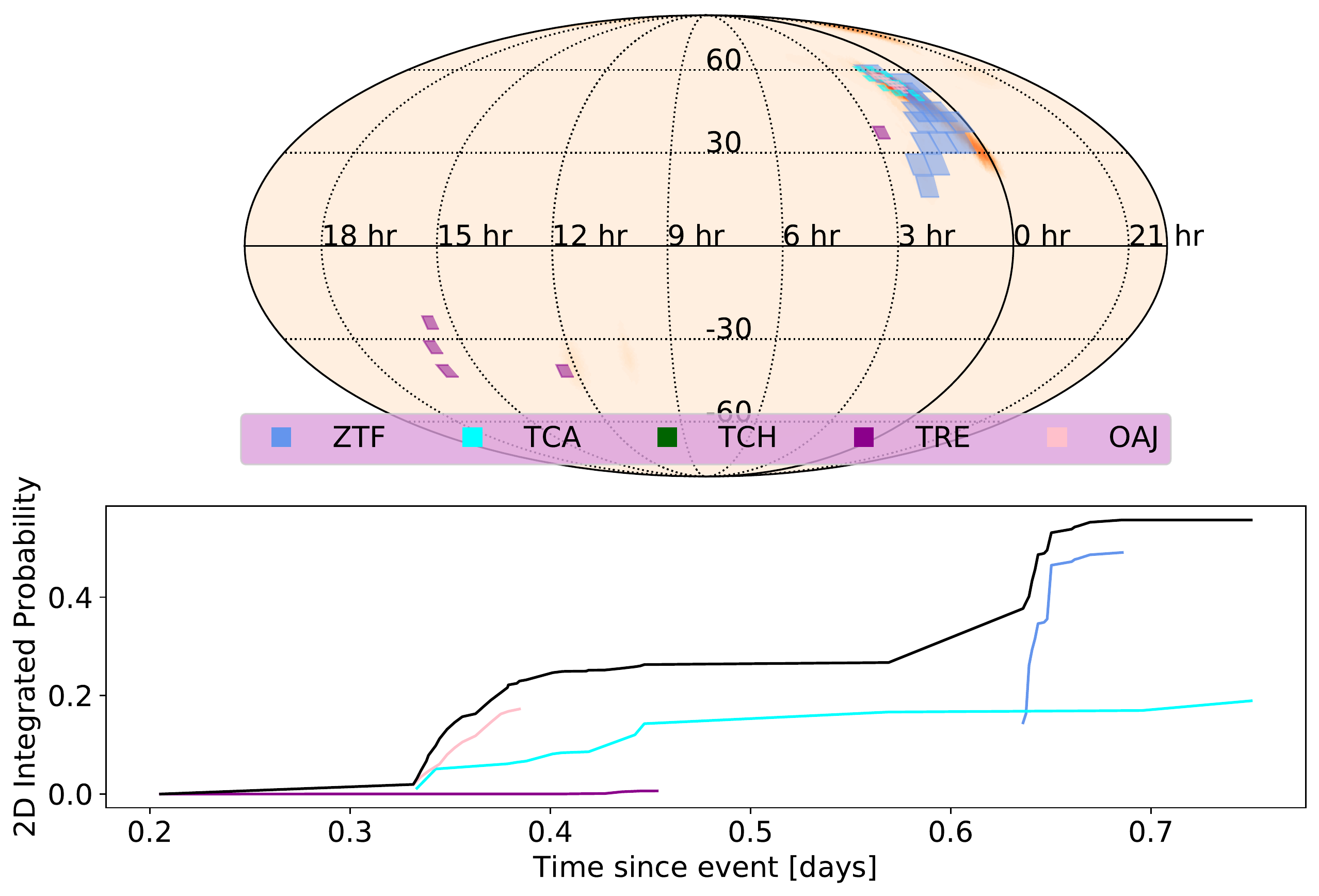} \\
\textbf{Within 48 hours of merger} \\
 \includegraphics[width=3.5in]{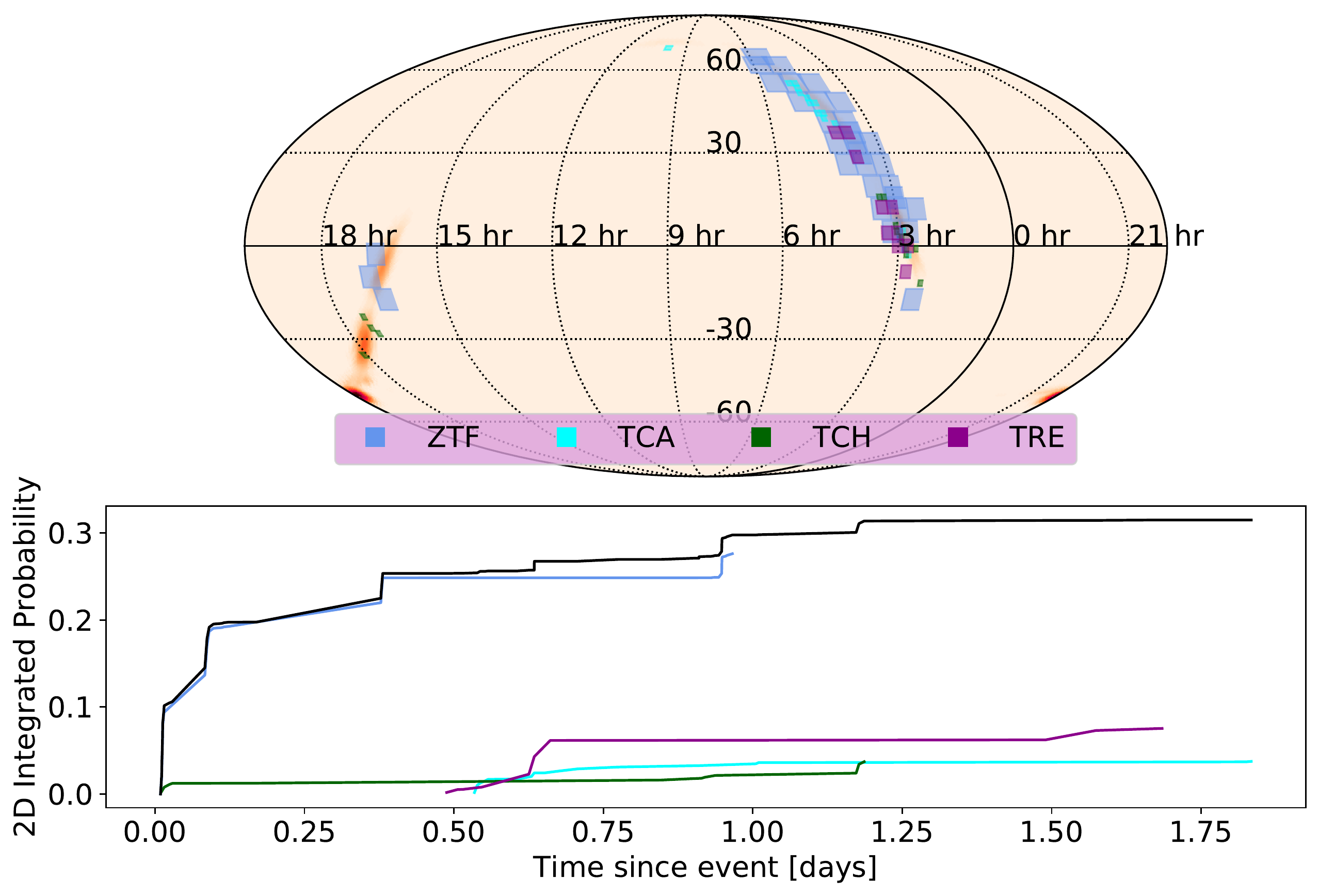}
 \includegraphics[width=3.5in]{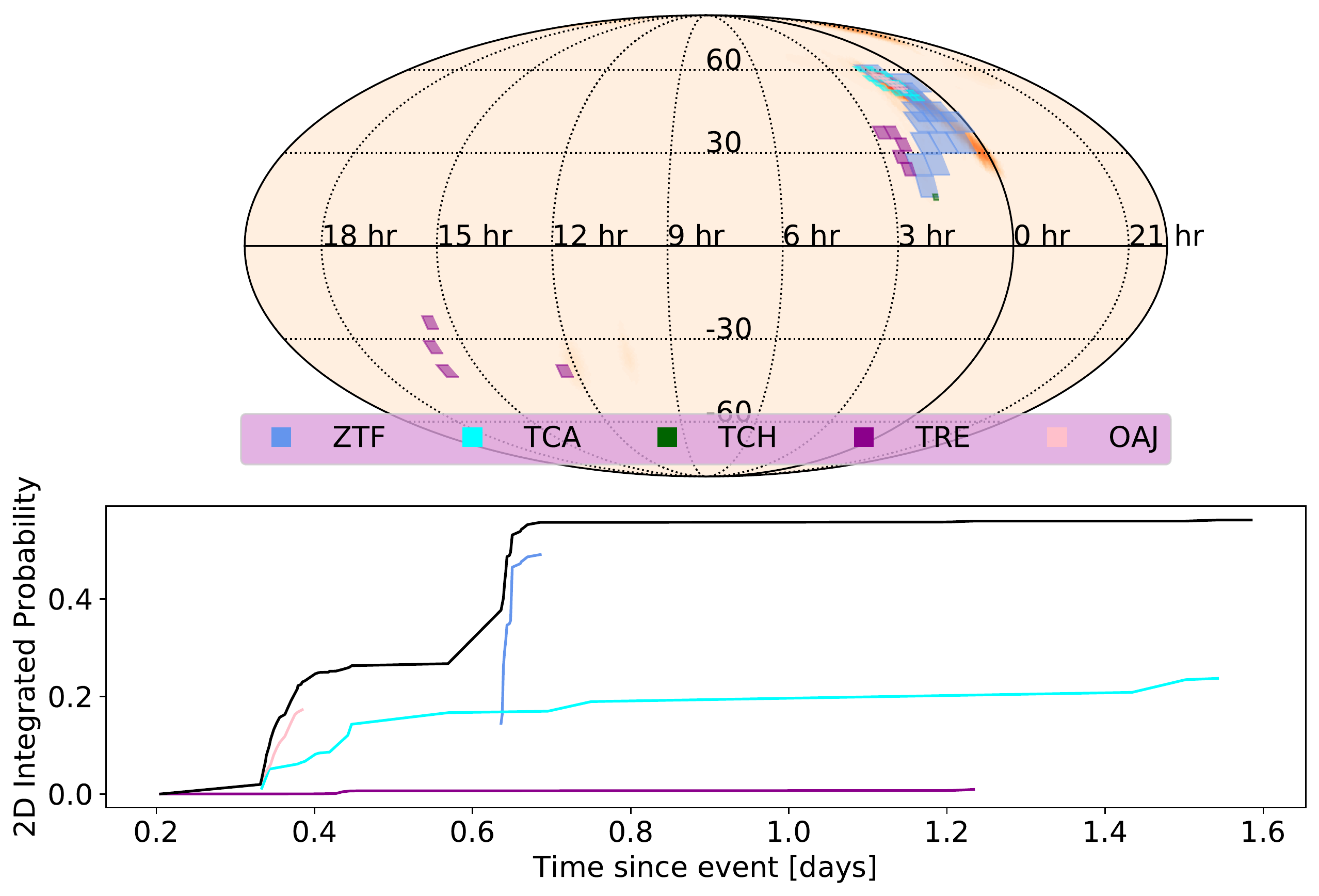}
 \caption{Coverage of the neutron star - black hole candidate S200115j (left column) and binary neutron star candidate S200213t (right column) within 12~hr (top row), 24~hr (middle row), and 48\,hr (bottom row) after the GW trigger time by ZTF \citep{gcn26767,gcn27051} and GRANDMA, including the TAROT (TCA, TCH and TRE) network and OAJ \citep{GRANDMA2020}. The LALInference localization probabilites are shown in shaded red. S200115j was detected at 2020-01-15 04:23:09.742 UTC, enabling immediate follow-up observations in South and North America (TCH and ZTF). S200213t was detected at 2020-02-13 at 04:10 UTC, offering only a few hours of observation for the European telescopes, such as for TCA, but a full night of observations with ZTF. OAJ could have begun observing immediately post-merger, but technical issues required human intervention and so the observations began only a few hours post-merger. We plot the integrated probability covered in the 2D skymap with solid lines. We show all telescopes combined in the black lines. The full list of observations is reported in Table~\ref{tab:Tableobs}.}
 \label{fig:S200213t_tiles}
\end{figure*}

In addition to the summaries below, we provide Table~\ref{tab:allevents}, displaying source properties based on publicly available information in GCNs and Table~\ref{tab:Tableobs}, displaying the results of follow-up efforts for the relevant candidates. All numbers listed regarding coverage of the localizations refer explicitly to the 90\% credible region. We treat S200115j as a BHNS candidate despite its official classification as a ``Mass-Gap'' event; it has p(HasRemnant) value close to 1, indicating the presence of a NS, but with a companion mass between 3 and 5 solar masses. In addition, we compare the limiting magnitudes and probabilities covered for S200115j and S200213t in Figure~\ref{fig:efficiency_2}, highlighted as example BHNS and BNS candidates with deep limits from a number of teams. As a point of reference, we include the apparent magnitude of an object with an absolute magnitude of $-16$ with distances ($\pm$\,1\,$\sigma$ error bars) consistent with the respective events. As a more physical visualization of the coordinated efforts that go into the follow-up process, we provide Figure~\ref{fig:S200213t_tiles}; this representation displays the tiles observed by various telescopes for the BNS merger candidate S200213t, along with a plot of the integrated probability and sky area that was covered over time by each of the telescopes. The black line is the combination of observations made by the telescopes indicated in the caption. These plots are also reminiscent of public, online visualization tools such as GWSky\footnote{https://github.com/ggreco77/GWsky}, the Transient Name Server (TNS)\footnote{\url{https://wis-tns.weizmann.ac.il}}, and the Gravitational Wave Treasure Map \citep{2020arXiv200100588W}.

\subsection{\textbf{S191205ah}}

LIGO/Virgo S191205ah was identified by the LIGO Hanford Observatory (H1), LIGO Livingston Observatory (L1), and Virgo Observatory (V1) at 2019-12-05
21:52:08 UTC \citep{gcn26350} with a false alarm rate of one in two years. It has been so far categorized as a BHNS signal $(93\%)$ with a small probability of being terrestrial $(7\%)$.
The distance is relatively far at $385 \pm 164$\,Mpc, and the event localization is coarse, covering nearly $6400$ square degrees. No update of the sky localisation and alert properties have been released by the LVC.

23 groups participated in the follow-up of the event including 3 neutrinos observatories (including IceCube and ANTARES; \citealt{gcn26352,gcn26349}), two VHE $\gamma$-ray observatories, eight $\gamma$-ray instruments, two X-ray telescopes and ten optical groups (see the \href{https://gcn.gsfc.nasa.gov/other/S191205ah.gcn3}{list of GCNs for S191205ah}). No candidates were found for the neutrinos, high-energy and $\gamma$-ray searches. Five of the optical groups have been engaged for the search of EM counterparts: GRANDMA network, MASTER network, SAGUARO, SVOM-GWAC, and the Zwicky Transient Facility (see Table~\ref{tab:Tableobs}). The MASTER-network led the way in covering a significant fraction of the localization area, observing $\approx56$\% down to 19 in a clear filter and within 144~h \citep{gcn26353}. Seven transient candidates were reported by the Zwicky Transient Facility \citep{gcn26416}, as well as four transient candidates reported by Gaia \citep{gcn26397}, and one candidate from the SAGUARO Collaboration \citep{gcn26360}, although none displayed particularly interesting characteristics encouraging further follow-up; all of the candidates for which spectra were obtained were ultimately ruled out as unrelated to S191205ah \citep{gcn26405,gcn26421,gcn26422}.

\subsection{\textbf{S191213g}}

LIGO/Virgo S191213g was identified by H1, L1, and V1 at 2019-12-13 04:34:08 UTC \citep{gcn26402}. It has been so far categorized as a BNS signal $(77\%)$ with a moderate probability of being terrestrial $(23\%)$, as well as a note that scattered light glitches in the LIGO detectors may have affected the estimated significance and sky position of the event.
As expected for BNS candidates, the distance is more nearby (initially $195 \pm 59$\,Mpc, later updated to be $201 \pm 81$\,Mpc with the LALInference map \citealt{gcn26417}). The updated map covered $\sim 4500$ square degrees. Since the updated skymap was released $\sim$\,1 day after trigger time, much of the observations made in the first night used the initial BAYESTAR map.

While it was the first BNS alert during the second half of the O3 campaign, the response to this alert was relatively tepid, likely due to the scattered light contamination. 
However, 53 report circulars have been distributed for this event due to the presence of an interesting transient found by the Pan-STARRS Collaboration PS19hgw/AT2019wxt, finally classified as supernovae IIb due to the photometry evolution and spectroscopy characterization \citep{gcn26485,gcn26508, GRANDMA2020}. 
In total, three neutrinos, one VHE, eight $\gamma$-rays, two X-rays, 19 optical and one radio groups participated to the S191213g campaign (see the \href{https://gcn.gsfc.nasa.gov/other/S191213g.gcn3}{list of GCNs for S191213g}). No significant neutrino, VHE and $\gamma$-ray GW counterpart was found in the archival analysis. A moderate fraction of the localization area was covered using a tiling approach (GRANDMA, Master-Network, ZTF)  (see Table~\ref{tab:Tableobs}). The MASTER-network covered $\approx41$\% within 144~h  down to 19 in clear \citep{gcn26400}, and the Zwicky Transient Facility covered $\approx28$\% down to 20.4 in $g$- and $r$-band \citep{gcn26424,gcn26437,Kasliwal2020}. 
The search yielded 19 candidates of interest from ZTF, as well as the transient counterpart AT2019wxt from the Pan-STARRS Collaboration \citep{gcn26485}. It was shown that all ZTF candidates were in fact unrelated with the GW candidate S191213g \citep{gcn26426,gcn26429,gcn26432}. %ZTF19acykzsk was interesting due to its red color ($g-r$=0.3) and absolute magnitude of $-15.5$ \citep{gcn26424}, but was later classified as a SN II at $z=0.02$ \citep{gcn26427}. ZTF19acyldun was detected one day after the merger, and the spectroscopic redshift of its host galaxy ($z=0.057$) was consistent with the distance estimate for the event \citep{gcn26437}; however, subsequent photometry showed that ZTF19acyldun's behaviour was not consistent with that of a decaying short GRB \citep{gcn26450}. 

In addition to searches by wide field of view telescopes, there was also  galaxy-targeted follow-up performed by the J-GEM Collaboration, observing 57 galaxies \citep{gcn26477}, and the GRANDMA citizen science program, observing 16 galaxies \citep{gcn26558} within the localization of S191213g. 

\subsection{\textbf{S200105ae}}

LIGO/Virgo S200105ae was identified by L1 (with V1 also observing) at 2020-01-05 16:24:26 UTC as a subthreshold event with a false alarm rate of 24 per year; if it is astrophysical, it is most consistent with being an BHNS. However, its significance is likely underestimated due to it being a single-instrument event. This candidate was most interesting due to the presence of chirp-like structure in the spectrograms \citep{gcn26640, gcn26657}. The first public notice was delivered 27.2~h after the GW trigger impacting significantly the follow-up campaign of the event. In addition, the most updated localization was very coarse, spanning $\sim7400$ square degrees with a distance of $283 \pm 74$\,Mpc \citep{gcn26688}.

S200105ae follow-up activity was comparable to S191205ah's: 25 circular reports were associated to the S200105ae in the GCN service with the search of counterpart engaged by two neutrinos, one VHE, seven $\gamma$-ray, one X-ray and five optical groups (see the \href{https://gcn.gsfc.nasa.gov/other/S200105ae.gcn3}{list of GCNs for S200105ae}). No significant neutrino, VHE and $\gamma$-ray GW counterpart was found in the archival analysis.
Various groups participated to the search of optical counterpart with ground-based observatories: GRANDMA, Master-Network, and the Zwicky Transient Facility (see Table~\ref{tab:Tableobs}). The alert space system for Gaia was also activated \citep{gcn26686}. The MASTER-network covered $\approx43$\% down to 19.5 in clear and within 144~h \citep{gcn26646}. The telescope network was already observing at the time of the trigger and because its routine observations were compatible with the sky localization of S200105ae, the delay was limited to 3\,hr. GRANDMA-TCA telescope was triggered as soon as the notice comes out, and the full GRANDMA network totalized 12.5 \% of the full LALInference skymap down to 17 mag in clear and within 60~h \citep{GRANDMA2020}. The Zwicky Transient Facility covered $\approx52$\% of the LALInference skymap down to 20.2 in both $g$- and $r$-bands \citep{gcn26673,Kasliwal2020} and with a delay of 10~h. There were 23 candidate transients reported by ZTF, as well as one candidate from the Gaia Alerts team \citep{gcn26686} out of which ZTF20aaervoa and ZTF20aaertpj were both quite interesting due to their red colors ($g-r$= 0.66 and 0.35 respectively), and absolute magnitudes ($-16.4$ and $-15.9$ respectively) \citep{gcn26673}. ZTF20aaervoa was soon classified as a SN IIp $\sim3$ days after maximum, and ZTF20aaertpj as a SN Ib close to maximum \citep{gcn26702,gcn26703}. 

\subsection{\textbf{S200115j}}

%\rednote{BE CAREFUL ABOUT DISCUSSION}

LIGO/Virgo S200115j, a MassGap signal $(99\%)$ with a very high probability $(99\%)$ of containing a NS as well, was identified by H1, L1, and V1 at 2020-01-15 04:23:09.742 UTC \citep{gcn26759}. As discussed before, it can be considered as a BHNS candidate. Due to its discovery by multiple detectors, the sky location is well-constrained; the most updated map spans $\sim765$ square degrees, with most of the probability shifting towards the southern lobe in comparison to the initial localization, and has a distance of $340 \pm 79$\, Mpc.

With a very high p$_{\mathrm{remnant}} > 99$\% \citep{gcn26807} and good localization, many space and ground instruments/telescopes followed up this signal: 33 circular reports were associated to the event in the GCN service with the search of counterpart engaged by two neutrinos, three VHE, five $\gamma$-ray, two X-ray and eight optical groups (see the \href{https://gcn.gsfc.nasa.gov/other/S200115j.gcn3}{list of GCNs for S200115j}). INTEGRAL was not active during the time of the event \citep{gcn26766} and so was unable to report any prompt short GRB emission. No significant neutrino, VHE and $\gamma$-ray GW counterpart was found in the archival analysis. Swift satellite was also pointed toward the best localization region for finding X-ray and UVOT counterpart. Some candidates were reported: one of them was detected in the optical by Swift/UVOT and the Zwicky Transient Facility, but was concluded to likely be due to AGN activity \citep{gcn26808,gcn26863}.

Various groups participated to the search of optical counterpart with ground observatories: GOTO, GRANDMA, Master-Network, Pan-Starrs, SVOM-GWAC and the Zwicky Transient Facility (see Table~\ref{tab:Tableobs}). GOTO \citep{gcn26794} covered $\approx50$\% down to 19.5 in $G$-band, starting almost immediately the observations, while the SVOM-GWAC team covered $\approx40$\% of the LALInference sky localization down to 16 in R-band using the SVOM-GWAC only 16h after the trigger time \citep{gcn26786}.

 In addition, a list of 20 possible host galaxies for the trigger was produced by convolving the GW localization with the 2MPZ galaxy catalog \citep{gcn26763,BiJa2014}; 12 of these galaxies were observed by GRAWITA \citep{gcn26823} in the r-sdss filter. %In addition, Swift-XRT observations covering a significant portion of the localization (after convolution with the 2MPZ galaxy catalog) resulted in multiple ``rank 2''\footnote{https://www.swift.ac.uk/ranks.php} X-ray sources \citep{gcn26798,gcn26855}. One of the Swift X-ray candidates was detected in the optical by Swift/UVOT and the Zwicky Transient Facility, but was concluded to likely be due to AGN activity \citep{gcn26808,gcn26863}. In addition, observations made with the PanSTARRS2 telescope yielded eight reported candidates \citep{gcn27082}, one of which showed a ~0.5mag rise, but none were obviously related to the event. INTEGRAL was not active during the time of the event \citep{gcn26766} and so was unable to report any prompt short GRB emission.

\subsection{\textbf{S200213t}}

S200213t was identified by H1, L1, and V1 at 2020-02-13 at 04:10:40 UTC \citep{gcn27042}. It has been categorized as a BNS signal $(63\%)$ with a moderate probability of being terrestrial $(37\%)$. The LALInference localization spanned $\sim2326$ square degrees, with a distance of $201 \pm 80$\, Mpc \citep{gcn27096}.
A total of 51 circular reports were associated to this event including two neutrinos, two VHE, eight $\gamma$-rays, two X-ray, and eleven optical groups (see the \href{https://gcn.gsfc.nasa.gov/other/S200213t.gcn3}{list of GCNs for S200213t}). Fermi and Swift were both transiting the South Atlantic Anomaly at the time of event, and so were unable to observe and report any GRBs coincident with S200213t \citep{gcn27056,gcn27058}.  No significant counterpart candidate was found during archival analysis: IceCube detected muon neutrino events, but it was shown that they have not originated from the GW source \citep{gcn27043}.

With a very high p$_{\mathrm{remnant}} > 99$\% and probable BNS classification, many telescopes followed-up this signal: DDOTI/OAN, GOTO, GRANDMA, MASTER and ZTF. DDTOI/OAN covered $\approx40$\% of the LALInference skymap starting less than 1h after the trigger time down to 19.2 in $w$-band \citep{gcn27061}, GRANDMA covered 32\% of the LALInference area within $\approx$ 26h down to 18 mag in clear (TCA) and down to 21 mag in R-band (OAJ). GOTO covered $\approx54$\% of bayestar skymap down to 18.4 in $G$-band \citep{gcn27069}. 15 candidate transients were reported by ZTF \citep{gcn27051,gcn27065,gcn27068}, as well as one by the MASTER-network \citep{gcn27077}. All were ultimately ruled out as possible counterparts to S200213t through either spectroscopy or due to pre-discovery detections \citep{gcn27060,gcn27063,gcn27074,gcn27075,gcn26839,gcn27085}. Galaxy targeted observations were conducted by several observatories: examples include KAIT, which observed 108 galaxies \citep{gcn27064}, Nanshan-0.6m, which observed a total of 120 galaxies \citep{gcn27070}, in addition to many other teams \citep{gcn27066,gcn27067}.

\section{Kilonova Modeling and possible ejecta mass limits}
\label{sec:limits}

Following \cite{CoDi2019b}, we will compare the upper limits described in Section~\ref{sec:EM_follow_up_campaigns} to different kilonova models. We seek to measure ``representative constraints,'' limited by the lack of field and time-dependent limits. To do so, we approximate the upper limits in a given passband as one-sided Gaussian distributions.
We take the sky-averaged distance in the GW localizations to determine the transformation from apparent to absolute magnitudes. To include the uncertainty in distance, we sample from a Gaussian distribution consistent with this uncertainty and add it to the model lightcurves.
In this analysis, we employ three kilonova models based on \cite{KaMe2017}, \cite{Bul2019}, and \cite{HoNa2019}, in order to compare any potential systematic effects. These models use similar heating rates \citep{MeMa2010,KoRo2012}, while using different treatments of the radiative transfer. 

\clearpage

\hspace{-0.5in}
\begin{sidewaysfigure}
\myRule[white]{0.75\textheight}{0.5\textheight}
%\begin{figure*}[t]
\centering
 \caption{Probability density for the total ejecta mass for the BNS events, S191213g and S200213t, and employed lightcurve models. From the left to the right, we show constraints as a function of lanthanide fraction for the \cite{KaMe2017} Model, as a function of inclination angle (from a polar orientation, system viewed face-on) for the \cite{Bul2019} Model, and as a function of the opacity of the 2-components, $\kappa_{\rm low}$ and $\kappa_{\rm high}$ for the \cite{HoNa2019} Model. The dashed lines are the upper limits that contain 90\% of the probability density. For S191213g, we use ZTF \citep{gcn26424,gcn26437} and the MASTER-Network \citep{gcn26400}. For S200213t, we use ZTF \citep{gcn27051} and GOTO \citep{gcn27069}. 
 See Table~\ref{tab:Tableobs} for further details.}
\hspace{10.5in} \textbf{Model I} \hspace{2.5in} \textbf{Model II} \hspace{2.5in} \textbf{Model III} \\
\vspace{0.1in}
 \textbf{S191213g} \\ 
 \includegraphics[width=3.0in,height=2.2in]{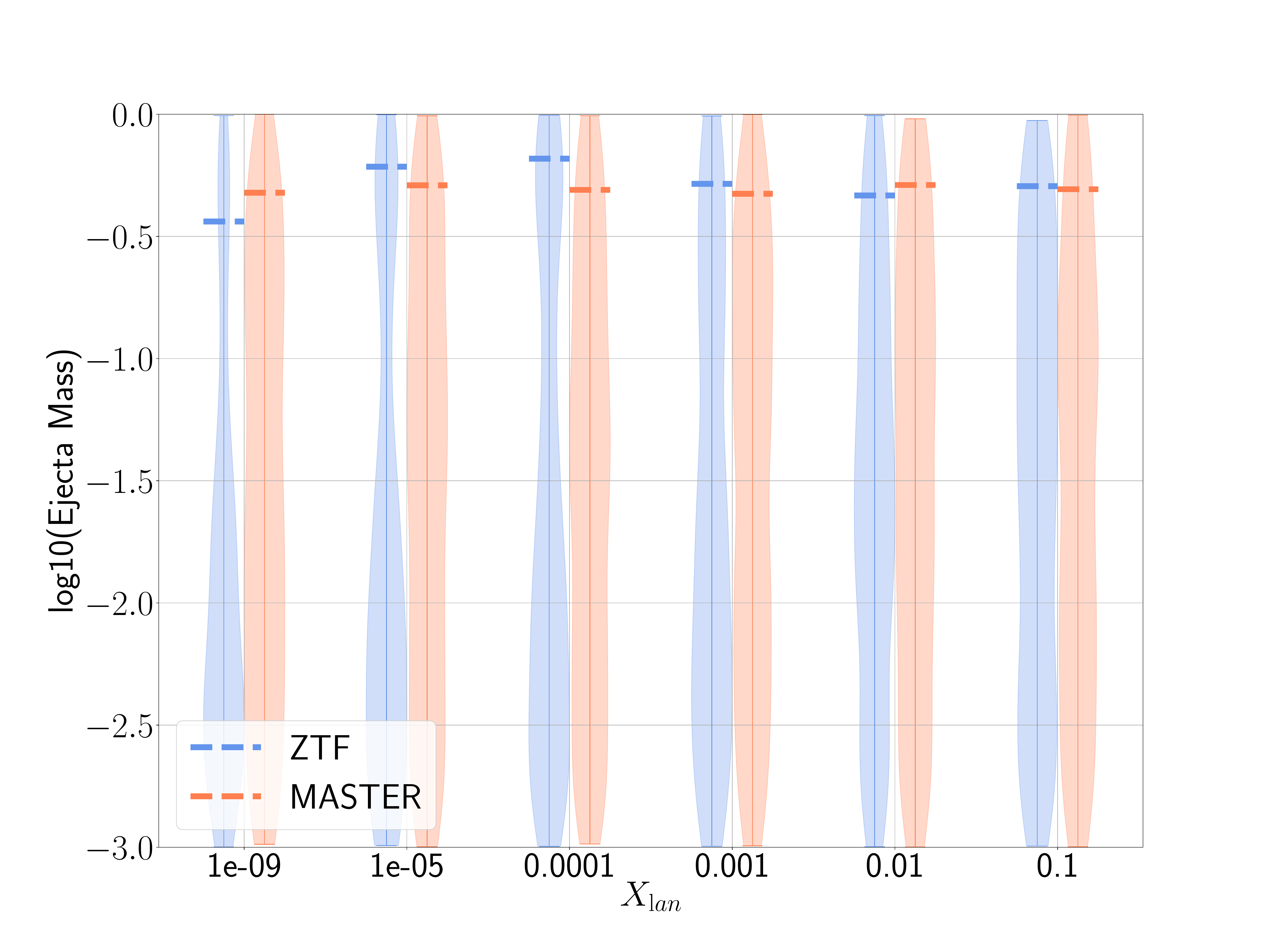}
 \includegraphics[width=3.0in,height=2.2in]{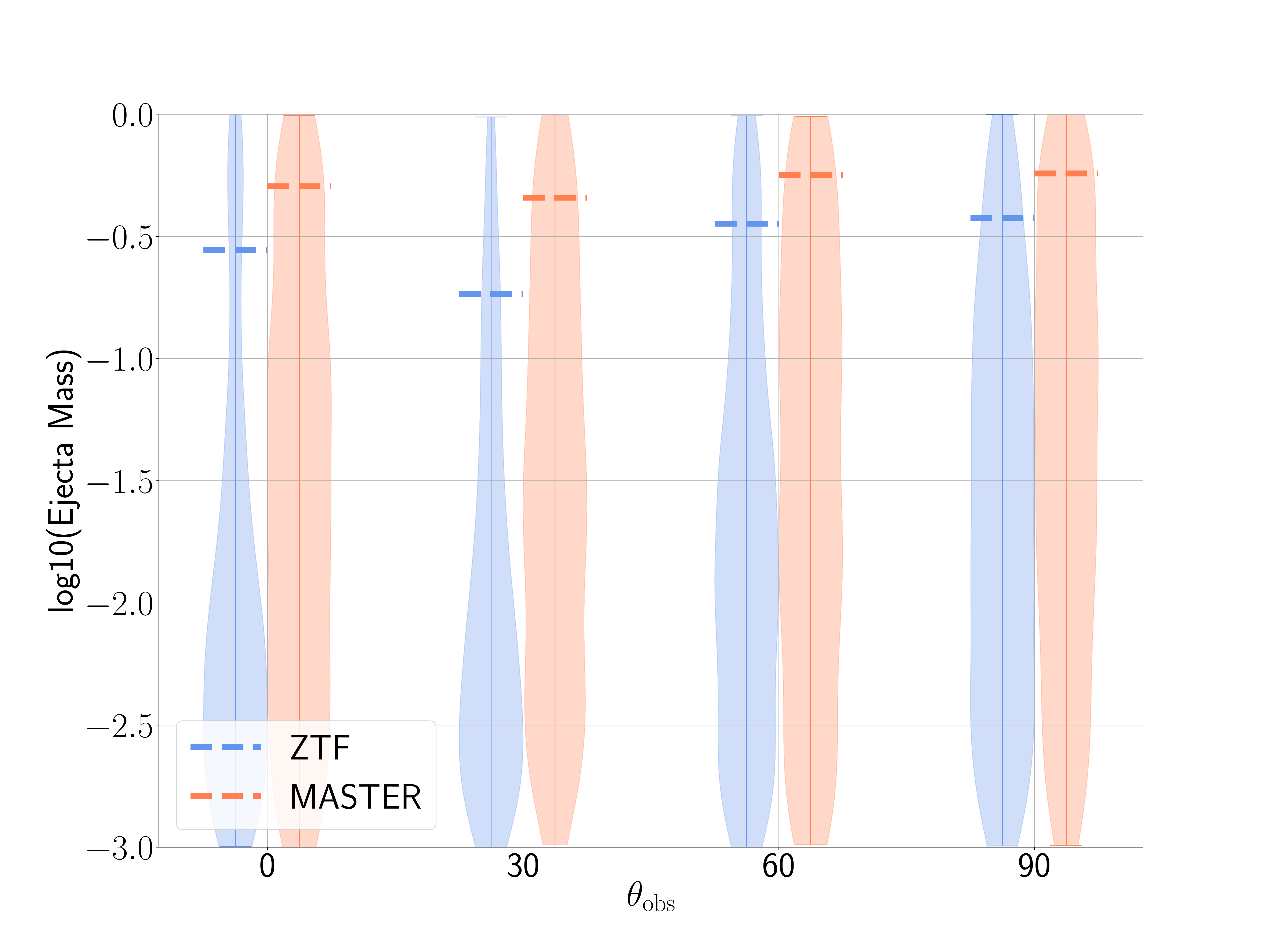}
 \includegraphics[width=3.0in,height=2.2in]{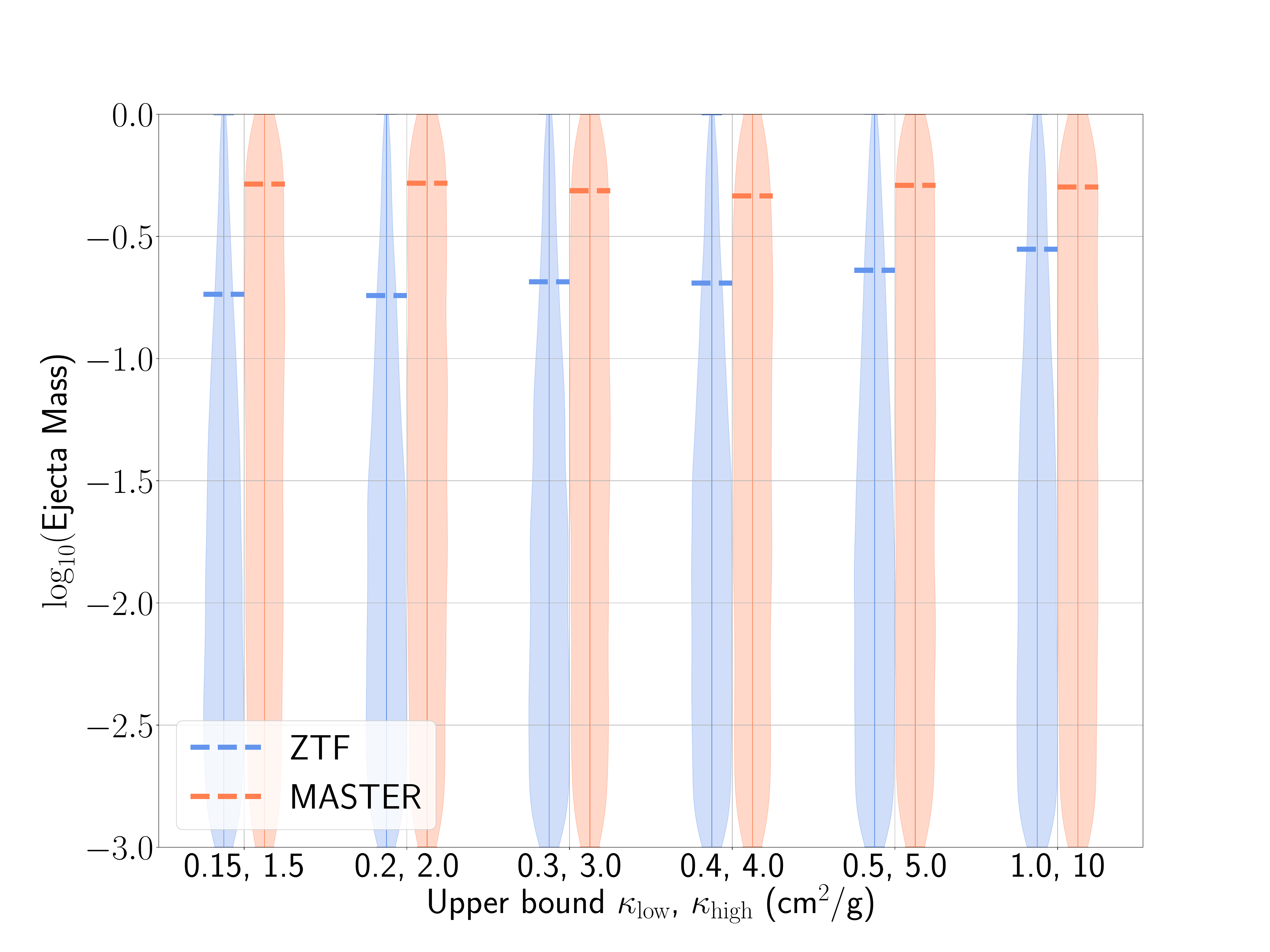}\\
 \textbf{S200213t} \\ 
 \includegraphics[width=3.0in,height=2.2in]{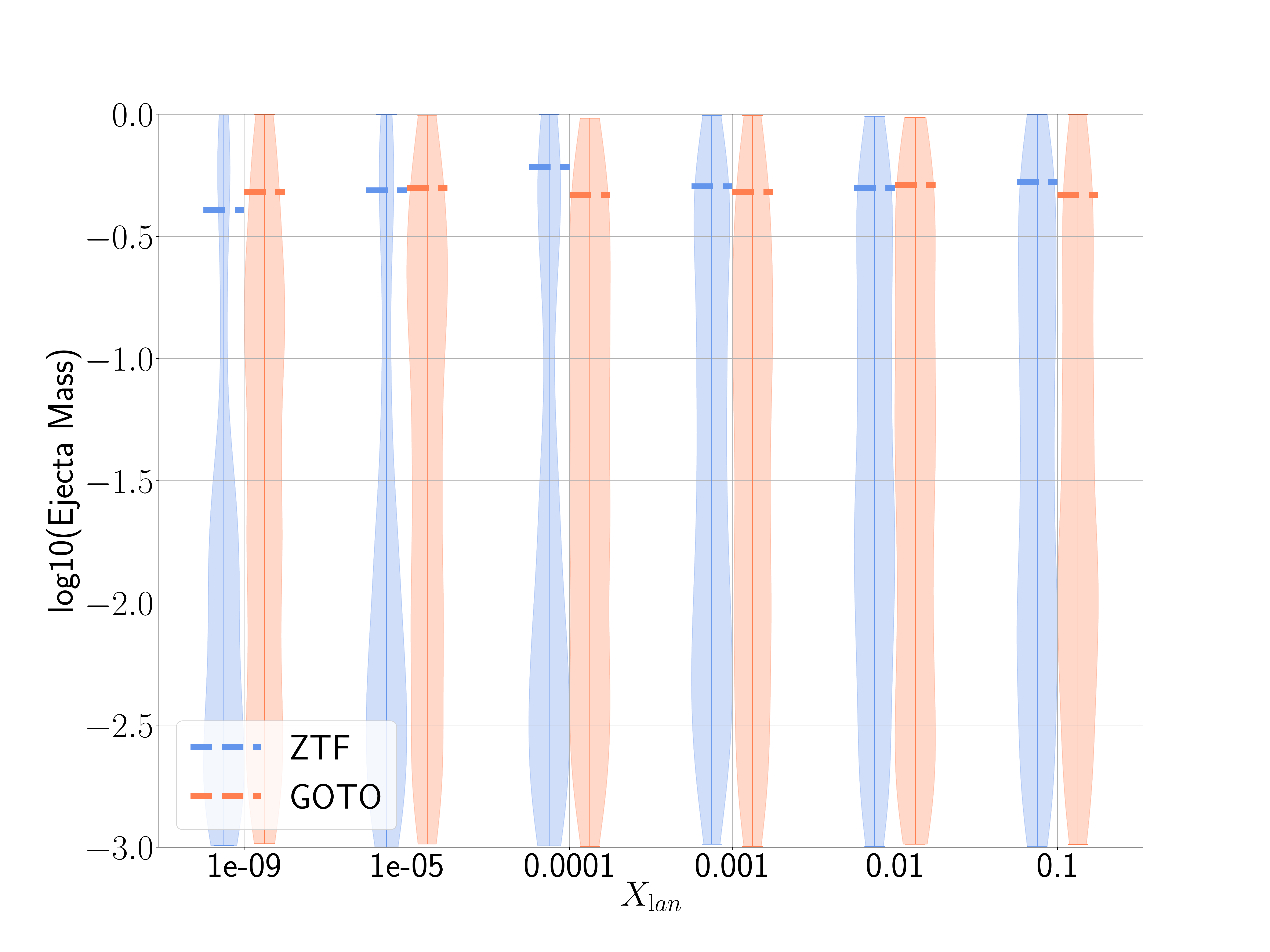}
 \includegraphics[width=3.0in,height=2.2in]{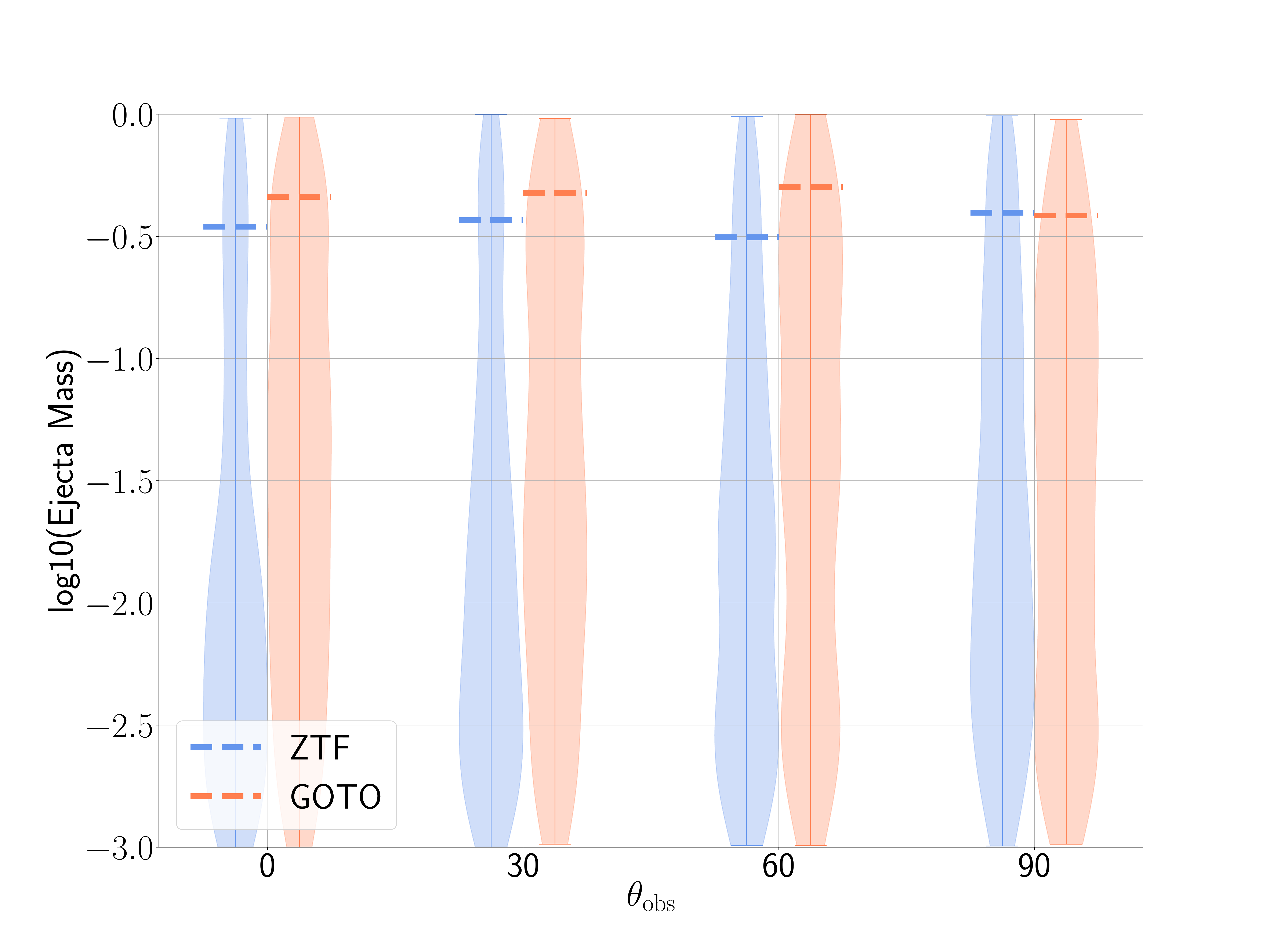}
 \includegraphics[width=3.0in,height=2.2in]{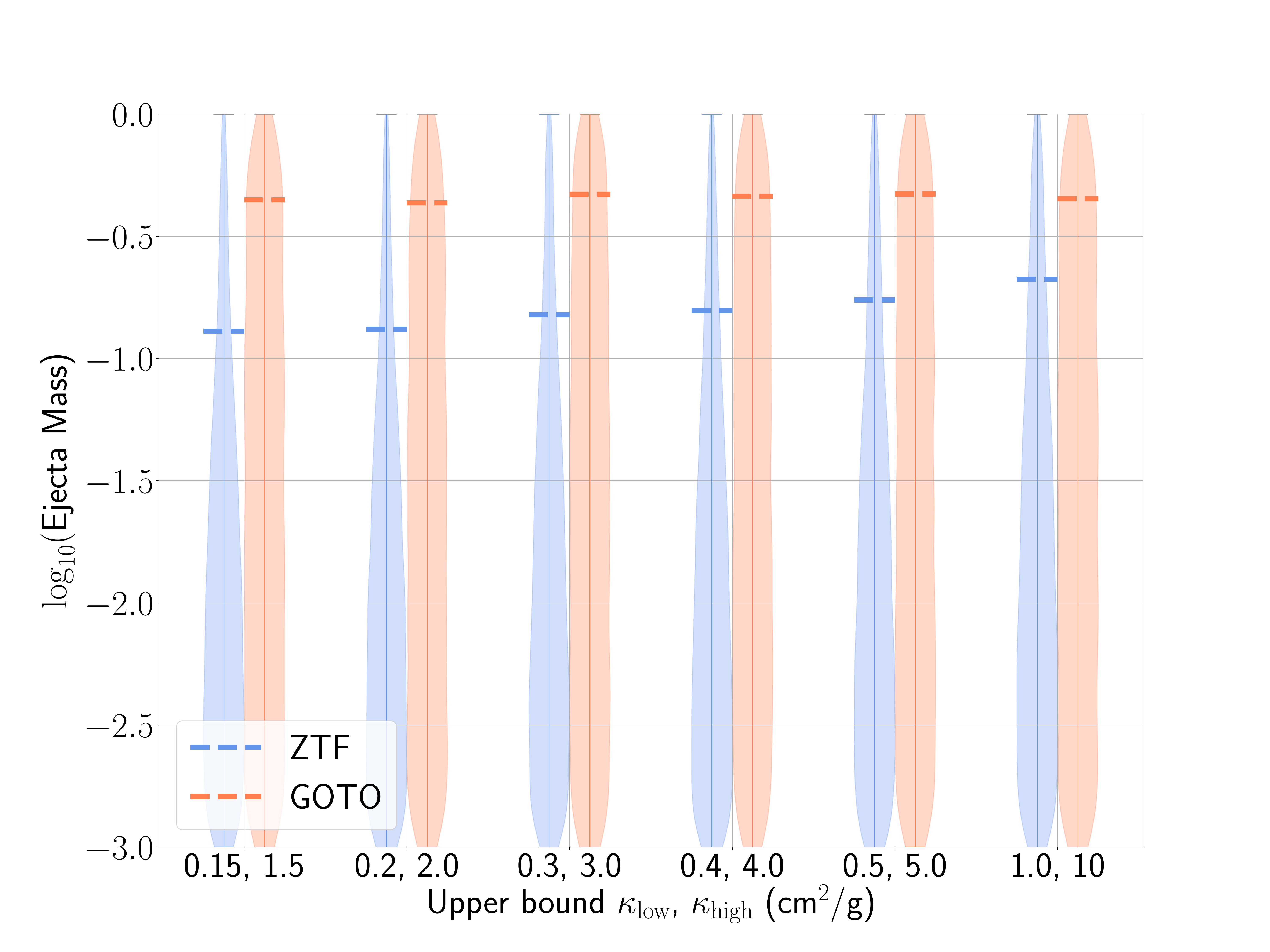}
 \label{fig:violin_constraints_BNS}
%\end{figure*}
\end{sidewaysfigure}
\hspace{0.5in}

\clearpage

\setlength{\voffset}{-0.5in}
\begin{sidewaysfigure}

%\begin{figure*}[t]
\centering
%\myRule[white]{0.75\textheight}{0.2\textheight}
%\begin{figure*}[t]
\centering
 \caption{Probability density for the total ejecta mass for the NSBH events, S191205ah, S200105ae, and S200115j, and employed lightcurve models. From the left to the right, we show constraints as a function of lanthanide fraction for the \cite{KaMe2017} Model, as a function of inclination angle (from a polar orientation, system viewed face-on) for the \cite{Bul2019} Model, and as a function of the opacity of the 2-components, $\kappa_{\rm low}$ and $\kappa_{\rm high}$ for the \cite{HoNa2019} Model. The dashed lines are the upper limits that contain 90\% of the probability density. For S191205ah, we use ZTF (left, \citealt{gcn26416}) and SAGUARO (right, \citealt{gcn26360}). 
 For S200105ae, we use ZTF \citep{gcn26673} and the MASTER-network \citep{gcn26646}. 
 For S200115j, we use ZTF \citep{gcn26767} and GOTO \citep{gcn26794}
 See Table~\ref{tab:Tableobs} for further details.}
\hspace{10.5in} \textbf{Model I} \hspace{2.5in} \textbf{Model II} \hspace{2.5in} \textbf{Model III} \\
\vspace{0.1in}
 \textbf{S191205ah} \\ 
 \includegraphics[width=3.0in,height=2.2in]{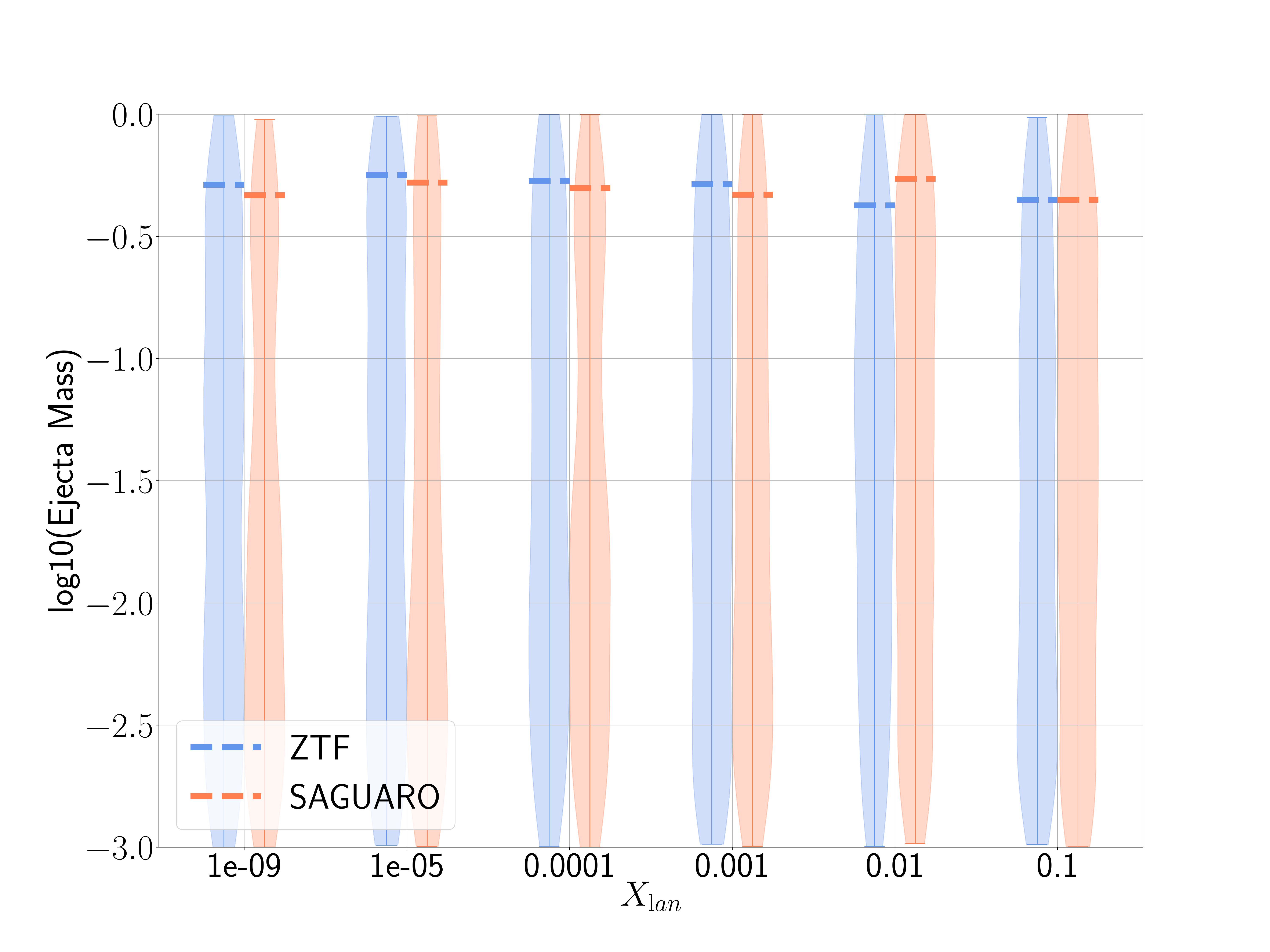}
 \includegraphics[width=3.0in,height=2.2in]{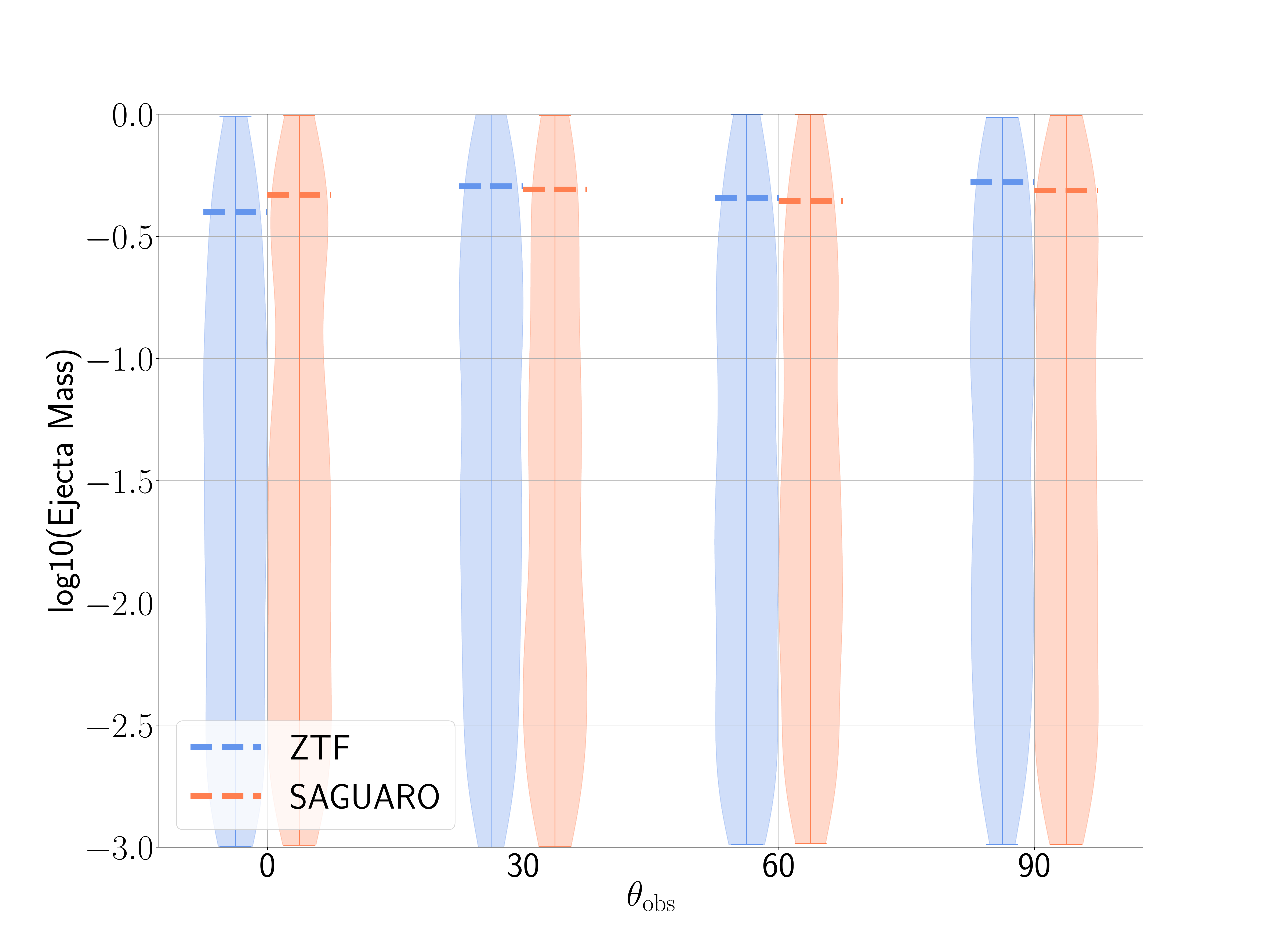}
 \includegraphics[width=3.0in,height=2.2in]{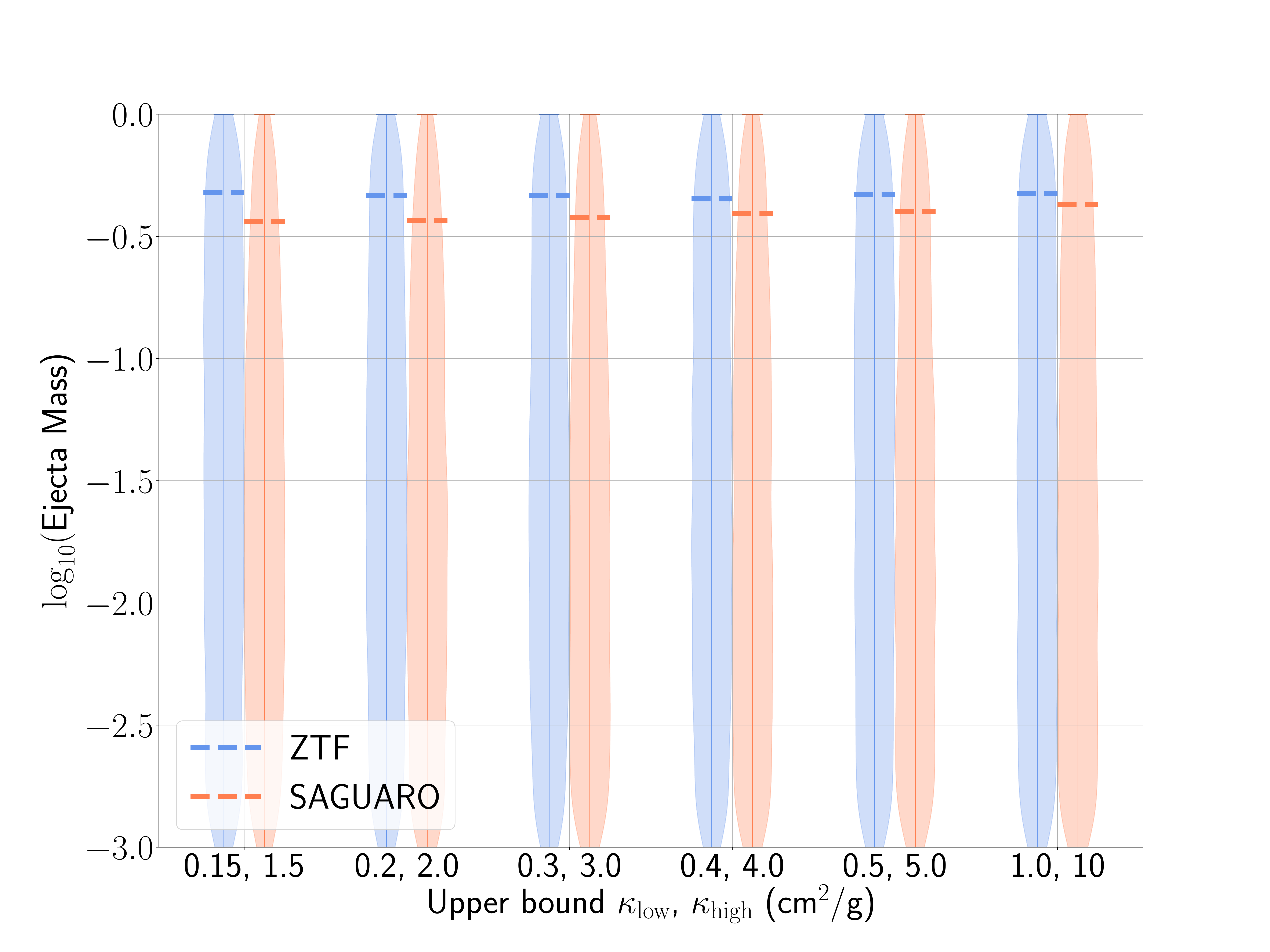}\\
 \textbf{S200105ae} \\ 
 \includegraphics[width=3.0in,height=2.2in]{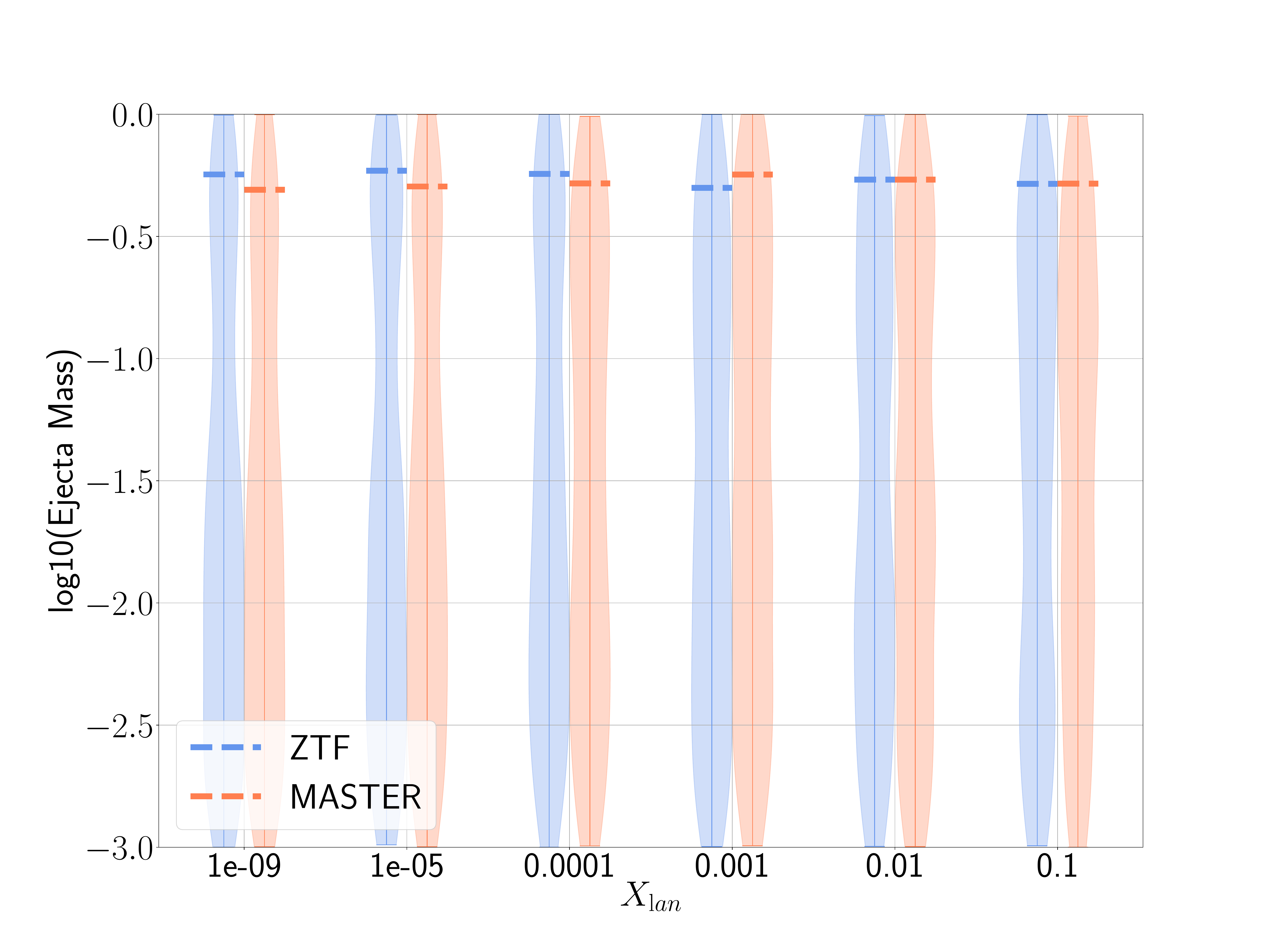}
 \includegraphics[width=3.0in,height=2.2in]{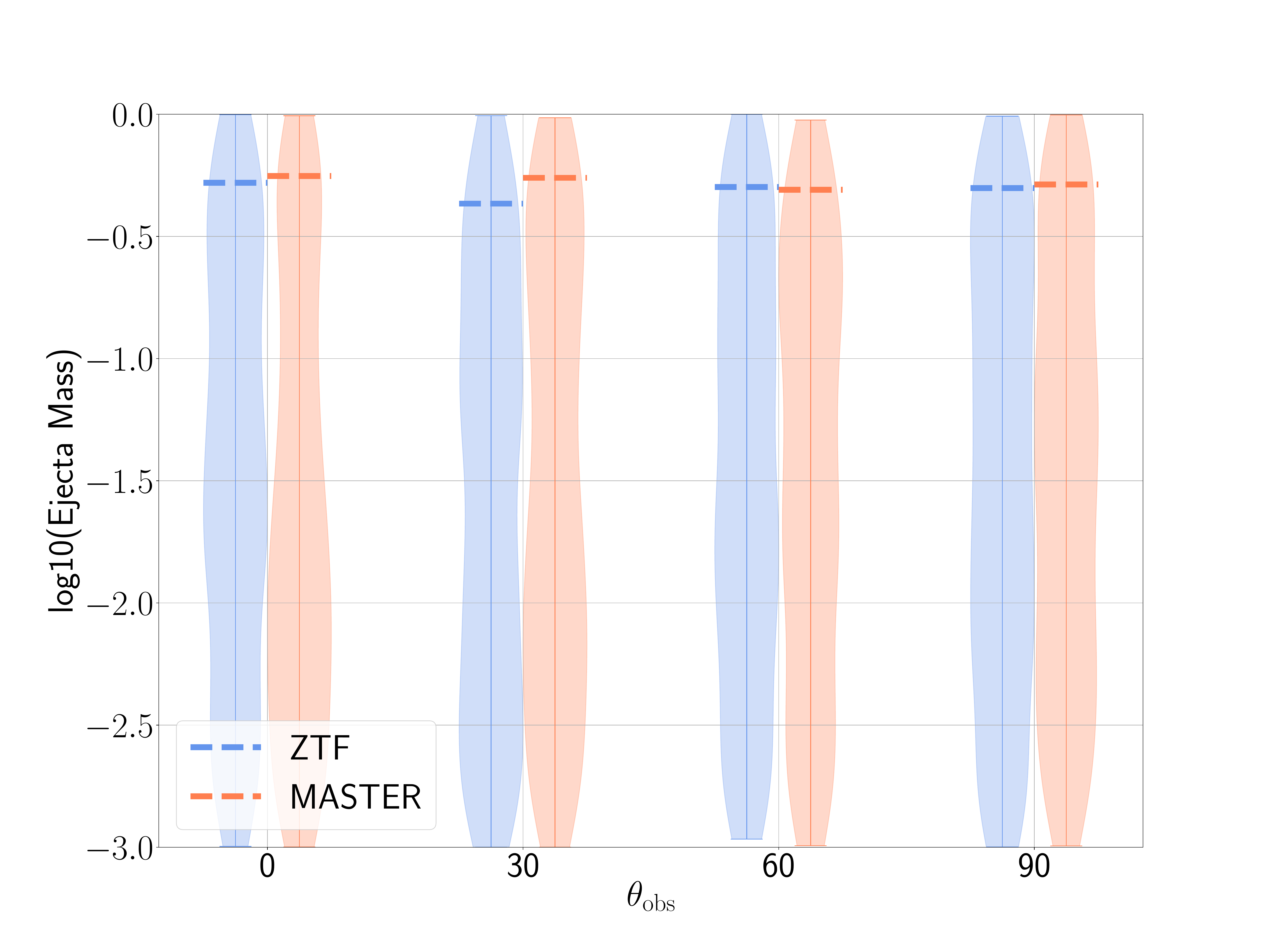}
 \includegraphics[width=3.0in,height=2.2in]{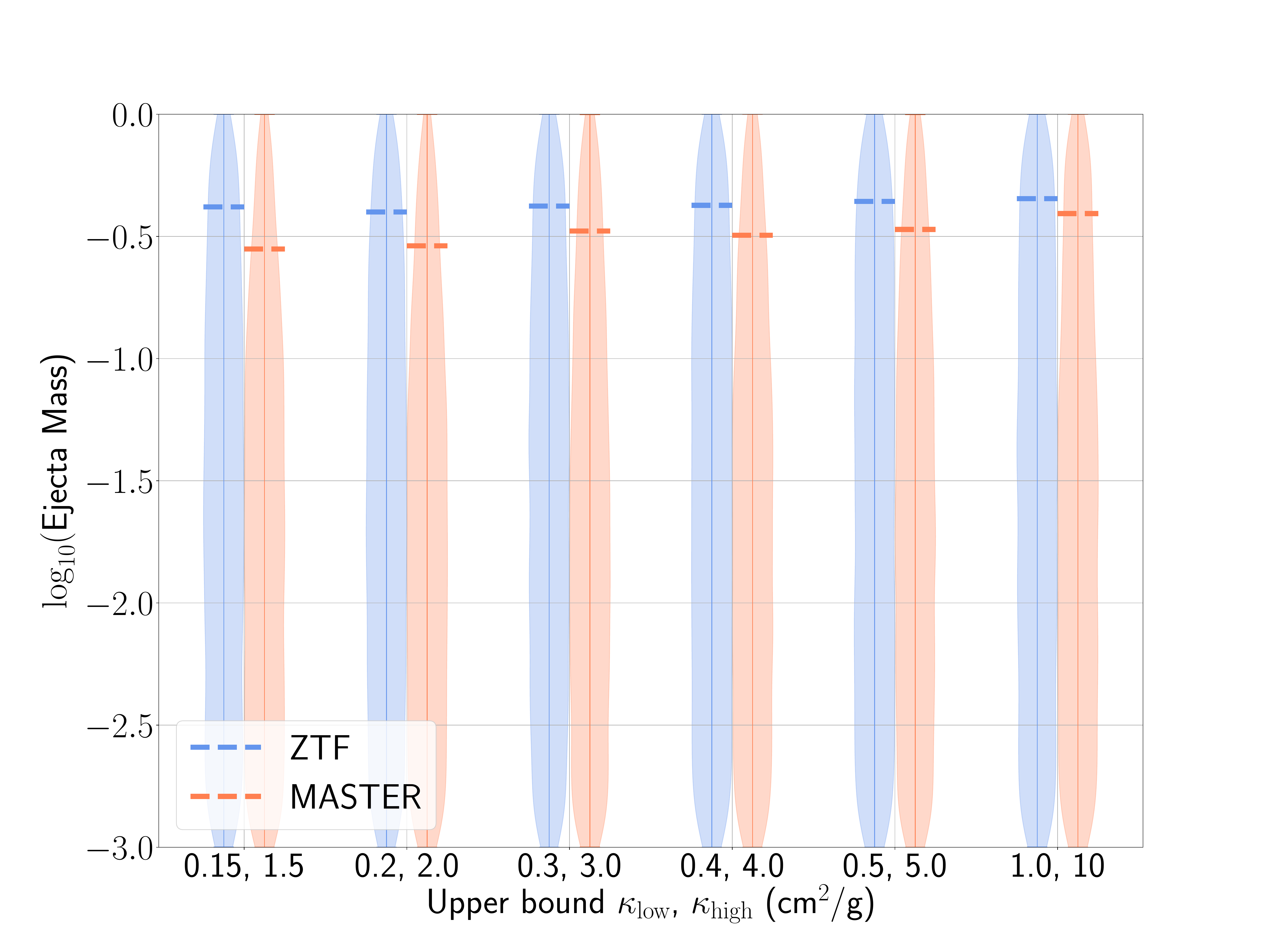}\\
 \textbf{S20015j} \\ 
 \includegraphics[width=3.0in,height=2.2in]{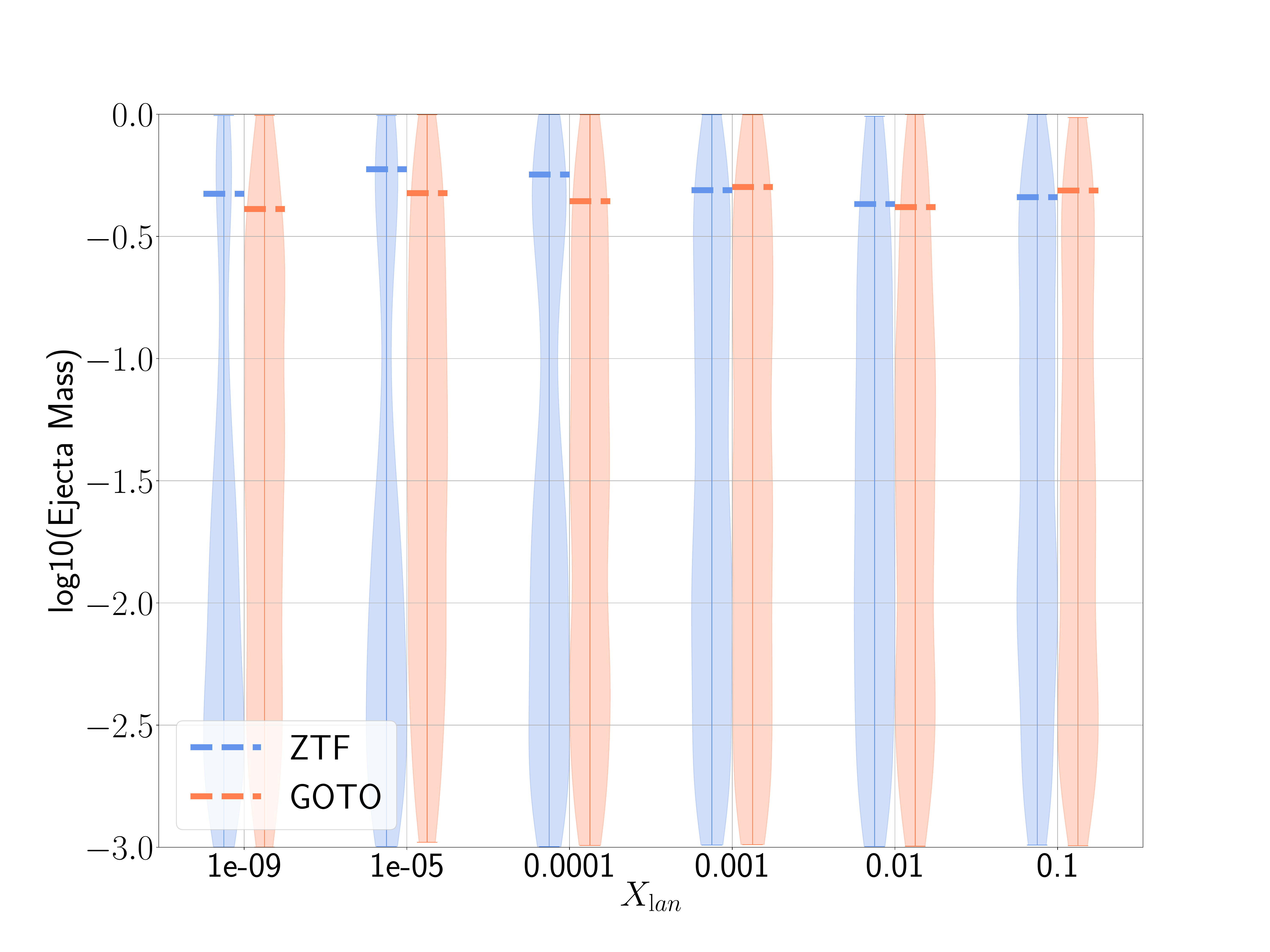}
 \includegraphics[width=3.0in,height=2.2in]{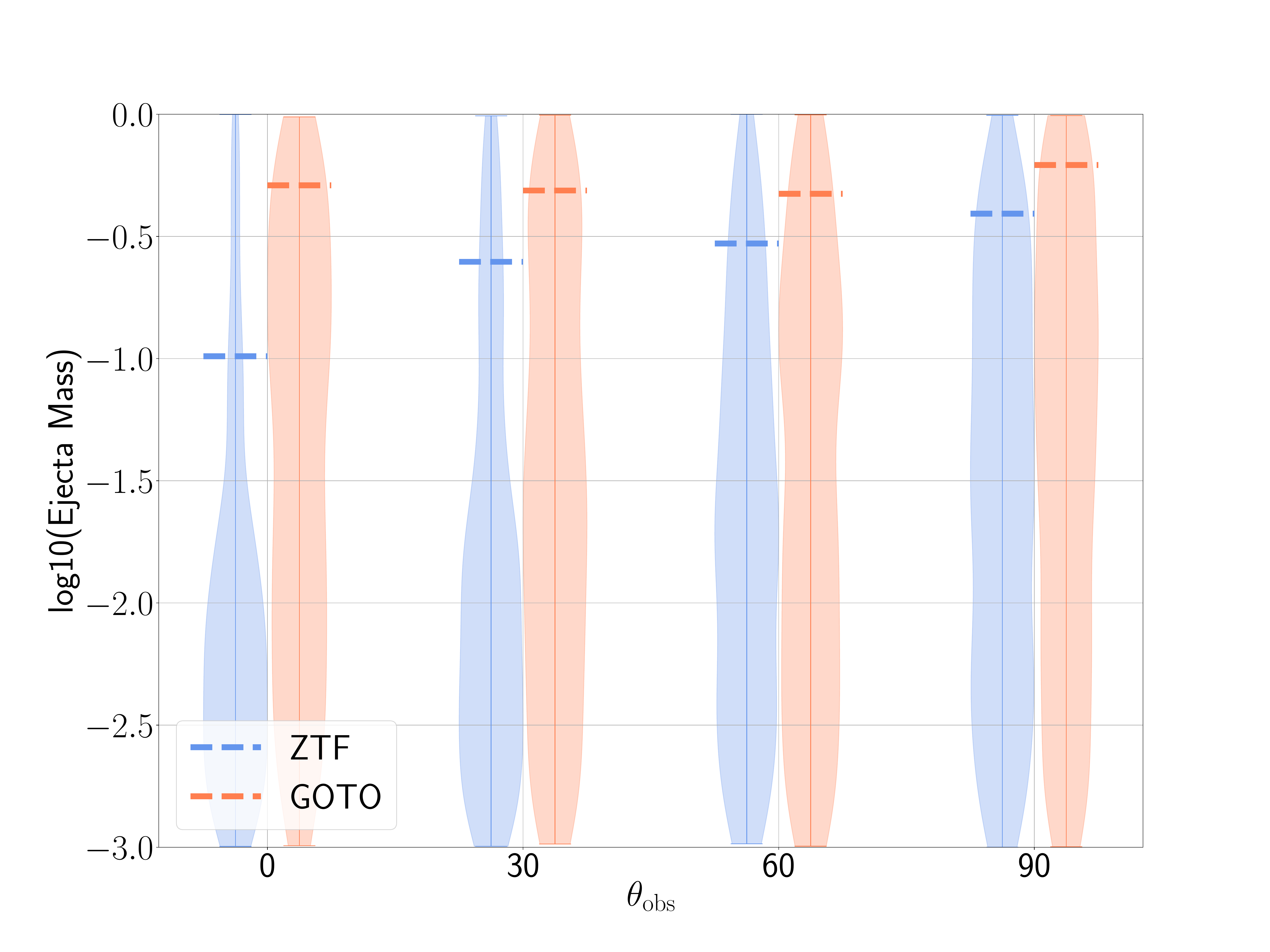}
 \includegraphics[width=3.0in,height=2.2in]{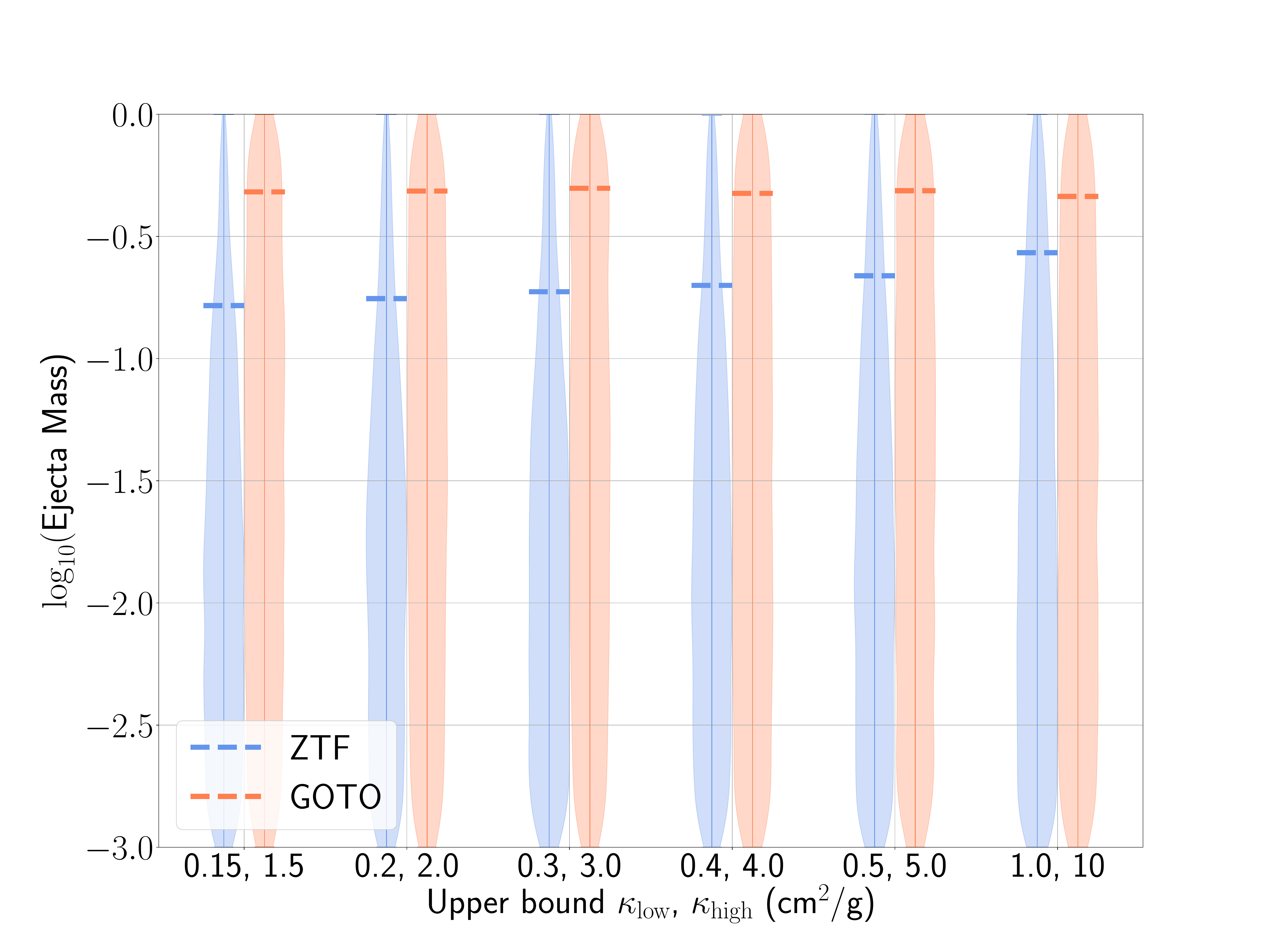} \\
 \label{fig:violin_constraints_NSBH}
%\end{figure*}
\myRule[white]{0.75\textheight}{0.4\textheight}
\end{sidewaysfigure}

\clearpage

We will show limits as a function of one parameter for each model chosen to maximize its impact on the predicted kilonova brightness and color, marginalizing out the other parameters when performing the sampling. 
For the models based on \cite{KaMe2017} and \cite{Bul2019}, as grid-based models, we interpolate these models by creating a surrogate model using a singular value decomposition (SVD) and Gaussian Process Regression (GPR) based interpolation \citep{DoFa2017} that allows us to create lightcurves for arbitrary ejecta properties within the parameter space of the model ~\citep{CoDi2018b,CoDi2018}. 
We refer the reader to \cite{CoDi2019b} for more details about the models, but we will also briefly describe them in the following for completeness.

Model I \citep{KaMe2017} depends on the ejecta mass $M_{\rm ej}$, the mass fraction of lanthanides $X_{\rm lan}$, and the ejecta velocity $v_{\rm ej}$. We allow the sampling to vary within $-3 \leq \log_{10} (M_{\rm ej}/M_\odot) \leq 0$ and $ 0 \leq v_{\rm ej} \leq 0.3$\,$c$, while restricting the lanthanide fraction to $X_{\rm lan}$ = [ $10^{-9}$, $10^{-5}$, $10^{-4}$, $10^{-3}$, $10^{-2}$, $10^{-1} ]$.

Model II \citep{Bul2019} assumes an axi-symmetric geometry with two ejecta components, one component representing the dynamical ejecta and one the post-merger wind ejecta. Model II depends on four parameters: the dynamical ejecta mass $M_{\rm ej,dyn}$, the post-merger wind ejecta mass $M_{\rm ej,pm}$, the half-opening angle of the lanthanide-rich dynamical-ejecta component $\phi$ and the inclination angle $\theta_{\rm obs}$ (with $\cos\theta_{\rm obs}=0$ and $\cos\theta_{\rm obs}=1$ corresponding to a system viewed edge-on and face-on, respectively). We refer the reader to \cite{DiCo2020} for a more detailed discussion of the ejecta geometry. In this study, we fix the dynamical ejecta mass to the best-fit value from \cite{DiCo2020}, $M_{\rm ej,dyn}=0.005\,M_\odot$, and allow the sampling to vary within $-3 \leq \log_{10} (M_{\rm ej,pm}/M_\odot) \leq 0$ and $0^\circ\leq \phi\leq90^\circ$, while restricting the inclination angle to $\theta_{\rm obs} = [0^\circ, 30^\circ, 60^\circ, 90^\circ]$. To facilitate comparison with the other models, we will provide constraints on the total ejecta mass $M_{\rm ej}=M_{\rm ej,dyn}+M_{\rm ej,pm}$ for Model II. We note that the model adopted here is more tailored to BNS than BHNS mergers given the relatively low dynamical ejecta mass, $M_{\rm ej,dyn}=0.005\,M_\odot$. However, for a given $M_{\rm ej,pm}$, the larger values of $M_{\rm ej,dyn}$ predicted in BHNS are expected to produce longer lasting kilonovae more easily detectable. Therefore, the ejecta mass upper limits derived below for BHNS systems should be considered conservative.

Model III \citep{HoNa2019} depends on the ejecta mass $M_{\rm ej}$, the dividing velocity between the inner and outer component $v_{\rm ej}$, the lower and upper limit of the velocity distribution $v_{\text{min}}$ and $v_{\text{max}}$, and the opacity of the 2-components, $\kappa_{\rm low}$ and $\kappa_{\rm high}$. We allow the sampling to vary $-3 \leq \log_{10} (M_{\rm ej}/M_\odot) \leq 0$, $ 0 \leq v_{\rm ej} \leq 0.3$\,$c$, $0.1 \leq v_{\text{min}}/v_{\text{ej}} \leq 1.0$ and $1.0 \leq v_{\text{max}}/v_{\text{ej}} \leq 2.0$. We restrict $\kappa_{\rm low}$ and $\kappa_{\rm high}$ to a set of representative values in the analysis, i.e. 0.15 and 1.5, 0.2 and 2.0, 0.3 and 3.0, 0.4 and 4.0, 0.5 and 5.0, and 1.0 and 10 cm$^2$/g.

Figure~\ref{fig:violin_constraints_BNS} shows the ejecta mass constraints for BNS events, S191213g and S200213t, while Figure~\ref{fig:violin_constraints_NSBH} shows them for NSBH events, S191205ah, S200105ae, and S200115j. We mark each $90\%$ confidence with a horizontal dashed line.
As a brief reminder, given that the entire localization region is not covered for these limits, and the limits implicitly assume that the region containing the counterpart was imaged, these should be interpreted as optimistic scenarios. It is also simplified to assume that the light curve can not exceed the stated limit at any point in time. Similar to what was found during the analysis of O3a \citep{CoDi2019b}, the constraints are not particularly strong, predominantly due to the large distances for many of the candidate events. Given the focus of these systems on the bluer optical bands, the constraints for the bluer kilonova models (low $X_{\rm lan}$, low $\theta_{\rm obs}$ and low $\kappa_{\rm low}/\kappa_{\rm high}$) tend to be stronger. 

\textbf{S191205ah:} The left column of Figure~\ref{fig:violin_constraints_NSBH} shows the ejecta mass constraints for S191205ah based on observations from ZTF (left, \citealt{gcn26416}) and SAGUARO (right, \citealt{gcn26360}). For all models we basically recover our prior, i.e., no constraint on the ejecta mass can be given. 

\textbf{S191213g:} The middle column of Figure~\ref{fig:violin_constraints_BNS} shows the ejecta mass constraints for S191213g based on the observations from ZTF \citep{gcn26424,gcn26437} and the MASTER-Network \citep{gcn26400}. Interestingly, 
Model II allows us for small values of $\theta_{\rm obs}$ (brighter kilonovae) to constrain ejecta masses above $\sim 0.3$~$M_\odot$, however for larger angles, no constraint can be made. For Model III we obtain even tighter ejecta mass limits between $0.2$~$M_\odot$ and $0.3$~$M_\odot$, where generally for potentially lower opacity ejecta we obtain better constraints. While $0.2$~$M_\odot$ rules out systems producing very large ejecta masses, e.g., highly unequal mass systems, AT2017gfo was triggered by only about a quarter of the ejecta mass and our best bound for GW190425~\citep{CoDi2019b} was a factor of a few smaller. 
Thus, we are overall unable to extract information that help us to constrain the properties of the GW trigger S191213g. 

\textbf{S200105ae:} The right column of Figure~\ref{fig:violin_constraints_NSBH} shows the ejecta mass constraints for S200105ae based on observations from ZTF \citep{gcn26673} and the MASTER-network \citep{gcn26646}. As for S191205ah our analysis recovers basically the prior and no additional information can be extracted.

\textbf{S200115j:} The left column of Figure~\ref{fig:violin_constraints_NSBH} shows the ejecta mass constraints for S200115j based on observations from ZTF \citep{gcn26767} and GOTO \citep{gcn26794}. Model II allows us for small values of $\theta_{\rm obs}$ (brighter kilonovae) to constrain ejecta masses above $\sim 0.1$~$M_\odot$, however for larger angle, no constraint can be made; similar constraints (ejecta masses below $0.15$~$M_\odot$) are also obtained with Model III. 
As for S191213g, the obtained bounds are not strong enough to reveal interesting properties about the source properties. 

\textbf{S200213t:} The right column of Figure~\ref{fig:violin_constraints_BNS} shows the ejecta mass constraints for S200213t based on observations from ZTF \citep{gcn27051} and GOTO \citep{gcn27069}. As for S191205ah and S191213g, our analysis recovers basically the prior and no additional information can be extracted for Model I and Model II.\footnote{While the 90\% indicates that the prior is recovered, the shape of the posterior distributions suggest that the parameter space is somewhat constrained, disfavoring the high ejecta masses somewhat, but not enough to affect the limits.} Model III allows us to rule out large ejecta masses~$> 0.15$~$M_\odot$ for low opacities.  

\textbf{Summary:}
 In conclusion, we find that for the follow-up surveys to the important triggers of O3b, the derived constraints on the ejecta mass are too weak to extract any information about the sources as it was possible for GW190425~\citep{CoDi2019b}. This is likely due to a number of different circumstances: a reduction number of observations from O3a to O3b, e.g., three GW events out of five were happening around 4~h UTC, leading to an important delay of observations for all facilities located in Asia and Europe. Furthermore, the distance to most of the events was quite far (around 200 Mpc) and there was the possibility that in many cases a non-astrophysical origin caused the GW alert. Also, the weather was particularly problematic for a number of the promising events (see above). Unfortunately, we also observed that some groups were less rigorous in their report compared to O3a and did not report all observations publicly, which clearly hinders the analysis outlined above. 
 Overall, some of the observational strategies were not optimal and motivates a more detailed discussion in Section~\ref{sec:strategies}. 
 
 While these analyses do not evaluate the joint constraints possible based on multiple systems, under the assumption that different telescopes observed the same portion of the sky in different bands (or at different times), it makes sense that improved constraints on physical parameters are possible. To demonstrate this, we show the ejecta mass constraints for GW190425 based on observations from ZTF (left, \cite{gcn24191}) and PS1 (right, \cite{gcn24210}) and the combination of the two. While the constraints for the low lanthanide fractions are stronger than available for the ``red kilonovae'' for all examples, the combination of $g$- and $r$-band observations from ZTF and $i$-band from PS1 yield stronger constraints across the board.

 \begin{figure}[t]
\centering
 \includegraphics[width=3.6in]{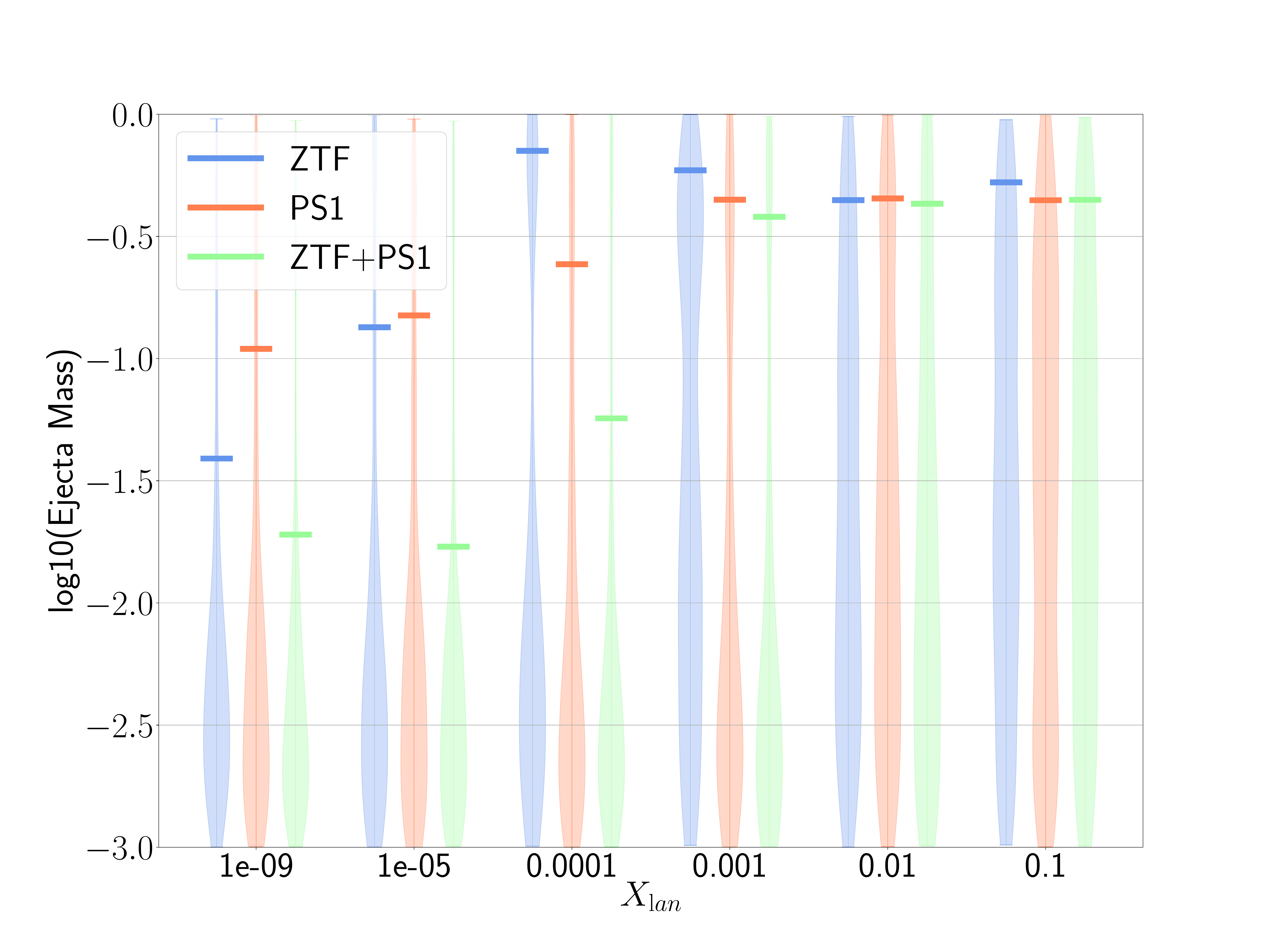}
 \caption{Probability density for the total ejecta mass for GW190425 based on the \cite{KaMe2017} model using the ZTF (left, \cite{gcn24191}), PS1 (right, \cite{gcn24210}), and joint ZTF and PS1 limits.}
 \label{fig:violin_constraints_GW190425}
\end{figure}
 
\section{Using the kilonova models to inform observational strategies}
\label{sec:strategies}

\begin{figure*}[t]
\centering
\textbf{OAJ \hspace{3.0in} PS1} \\ 
 \includegraphics[width=3.5in,height=2.0in]{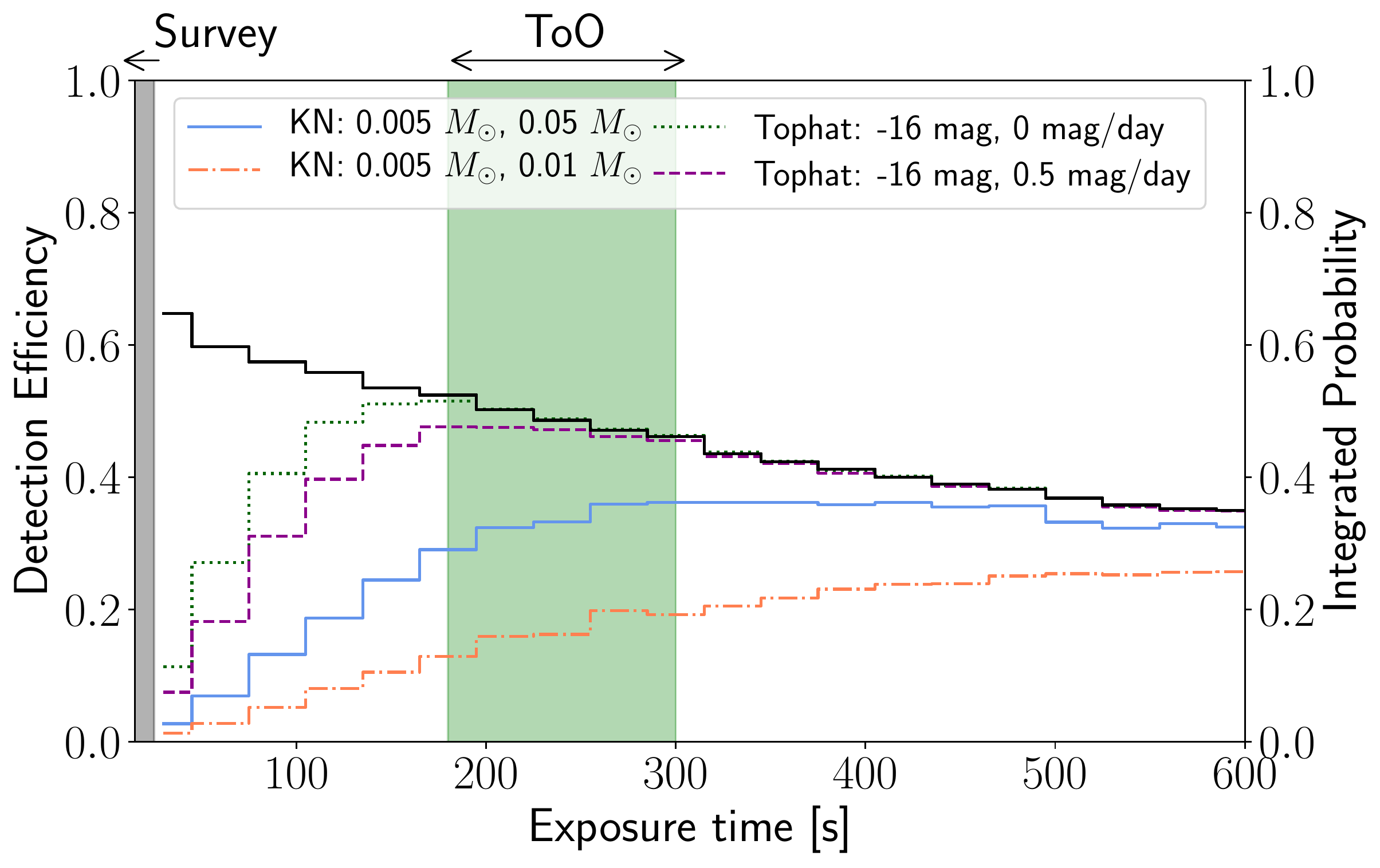}
 \includegraphics[width=3.5in,height=2.0in]{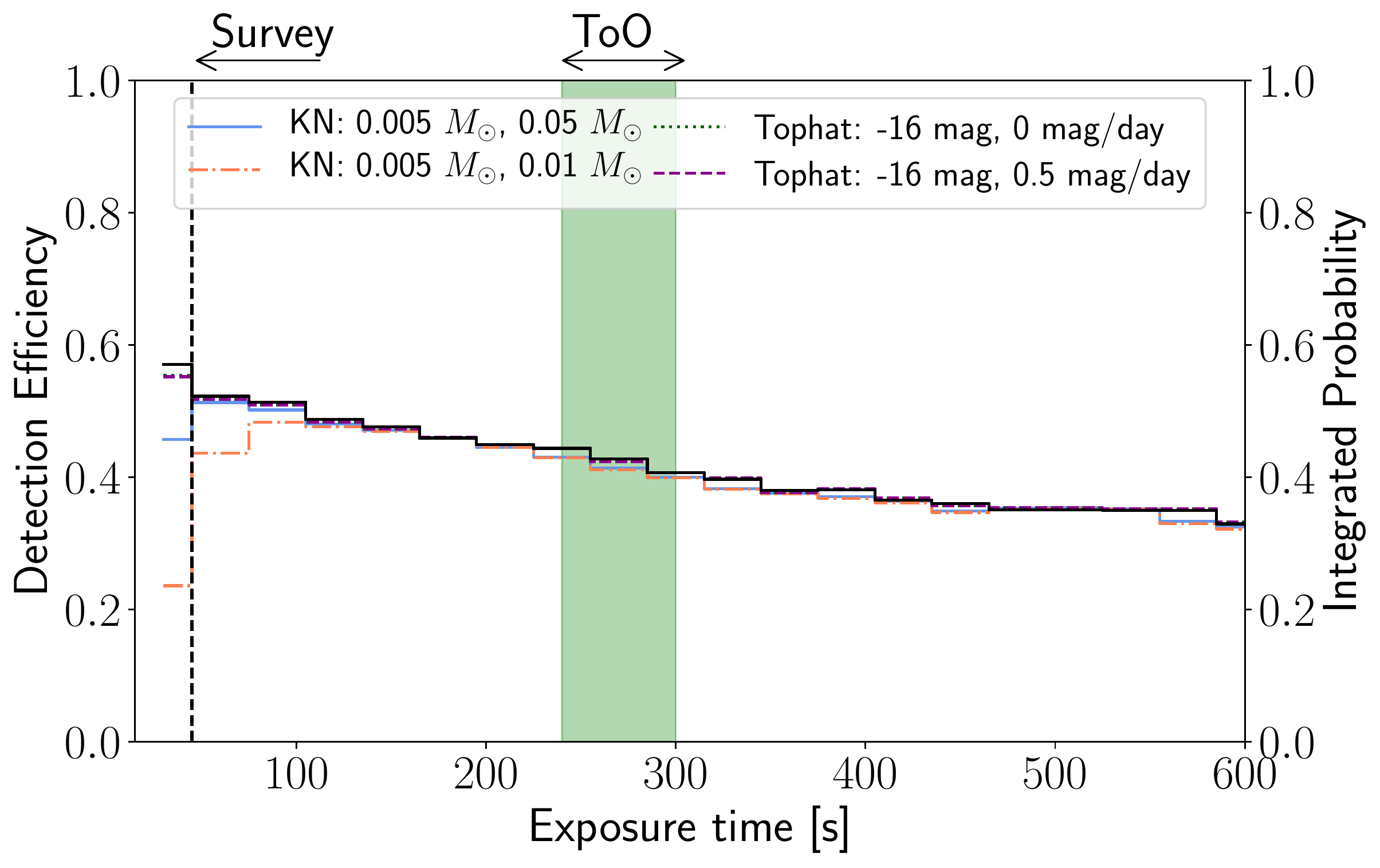}\\
\textbf{ZTF \hspace{2.7in} ZTF+PS1+OAJ} \\ 
 \includegraphics[width=3.5in,height=2.0in]{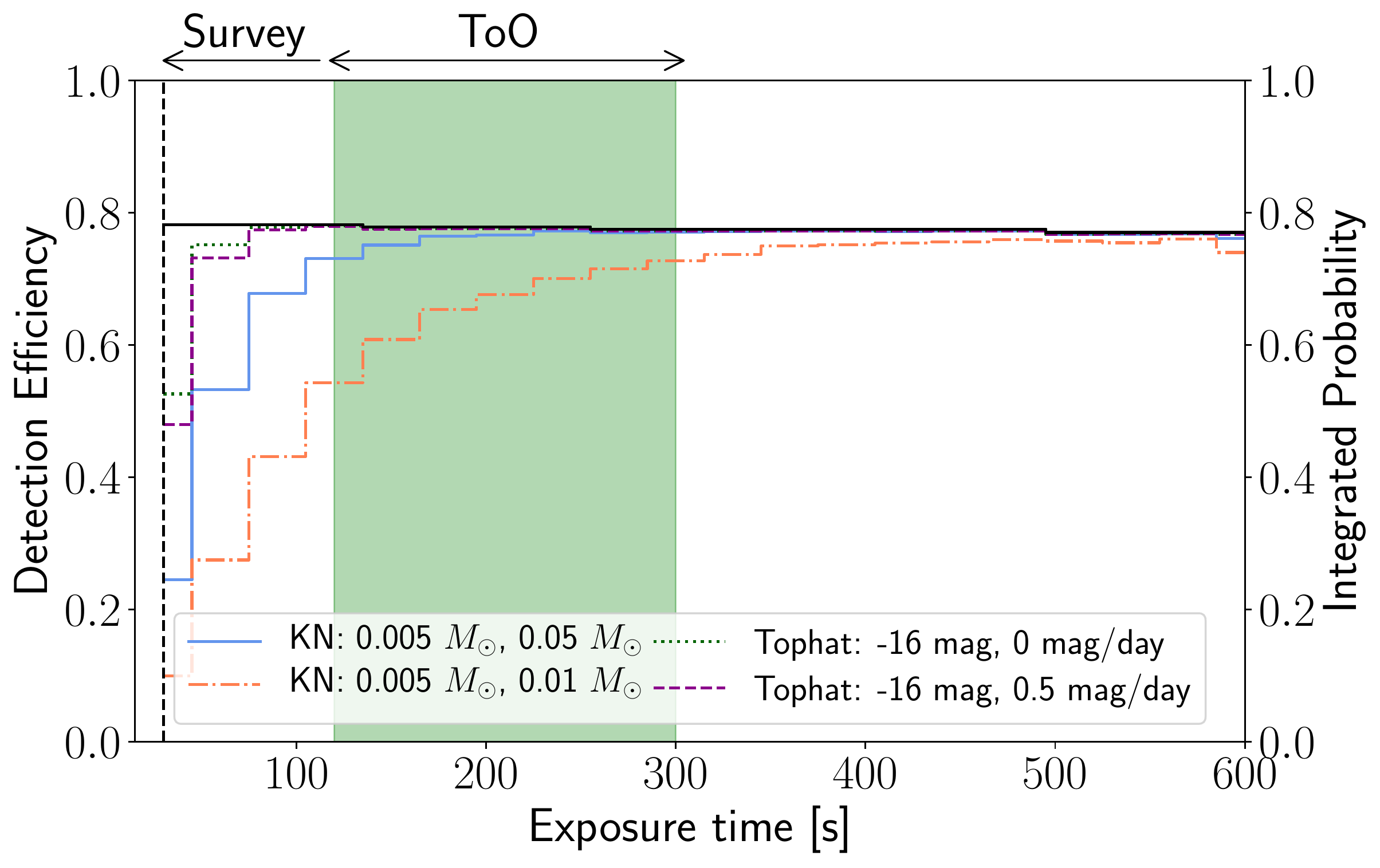}
 \includegraphics[width=3.5in,height=2.0in]{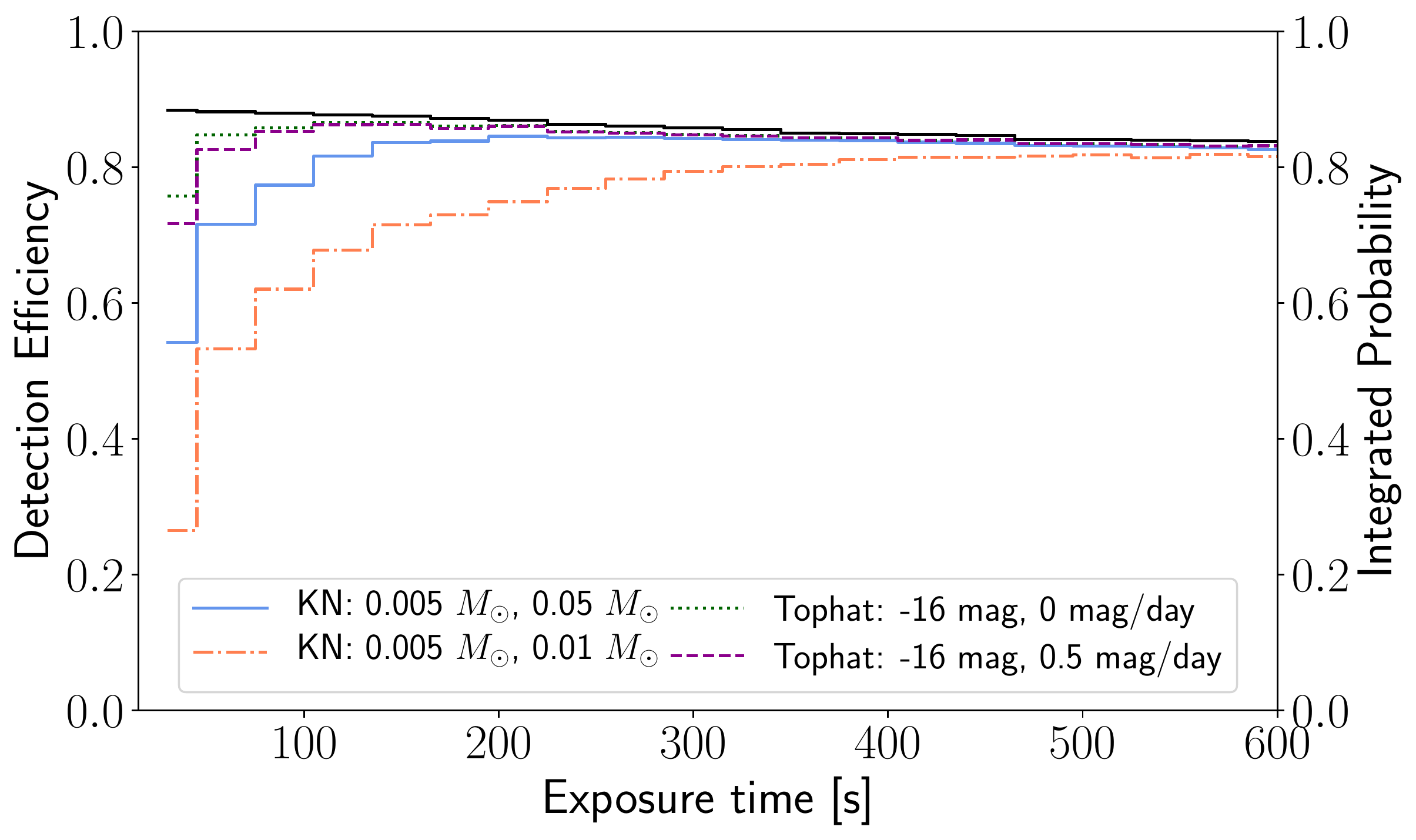}\\
 \caption{Efficiency of recoveries for S200213t (focusing on $g$- and $r$-band observations). 
 We include a model with a constant absolute magnitude of $-16$ with 0\,mag/day-decay, 
 a model with a base absolute magnitude of $-16$ and decay rate of 0.5\,mag/day, 
and two kilonova models \citep{Bul2019}, one with dynamical ejecta of $M_{\rm ej,dyn}=0.005\,M_\odot$ and post-merger wind ejecta $M_{\rm ej,pm}=0.01\,M_\odot$ and the other one with $M_{\rm ej,dyn}=0.005\,M_\odot$ and $M_{\rm ej,pm}=0.05\,M_\odot$. We show the integrated probability of the most updated sky localization area of S200213t covered by observations made within 72 hours of the event in a solid black line; we note that this is the same integrated probability for the schedule in all four models, and the detection efficiency and integrated probability should converge to the same values in cases where all kilonovae within a specific portion of the 2-D localization are detectable. The maximum coverage reachable for the three sites is 65\% for OAJ, 78\% for ZTF, 57\% for PS1, and 88\% for the network.  We also show the nominal survey exposure times in vertical dashed lines (for OAJ, we show a gray band indicating the range of survey times employed, which changes based on atmospheric and moon conditions) and range of ToO observation exposure times (120-300\,s) for comparison. We include analyses using OAJ (top left), PS1 (top right), ZTF (bottom left) individually, and a joint analysis of the three.}
 \label{fig:efficiency}
\end{figure*}

Given the relatively poor limits on the ejecta masses, we are interested in understanding how optimized scheduling strategies can aid in obtaining higher detection efficiencies of kilonova counterparts. 
Similar but slightly stronger constraints were obtained during the analysis of the first six months of O3 \citep{CoDi2019b}, where we advocated for longer observations at the cost of a smaller sky coverage.

For our investigation, we use the codebase \texttt{gwemopt}\footnote{\url{https://github.com/mcoughlin/gwemopt}} (Gravitational-Wave ElectroMagnetic OPTimization) \citep{CoTo2018}, which has been developed to schedule Target of Opportunity (ToO) telescope observations after the detection of possible multi-messenger signals, including neutrinos, gravitational waves, and $\gamma$-ray bursts. There are three main aspects to this scheduling: tiling, time allocation, and scheduling of the requested observations. Multi-telescope, network-level observations \citep{CoAn2019} and improvements for scheduling in the case of multi-lobed maps \citep{AlCo2020} are the most recent developments in these areas. We note that \texttt{gwemopt} naturally accounts for slew and read out times based on telescope-specific configuration parameters, which are important to account for inefficiencies in either long slews or when requesting short exposure times.

We now perform a study employing these latest scheduling improvements to explore realistic schedules, analyzing them with respect to exposure time in order to determine the time-scales required to make kilonova detections. We will use four different types of lightcurve models to explore this effect.
The first is based on a ``top hat'' model, where a specific absolute magnitude is taken as constant over the course of the observations; in this paper, we take an absolute magnitude (in all bands) of $-16$, which is roughly the peak magnitude of AT2017gfo \citep{ArHo2017}.
The second is similar: a base absolute magnitude of $-16$ is taken at the start of observation, but the magnitude decays linearly over time at a decay rate of 0.5\,mag/day.
These agnostic models depend only on the intrinsic luminosity and luminosity evolution of the source.
The third and fourth model types are derived from our Model II \citep{Bul2019}.
We use two different values of the post-merger wind ejecta component to explore the dependence on the amount of ejecta, one with dynamical ejecta $M_{\rm ej,dyn}=0.005\,M_\odot$ and post-merger wind ejecta $M_{\rm ej,pm}=0.01\,M_\odot$ and the other with $M_{\rm ej,dyn}=0.005\,M_\odot$ and $M_{\rm ej,pm}=0.05\,M_\odot$, similar to that found for AT2017gfo \citep{DiCo2020}. As mentioned in Section~\ref{sec:limits}, dynamical ejecta masses of $M_{\rm ej,dyn}=0.005\,M_\odot$ are more typical for BNS than BHNS mergers, and therefore we restrict our analysis to a BNS event (see below).

Figure~\ref{fig:efficiency} shows the efficiency of transient discovery for these models as a function of exposure time for a BNS event occurring at a distance similar to that of S200213t, 224 $\pm$ 90\,Mpc. We inject kilonovae according to the 3D probability distribution in the final LALInference localization of S200213t and generate a set of tilings for each telescope (with fixed exposure times) through scheduling algorithms. Here, the detection efficiency corresponds to the total number of detected kilonovae divided by the total number of simulated kilonovae, which is a proxy for the probability that the telescope covered the correct sky location during observations to a depth sufficient to detect the transient.
% i.e., we present the 
% probability that the correct sky location has been covered during the observation 
% and that the sensitivity reached the necessary level to detect the transient. 

We show the total integrated probability that the event was part of the covered sky area as a black line, and the probabilities for all four different lightcurve models as colored lines.\footnote{For intuition purposes: a tourist observing the full night sky at Mauna Kea in Hawaii would have reached 70\% for the integrated probability, but a detection efficiency of 0\% (since the typical depth reached by the human eye is about 7~mag), whereas a $\sim$\,one arcminute field observed by Keck, a 10\,m-class telescope on the mountain near to them, would have reached the necessary sensitivity but covered close to 0\% of the integrated probability.}
For our study, we use OAJ (top left), PS1 (top right), and ZTF (bottom left), and a network consisting of all three telescopes. 
As expected, there is a trade-off between exposure time and the ability to effectively cover a large sky area. Both of these contribute to the overall detection efficiency, given that the depths required for discovery are quite significant. In order to rule out moving objects (e.g., asteroids) during the transient-filtering process, it is important to have at least 30\, min gaps between multi-epoch observations; opting for longer exposure times can render this close to impossible, and hinder achieving coverage of the 90\% credible region during the first 24~hours, especially for larger localizations.
There are also observational difficulties, as field star-based guiding is not available on all telescopes, so some systems are not able to exceed exposure durations of a few minutes without sacrificing image quality.
Therefore, we are interested in pinpointing the approximate peaks in efficiency so as to find a balance between the depth and coverage attained, and ultimately increase the possibility of a kilonova detection. It is important to note that the comparably ``close'' distance of S200213t (listed in Section~\ref{sec:EM_follow_up_campaigns}) must be taken into account in this analysis, as farther events will likely favor relatively longer exposure times to achieve the depth required. In addition to exposure time, visibility constraints also contribute to the maximum probability coverage observable from a given site.

Only taking into consideration the single-telescope observations shown in Figure~\ref{fig:efficiency}, we find that as expected, the peak differs considerably depending on the telescope, by virtue of its configuration. The results with PS1, for example, are illustrative of its lower field of view in combination with its higher limiting magnitude of 21.5 (assuming optimal conditions), leading to both a quick decline in coverage for longer exposure times, and sufficient depth achieved at shorter exposure times. As a result, the efficiency peaks at a much earlier range of $\sim$\,30-100s for this event. In the case of OAJ, the similar field of view to PS1 but relatively lower limiting magnitude supports opting for exposure times of $\sim$\,160 - 300s ---in which one expects to reach $\sim$\,20.8 - $\sim$\,21.5\,mag--- to not lose out on coverage to the point of jeopardizing the detection efficiency for this skymap.
ZTF's 47-square-degree field of view, however, allows for longer exposure times to be explored while maintaining an increase in efficiency. Generally, ZTF ToO follow-ups have used $\sim$\,120 - 300\,s exposures \citep{CoAh2019b}, expected to reach $\sim$\,21.5 $-$ $\sim$\,22.4\,mag, but going for even longer exposure times appears beneficial to optimizing counterpart detection for ZTF. The bottom right panel, which shows the joint analysis, aptly re-emphasizes the potential benefit of multi-telescope coordination through the gain in detection efficiency due to the ability to more effectively cover a large sky area; additionally, since achieving significant coverage is no longer an issue, pushing for longer exposure times will only positively affect the chances of detecting a transient counterpart.
%While the exact peak varies slightly according to the model, it is broadly located at $\sim$\,10\,min exposures for the representative BNS event (S200213t) and at  $\sim$\,6-8\,min for the representative BHNS event (S200105ae), depending on the limiting magnitude adopted for the analysis.

As grounds for comparison, we also performed identical simulations for BNS event GW190425 \citep{SiEA2019b} (with a sizable updated localization of $\sim$\,7500 square degrees) in order to investigate the effects of the skymap's size on the peak efficiencies and the corresponding exposure times. The results are compared using a single telescope configuration (ZTF) vs. a multi-telescope configuration (ZTF, PS1, and OAJ) for different lightcurve models. For the Tophat model with a decay rate of 0.5 mag/day, the detection efficiency peaked at $\sim$\,70s for ZTF, with both the integrated probability and detection efficiency at 27\%. However, under identical conditions, the telescope network configuration peaked at a detection efficiency and integrated probability of 34\% at $\sim$\,40s. 
Using the synthetic lightcurve adopted from Model II, with dynamical ejecta of $M_{\rm ej,dyn}=0.005\,M_\odot$ and post-merger wind ejecta of $M_{\rm ej,pm}=0.01\,M_\odot$, ZTF attained a peak efficiency of 25\% at $\sim$\,170s. On the other hand, the telescope network resulted in a higher detection efficiency of 29\% at $\sim$\,100s due to the increased coverage. It is clear that regardless of the model adopted, there is some benefit in utilizing telescope networks to optimize the search for counterparts, especially in the case of such large localizations; however, truly maximizing this benefit requires the ability to optimize exposure times on a field-by-field (or at least, telescope-by-telescope) basis.
This also requires that the telescopes coordinate their observations, or in other words, optimize their joint observation schedules above and beyond optimization of individual observation schedules.\\

\begin{figure*}[t]
\centering
 \includegraphics[width=3.5in,height=2.0in]{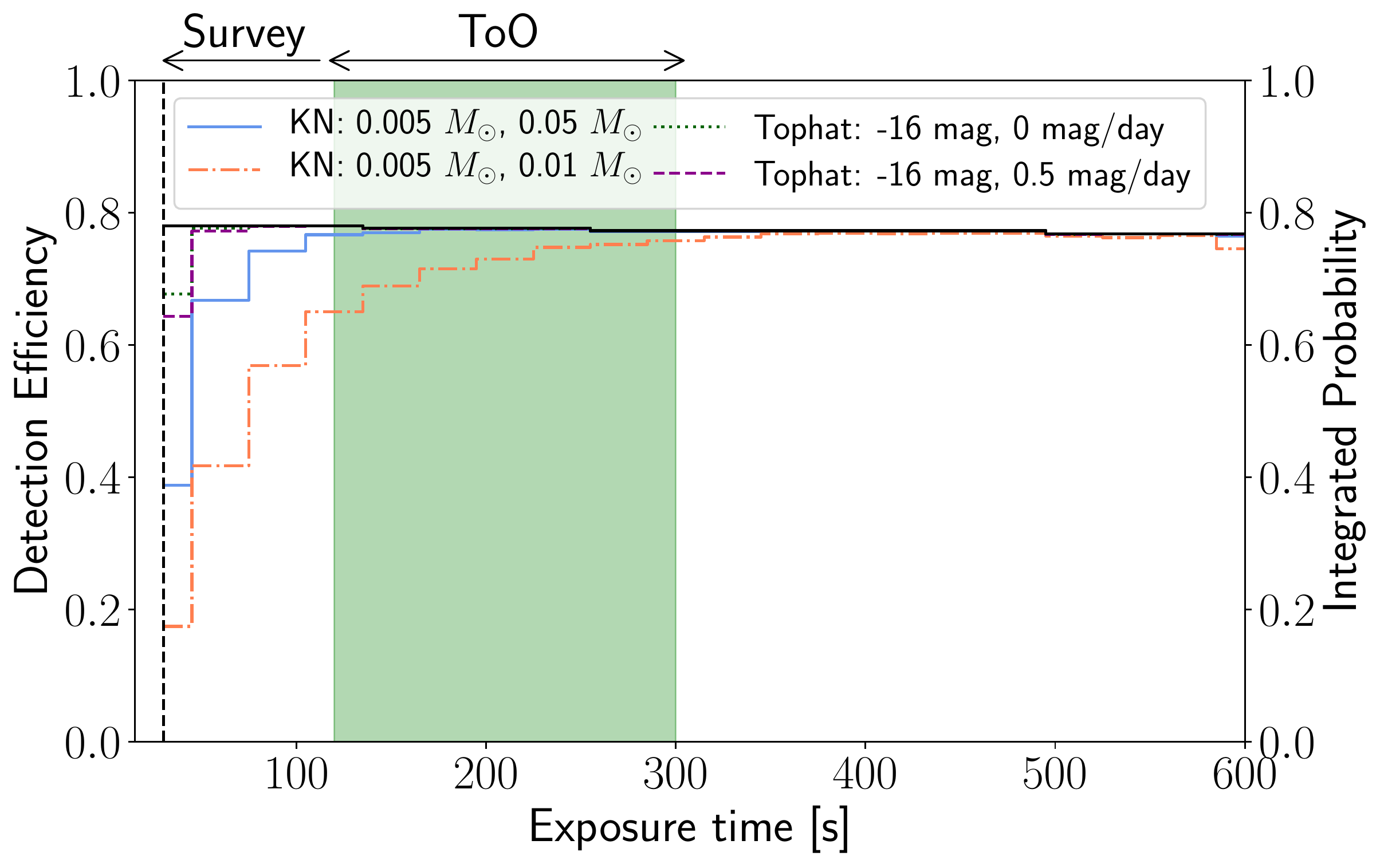} \includegraphics[width=3.5in,height=2.0in]{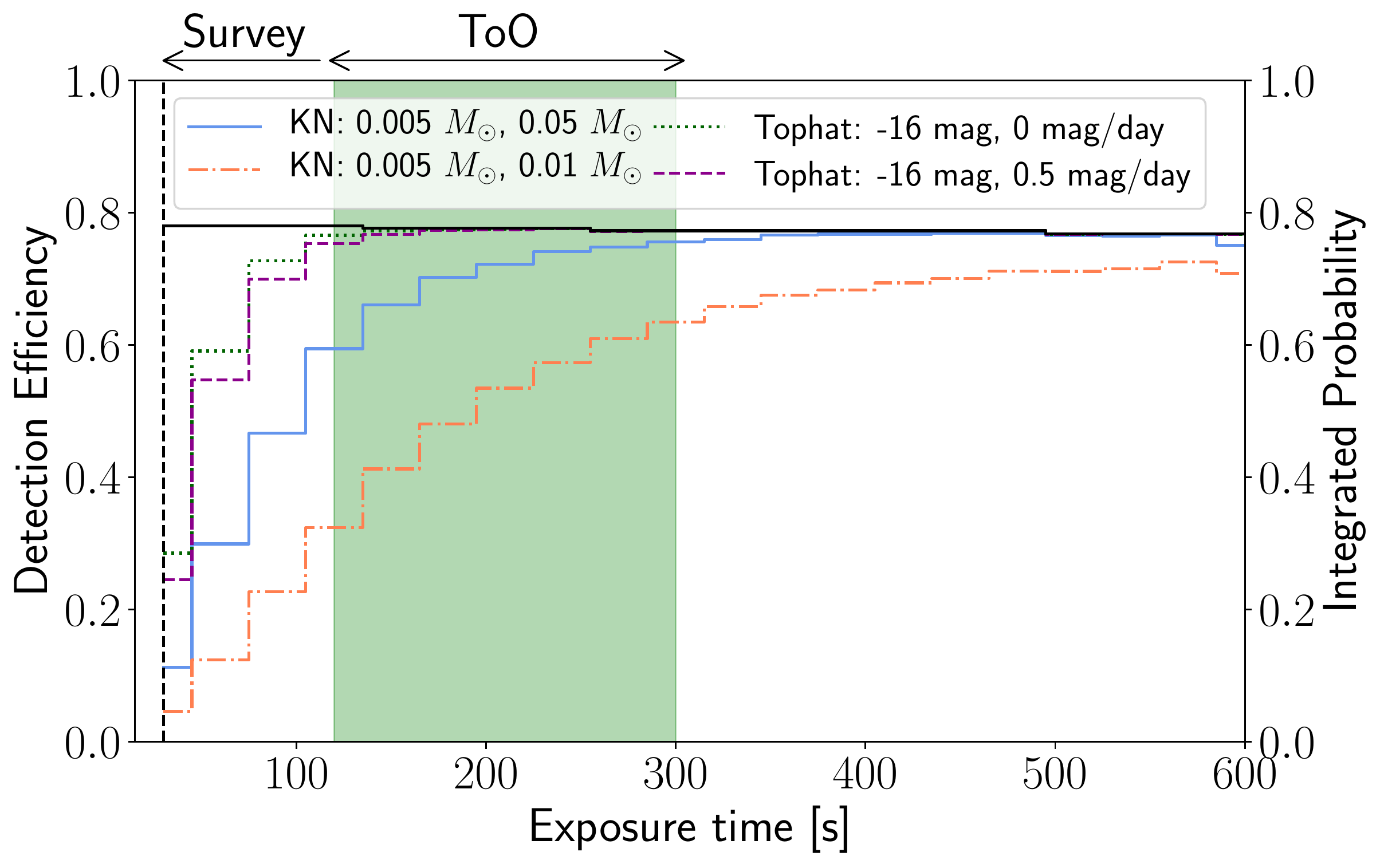}\\
 \caption{Efficiency of recoveries for S200213t for a model with a constant absolute magnitude of $-16$ (Tophat), a model with a base absolute magnitude of $-16$ and decay rate of 0.5\,mag/day (Tophat), and two kilonova models \citep{Bul2019}, one with dynamical ejecta $M_{\rm ej,dyn}=0.005\,M_\odot$ and post-merger wind ejecta $M_{\rm ej,pm}=0.01\,M_\odot$ and the other one with $M_{\rm ej,dyn}=0.005\,M_\odot$ and $M_{\rm ej,pm}=0.05\,M_\odot$, similar to that found for AT2017gfo \citep{DiCo2020}. 
 We also show the nominal ZTF survey exposure time (30\,s) and range of ToO observation exposure times (120-300\,s) for comparison. On the left is for a limiting magnitude of 20.5, corresponding to 16th percentile night, while on the right, the limiting magnitude is 19.5, corresponding to a 84th percentile night.}
 \label{fig:efficiency_bad_good}
\end{figure*}

Finally, we want to show the impact of observation conditions on the peak detection efficiencies and the corresponding exposure times in Fig.~\ref{fig:efficiency_bad_good}. 
We uses two baselines for ZTF magnitude limits, with one corresponding to 19.5, the median $-1\sigma$ and the other to 20.5, the median $+1\sigma$. Our analysis shows that for good conditions (left panel), the performance for ToOs is reasonable, although especially optimal towards the upper end of the 120 - 300\,s range. For relatively poor conditions (right panel), longer exposure times are required, which is now possible due to the significant work that has gone into improving ZTF references to adequate depths for these deeper observations. One more point of consideration is the distance information for the event; a kilonova with twice the luminosity distance will produce four times less flux, and this will affect the depth required to possibly detect the transient. This aspect of the analysis does not overshadow the importance of prioritizing longer exposure times (in particular under bad observational conditions). We note that the quoted limits for S200213t are $\sim$\,20.7\,mag in 120\,s from ZTF \citep{gcn27051}; this corresponds to $\sim$\,19.2 expected for 30\,s exposures, and therefore sub-optimal conditions.

%The aforementioned trade-off between coverage and depth is quite evident with a noticeable decrease in efficiency after around 300\,s for all models considered; this is likely due to difficulty in maintaining consistent coverage over the coarse localization for longer exposure times.

\section{Summary}
\label{sec:summary}

In this paper, we have presented a summary of the searches for EM counterparts during the second half of the third observing run of Advanced LIGO and Advanced Virgo; we focus on the gravitational-wave event candidates which are likely to be the coalescence of compact binaries with at least one neutron star component. 
We used three different, independent kilonova models \cite{KaMe2017, Bul2019, HoNa2019} to explore potential ejecta mass limits based on the non-detection of kilonova counterparts of the five potential GW events S191205ah, S191213, S200105ae, S200115j, and S200213t by comparing apparent magnitude limits from optical survey systems to the gravitational-wave distances. While the models differ in their radiative transfer treatment, our results show that the publicly-available observations do not provide any strong constraints on the quantity of mass ejected during the possible events, assuming the source was covered by those observations. The most constraining measurement is obtained for S200115j thanks to the observations of ZTF and GOTO; the model of \citet{Bul2019} excludes an ejecta of more than 0.1\,$M_\odot$ for some viewing angles. 
In general, the reduced number of observations between O3a and O3b, the delay of observations, the shallower depth of observations, and large distances of the candidates, which yield faint kilonovae, explain the minimal constraints for the compact binary candidates. However, it shows the benefit of a systematic diagnostic about quantity of ejecta thanks to the observations, as was done in the analysis of O3a \citep{CoDi2019b}. Although the strategy of follow-up employed by the various teams and their instrument capabilities did not evolve significantly in the eleven months of O3, it is clear that a global coordination of the observations would yield expected gains in efficiency, both in terms of coverage and sensitivity.

Given the uninformative constraints, we explored the depths that would be required to improve the detection efficiencies at the cost of coverage of the sky location areas for both single telescopes and network level observations. We find that exposure times of $\sim$\,3-10\,min would be useful for ZTF to maximize its sensitivity for the events discussed here, depending on the model and atmospheric conditions, which is a factor of 6-20\,$\times$ longer than survey observations, and up to a factor of 2\,$\times$ longer than for current ToO observations; the result is similar for OAJ. For PS1, on the other hand, its larger aperture leads to the conclusion that its natural survey exposure time is about right for events in the BNS distance range. Our results also highlight the advantages of telescope networks in increasing coverage of the localization and thereby allowing for longer exposure times to be used, thus leading to a corresponding increase in detection efficiencies.

It is also important to connect our results to conclusions drawn in other works: \citealt{CaBu2020} showed that detections of a AT2017gfo-like light curve at 200 Mpc requires observations down to limiting magnitudes of 23\,mag for lanthanide-rich viewing angles and 22\,mag for lanthanide-free viewing angles. The authors point out that because the optical lightcurves of kilonovae become red in a matter of few days, observing in red filters, such as inclusion of $i$-band observations, results in almost double the detections as compared to observations in $g$- and $r$-band only. They propose that observations of rapid decay in blue bands, followed by longer observations in redder bands is therefore an ideal strategy for searching for kilonovae. This strategy can be combined with the exposure time measurements here to create more optimized schedules. \citealt{Kasliwal2020} also demonstrate that under the assumption that the GW events are astrophysical, strong constraints on kilonova luminosity functions are possible by taking multiple events and considering them together, even when the probabilities and depths covered on individual events are not always strong. This motivates future work where ejecta mass constraints can be made on a population basis by considering the joint constraints over all events.

Building in field-dependent exposure times will be critical for improving the searches for counterparts. While our estimates are clearly model dependent (e.g., by assuming an absolute magnitude, a decay rate for candidate counterparts, and a particular kilonova model), it is clear that deeper observations are required, especially with the future upgrades of the GW detectors, to improve detection efficiencies when the localization area and telescope configuration allow for it. Telescope upgrades alone do not guarantee success, as detecting more marginal events at further distances will not necessarily yield better covered skymaps. Smaller localizations from highly significant, nearby events are key, perhaps with the inclusion of more information to differentiate those most likely to contain counterpart, such as the chirp mass \citep{MaMe2019}, to support the follow-up. 

\section*{Acknowledgements}

SA is supported by the CNES Postdoctoral Fellowship at Laboratoire Astroparticle et Cosmologie. MB acknowledges support from the G.R.E.A.T research environment funded by the Swedish National Science Foundation.
MWC acknowledges support from the National Science Foundation with grant number PHY-2010970.
FF gratefully acknowledges support from NASA through grant 80NSSC18K0565, from the NSF through grant PHY-1806278, and from the DOE through CAREER grant DE-SC0020435.
SGA acknowledges support from the GROWTH (Global Relay of Observatories Watching Transients Happen) project funded by the National Science Foundation under PIRE Grant No 1545949. 
G.R. and S.N. are grateful for support from the Nederlandse Organisatie voor Wetenschappelijk Onderzoek (NWO) through the VIDI and Projectruimte grants (PI Nissanke).
The lightcurve fitting / upper limits code used here is available at: \url{https://github.com/mcoughlin/gwemlightcurves}. We also thank Kerry Paterson, Samuel Dilon Wyatt and Owen McBrien for giving explanations of their observations.

\section*{Data Availability}
The data underlying this article are derived from public code found here: \href{https://github.com/mcoughlin/gwemlightcurves}{https://github.com/mcoughlin/gwemlightcurves}. The simulations resulting will be shared on reasonable request to the corresponding author.

\bibliographystyle{mnras}
\bibliography{references}

\begin{thebibliography}{}
\makeatletter
\relax
\def\mn@urlcharsother{\let\do\@makeother \do\$\do\&\do\#\do\^\do\_\do\%\do\~}
\def\mn@doi{\begingroup\mn@urlcharsother \@ifnextchar [ {\mn@doi@}
  {\mn@doi@[]}}
\def\mn@doi@[#1]#2{\def\@tempa{#1}\ifx\@tempa\@empty \href
  {http://dx.doi.org/#2} {doi:#2}\else \href {http://dx.doi.org/#2} {#1}\fi
  \endgroup}
\def\mn@eprint#1#2{\mn@eprint@#1:#2::\@nil}
\def\mn@eprint@arXiv#1{\href {http://arxiv.org/abs/#1} {{\tt arXiv:#1}}}
\def\mn@eprint@dblp#1{\href {http://dblp.uni-trier.de/rec/bibtex/#1.xml}
  {dblp:#1}}
\def\mn@eprint@#1:#2:#3:#4\@nil{\def\@tempa {#1}\def\@tempb {#2}\def\@tempc
  {#3}\ifx \@tempc \@empty \let \@tempc \@tempb \let \@tempb \@tempa \fi \ifx
  \@tempb \@empty \def\@tempb {arXiv}\fi \@ifundefined
  {mn@eprint@\@tempb}{\@tempb:\@tempc}{\expandafter \expandafter \csname
  mn@eprint@\@tempb\endcsname \expandafter{\@tempc}}}

\bibitem[\protect\citeauthoryear{{Aasi et al}}{{Aasi et al}}{2015}]{aLIGO}
{Aasi et al} 2015, Classical and Quantum Gravity, 32, 074001

\bibitem[\protect\citeauthoryear{{Abbott} B.~P.}{{Abbott}
  B.~P.}{2017}]{AbEA2017b}
{Abbott} B.~P. e.~a.,  2017, \mn@doi [Phys. Rev. Lett.]
  {10.1103/PhysRevLett.119.161101}, 119, 161101

\bibitem[\protect\citeauthoryear{{Abbott et al.}}{{Abbott et
  al.}}{2017a}]{AbEA2017g}
{Abbott et al.} 2017a, Nature, 551, 85

\bibitem[\protect\citeauthoryear{{Abbott} et~al.}{{Abbott}
  et~al.}{2017b}]{GBM:2017lvd}
{Abbott} B.~P.,  et~al., 2017b, \mn@doi [Astrophys. J.]
  {10.3847/2041-8213/aa91c9}, 848, L12

\bibitem[\protect\citeauthoryear{{Abbott et al.}}{{Abbott et
  al.}}{2017c}]{AbEA2017f}
{Abbott et al.} 2017c, The Astrophysical Journal Letters, 850, L39

\bibitem[\protect\citeauthoryear{Abbott et~al.,}{Abbott
  et~al.}{2019a}]{LIGOScientific:2018mvr}
Abbott B.~P.,  et~al., 2019a, \mn@doi [Phys. Rev. X]
  {10.1103/PhysRevX.9.031040}, 9, 031040

\bibitem[\protect\citeauthoryear{{Abbott} et~al.}{{Abbott}
  et~al.}{2019b}]{Abbott:2018wiz}
{Abbott} B.~P.,  et~al., 2019b, \mn@doi [Phys. Rev.]
  {10.1103/PhysRevX.9.011001}, X9, 011001

\bibitem[\protect\citeauthoryear{Abbott et~al.,}{Abbott
  et~al.}{2019c}]{AbEA2019b}
Abbott B.,  et~al., 2019c, \mn@doi [Physical Review Letters]
  {10.1103/physrevlett.123.011102}, 123

\bibitem[\protect\citeauthoryear{Abbott et~al.,}{Abbott
  et~al.}{2019d}]{AbEA2019}
Abbott B.~P.,  et~al., 2019d, \mn@doi [The Astrophysical Journal]
  {10.3847/1538-4357/ab0e8f}, 875, 161

\bibitem[\protect\citeauthoryear{Abbott et~al.,}{Abbott
  et~al.}{2020a}]{AbEA2020}
Abbott R.,  et~al., 2020a, \mn@doi [The Astrophysical Journal]
  {10.3847/2041-8213/ab960f}, 896, L44

\bibitem[\protect\citeauthoryear{Abbott,   et~al.}{Abbott
  et~al.}{2020b}]{AbEA2019_GW190425}
Abbott B.~P.,    et~al., 2020b, arXiv, 2001.01761

\bibitem[\protect\citeauthoryear{{Acernese et al}}{{Acernese et
  al}}{2015}]{adVirgo}
{Acernese et al} 2015, Classical and Quantum Gravity, 32, 024001

\bibitem[\protect\citeauthoryear{{Ackley} et~al.}{{Ackley}
  et~al.}{2019a}]{gcn25337}
{Ackley} K.,  et~al., 2019a, GRB Coordinates Network, 25337

\bibitem[\protect\citeauthoryear{{Ackley} et~al.}{{Ackley}
  et~al.}{2019b}]{gcn25654}
{Ackley} K.,  et~al., 2019b, GRB Coordinates Network, 25654

\bibitem[\protect\citeauthoryear{Ackley et~al.,}{Ackley
  et~al.}{2020}]{AcAm2020}
Ackley K.,  et~al., 2020, Observational constraints on the optical and
  near-infrared emission from the neutron star-black hole binary merger
  S190814bv (\mn@eprint {arXiv} {2002.01950})

\bibitem[\protect\citeauthoryear{Agathos, Zappa, Bernuzzi, Perego, Breschi  \&
  Radice}{Agathos et~al.}{2019}]{Agathos:2019sah}
Agathos M.,  Zappa F.,  Bernuzzi S.,  Perego A.,  Breschi M.,   Radice D.,
  2019

\bibitem[\protect\citeauthoryear{{Ageron}, {Baret}, {Coleiro}, {Colomer},
  {Dornic}, {Kouchner}  \& {Pradier}}{{Ageron} et~al.}{2019}]{gcn26352}
{Ageron} M.,  {Baret} B.,  {Coleiro} A.,  {Colomer} M.,  {Dornic} D.,
  {Kouchner} A.,   {Pradier} T.,  2019, GRB Coordinates Network, 26352

\bibitem[\protect\citeauthoryear{{Alan} et~al.}{{Alan} et~al.}{2020}]{gcn27061}
{Alan} A.,  et~al., 2020, GRB Coordinates Network, 27061

\bibitem[\protect\citeauthoryear{Almualla, Coughlin, Anand, Alqassimi, Guessoum
   \& Singer}{Almualla et~al.}{2020}]{AlCo2020}
Almualla M.,  Coughlin M.~W.,  Anand S.,  Alqassimi K.,  Guessoum N.,   Singer
  L.~P.,  2020, \mn@doi [Monthly Notices of the Royal Astronomical Society]
  {10.1093/mnras/staa1498}, 495, 4366–4371

\bibitem[\protect\citeauthoryear{{Anand} et~al.}{{Anand}
  et~al.}{2019}]{gcn25706}
{Anand} S.,  et~al., 2019, GRB Coordinates Network, 25706

\bibitem[\protect\citeauthoryear{Andreoni et~al.}{Andreoni
  et~al.}{2019a}]{AnGo2019}
Andreoni I.,  et~al., 2019a, \mn@doi [Astrophys. J.]
  {10.3847/2041-8213/ab3399}, 881, L16

\bibitem[\protect\citeauthoryear{{Andreoni} et~al.}{{Andreoni}
  et~al.}{2019b}]{gcn26416}
{Andreoni} I.,  et~al., 2019b, GRB Coordinates Network, 26416

\bibitem[\protect\citeauthoryear{{Andreoni} et~al.}{{Andreoni}
  et~al.}{2019c}]{gcn26424}
{Andreoni} I.,  et~al., 2019c, GRB Coordinates Network, 26424

\bibitem[\protect\citeauthoryear{{Andreoni} et~al.}{{Andreoni}
  et~al.}{2019d}]{gcn26432}
{Andreoni} I.,  et~al., 2019d, GRB Coordinates Network, 26432

\bibitem[\protect\citeauthoryear{{Andreoni} et~al.,}{{Andreoni}
  et~al.}{2020a}]{Andreoni2020}
{Andreoni} I.,  et~al., 2020a, \mn@doi [\apj] {10.3847/1538-4357/ab6a1b}, \href
  {https://ui.adsabs.harvard.edu/abs/2020ApJ...890..131A} {890, 131}

\bibitem[\protect\citeauthoryear{{Andreoni} et~al.}{{Andreoni}
  et~al.}{2020b}]{gcn26863}
{Andreoni} I.,  et~al., 2020b, GRB Coordinates Network, 26863

\bibitem[\protect\citeauthoryear{{Andreoni} et~al.}{{Andreoni}
  et~al.}{2020c}]{gcn27065}
{Andreoni} I.,  et~al., 2020c, GRB Coordinates Network, 27065

\bibitem[\protect\citeauthoryear{{Andreoni} et~al.}{{Andreoni}
  et~al.}{2020d}]{gcn27075}
{Andreoni} I.,  et~al., 2020d, GRB Coordinates Network, 27075

\bibitem[\protect\citeauthoryear{{Antier} et~al.,}{{Antier}
  et~al.}{2019}]{GRANDMAO3A}
{Antier} S.,  et~al., 2019, \mn@doi [\mnras] {10.1093/mnras/stz3142}, \href
  {https://ui.adsabs.harvard.edu/abs/2019MNRAS.tmp.2740A} {p.~2740}

\bibitem[\protect\citeauthoryear{Antier et~al.,}{Antier
  et~al.}{2020}]{GRANDMA2020}
Antier S.,  et~al., 2020, GRANDMA Observations of Advanced LIGO's and Advanced
  Virgo's Third Observational Campaign (\mn@eprint {arXiv} {2004.04277})

\bibitem[\protect\citeauthoryear{{Arcavi et al.}}{{Arcavi et
  al.}}{2017}]{ArHo2017}
{Arcavi et al.} 2017, Nature, 551, 64 EP

\bibitem[\protect\citeauthoryear{Baker, Bellini, Ferreira, Lagos, Noller  \&
  Sawicki}{Baker et~al.}{2017}]{BaBe17}
Baker T.,  Bellini E.,  Ferreira P.~G.,  Lagos M.,  Noller J.,   Sawicki I.,
  2017, \mn@doi [Phys. Rev. Lett.] {10.1103/PhysRevLett.119.251301}, 119,
  251301

\bibitem[\protect\citeauthoryear{{Bauswein}, {Baumgarte}  \&
  {Janka}}{{Bauswein} et~al.}{2013}]{Bauswein:2013jpa}
{Bauswein} A.,  {Baumgarte} T.~W.,   {Janka} H.~T.,  2013, \mn@doi [Phys. Rev.
  Lett.] {10.1103/PhysRevLett.111.131101}, 111, 131101

\bibitem[\protect\citeauthoryear{{Bauswein} et~al.}{{Bauswein}
  et~al.}{2017}]{BaJu2017}
{Bauswein} A.,  et~al., 2017, The Astrophysical Journal Letters, 850, L34

\bibitem[\protect\citeauthoryear{Bernuzzi et~al.,}{Bernuzzi
  et~al.}{2020}]{Beea20}
Bernuzzi S.,  et~al., 2020, Accretion-induced prompt black hole formation in
  asymmetric neutron star mergers, dynamical ejecta and kilonova signals
  (\mn@eprint {arXiv} {2003.06015})

\bibitem[\protect\citeauthoryear{{Bhalerao} et~al.}{{Bhalerao}
  et~al.}{2019}]{gcn24258}
{Bhalerao} V.,  et~al., 2019, GRB Coordinates Network, 24258

\bibitem[\protect\citeauthoryear{{Bhalerao} et~al.}{{Bhalerao}
  et~al.}{2020}]{gcn26767}
{Bhalerao} V.,  et~al., 2020, GRB Coordinates Network, 26767

\bibitem[\protect\citeauthoryear{Bilicki, Jarrett, Peacock, Cluver  \&
  Steward}{Bilicki et~al.}{2014}]{BiJa2014}
Bilicki M.,  Jarrett T.~H.,  Peacock J.~A.,  Cluver M.~E.,   Steward L.,  2014,
  The Astrophysical Journal Supplement Series, 210, 9

\bibitem[\protect\citeauthoryear{{Brennan} et~al.}{{Brennan}
  et~al.}{2019}]{gcn26429}
{Brennan} S.,  et~al., 2019, GRB Coordinates Network, 26429

\bibitem[\protect\citeauthoryear{{Bulla}}{{Bulla}}{2019}]{Bul2019}
{Bulla} M.,  2019, \mn@doi [\mnras] {10.1093/mnras/stz2495}, \href
  {https://ui.adsabs.harvard.edu/abs/2019MNRAS.489.5037B} {489, 5037}

\bibitem[\protect\citeauthoryear{{Burrows} et~al.,}{{Burrows}
  et~al.}{2006}]{2006ApJ...653..468B}
{Burrows} D.~N.,  et~al., 2006, \mn@doi [\apj] {10.1086/508740}, \href
  {https://ui.adsabs.harvard.edu/abs/2006ApJ...653..468B} {653, 468}

\bibitem[\protect\citeauthoryear{Capano et~al.,}{Capano
  et~al.}{2020}]{Capano:2019eae}
Capano C.~D.,  et~al., 2020, \mn@doi [Nature Astronomy]
  {10.1038/s41550-020-1014-6}

\bibitem[\protect\citeauthoryear{Carracedo, Bulla, Feindt  \& Goobar}{Carracedo
  et~al.}{2020}]{CaBu2020}
Carracedo A.~S.,  Bulla M.,  Feindt U.,   Goobar A.,  2020, Detectability of
  kilonovae in optical surveys: $post$-$mortem$ examination of the LVC O3 run
  follow-up (\mn@eprint {arXiv} {2004.06137})

\bibitem[\protect\citeauthoryear{{Castro-Tirado} et~al.}{{Castro-Tirado}
  et~al.}{2019a}]{gcn26405}
{Castro-Tirado} A.,  et~al., 2019a, GRB Coordinates Network, 26405, 1

\bibitem[\protect\citeauthoryear{{Castro-Tirado} et~al.}{{Castro-Tirado}
  et~al.}{2019b}]{gcn26421}
{Castro-Tirado} A.,  et~al., 2019b, GRB Coordinates Network, 26421, 1

\bibitem[\protect\citeauthoryear{{Castro-Tirado} et~al.}{{Castro-Tirado}
  et~al.}{2019c}]{gcn26422}
{Castro-Tirado} A.,  et~al., 2019c, GRB Coordinates Network, 26422, 1

\bibitem[\protect\citeauthoryear{{Castro-Tirado} et~al.}{{Castro-Tirado}
  et~al.}{2020a}]{gcn26702}
{Castro-Tirado} A.,  et~al., 2020a, GRB Coordinates Network, 26702

\bibitem[\protect\citeauthoryear{{Castro-Tirado} et~al.}{{Castro-Tirado}
  et~al.}{2020b}]{gcn26703}
{Castro-Tirado} A.,  et~al., 2020b, GRB Coordinates Network, 26703

\bibitem[\protect\citeauthoryear{{Castro-Tirado} et~al.}{{Castro-Tirado}
  et~al.}{2020c}]{gcn27060}
{Castro-Tirado} A.,  et~al., 2020c, GRB Coordinates Network, 27060

\bibitem[\protect\citeauthoryear{{Castro-Tirado} et~al.}{{Castro-Tirado}
  et~al.}{2020d}]{gcn27063}
{Castro-Tirado} A.,  et~al., 2020d, GRB Coordinates Network, 27063

\bibitem[\protect\citeauthoryear{Chatterjee, Ghosh, Brady, Kapadia, Miller,
  Nissanke  \& Pannarale}{Chatterjee et~al.}{2019}]{ChSh2019}
Chatterjee D.,  Ghosh S.,  Brady P.~R.,  Kapadia S.~J.,  Miller A.~L.,
  Nissanke S.,   Pannarale F.,  2019, A Machine Learning Based Source Property
  Inference for Compact Binary Mergers (\mn@eprint {arXiv} {1911.00116})

\bibitem[\protect\citeauthoryear{{Connaughton} \& {the GBM Team}}{{Connaughton}
  \& {the GBM Team}}{2012}]{Connaughton12}
{Connaughton} V.,  {the GBM Team} 2012, arXiv e-prints, \href
  {https://ui.adsabs.harvard.edu/abs/2012arXiv1202.5534C} {p. arXiv:1202.5534}

\bibitem[\protect\citeauthoryear{Coughlin et~al.,}{Coughlin
  et~al.}{2018a}]{CoTo2018}
Coughlin M.~W.,  et~al., 2018a, \mn@doi [Monthly Notices of the Royal
  Astronomical Society] {10.1093/mnras/sty1066}, 478, 692

\bibitem[\protect\citeauthoryear{{Coughlin} et~al.,}{{Coughlin}
  et~al.}{2018b}]{CoDi2018}
{Coughlin} M.~W.,  et~al., 2018b, \mn@doi [Monthly Notices of the Royal
  Astronomical Society] {10.1093/mnras/sty2174}, 480, 3871

\bibitem[\protect\citeauthoryear{Coughlin et~al.,}{Coughlin
  et~al.}{2019a}]{CoAn2019}
Coughlin M.~W.,  et~al., 2019a, \mn@doi [Monthly Notices of the Royal
  Astronomical Society] {10.1093/mnras/stz2485}

\bibitem[\protect\citeauthoryear{Coughlin, Dietrich, Margalit  \&
  Metzger}{Coughlin et~al.}{2019b}]{CoDi2018b}
Coughlin M.~W.,  Dietrich T.,  Margalit B.,   Metzger B.~D.,  2019b, \mn@doi
  [Monthly Notices of the Royal Astronomical Society: Letters]
  {10.1093/mnrasl/slz133}, 489, L91

\bibitem[\protect\citeauthoryear{Coughlin et~al.,}{Coughlin
  et~al.}{2019c}]{CoDi2019b}
Coughlin M.~W.,  et~al., 2019c, \mn@doi [Monthly Notices of the Royal
  Astronomical Society] {10.1093/mnras/stz3457}, 492, 863–876

\bibitem[\protect\citeauthoryear{Coughlin et~al.,}{Coughlin
  et~al.}{2019d}]{CoAh2019b}
Coughlin M.~W.,  et~al., 2019d, \mn@doi [The Astrophysical Journal]
  {10.3847/2041-8213/ab4ad8}, 885, L19

\bibitem[\protect\citeauthoryear{Coughlin et~al.,}{Coughlin
  et~al.}{2020}]{CoDi2019}
Coughlin M.~W.,  et~al., 2020, \mn@doi [Phys. Rev. Research]
  {10.1103/PhysRevResearch.2.022006}, 2, 022006

\bibitem[\protect\citeauthoryear{{Coulter} et~al.,}{{Coulter}
  et~al.}{2017}]{2017Sci...358.1556C}
{Coulter} D.~A.,  et~al., 2017, \mn@doi [Science] {10.1126/science.aap9811},
  \href {http://adsabs.harvard.edu/abs/2017Sci...358.1556C} {358, 1556}

\bibitem[\protect\citeauthoryear{Creminelli \& Vernizzi}{Creminelli \&
  Vernizzi}{2017}]{CrVe17}
Creminelli P.,  Vernizzi F.,  2017, \mn@doi [Phys. Rev. Lett.]
  {10.1103/PhysRevLett.119.251302}, 119, 251302

\bibitem[\protect\citeauthoryear{{Cutter} et~al.}{{Cutter}
  et~al.}{2020}]{gcn27069}
{Cutter} R.,  et~al., 2020, GRB Coordinates Network, 27069

\bibitem[\protect\citeauthoryear{Darbha \& Kasen}{Darbha \&
  Kasen}{2020}]{DaKa20}
Darbha S.,  Kasen D.,  2020

\bibitem[\protect\citeauthoryear{{De} et~al.,}{{De} et~al.}{2019}]{gcn24187}
{De} K.,  et~al., 2019, GRB Coordinates Network, 24187

\bibitem[\protect\citeauthoryear{{Dhawan}, {Bulla}, {Goobar}, {Sagu{\'e}s
  Carracedo}  \& {Setzer}}{{Dhawan} et~al.}{2019}]{DhBu2019}
{Dhawan} S.,  {Bulla} M.,  {Goobar} A.,  {Sagu{\'e}s Carracedo} A.,   {Setzer}
  C.~N.,  2019, arXiv e-prints, \href
  {https://ui.adsabs.harvard.edu/abs/2019arXiv190913810D} {p. arXiv:1909.13810}

\bibitem[\protect\citeauthoryear{{Dichiara} et~al.}{{Dichiara}
  et~al.}{2019}]{gcn25352}
{Dichiara} S.,  et~al., 2019, GRB Coordinates Network, 25352

\bibitem[\protect\citeauthoryear{Dietrich \& Ujevic}{Dietrich \&
  Ujevic}{2017}]{DiUj2017}
Dietrich T.,  Ujevic M.,  2017, \mn@doi [Class. Quant. Grav.]
  {10.1088/1361-6382/aa6bb0}, 34, 105014

\bibitem[\protect\citeauthoryear{Dietrich, Coughlin, Pang, Bulla, Heinzel,
  Issa, Tews  \& Antier}{Dietrich et~al.}{2020}]{DiCo2020}
Dietrich T.,  Coughlin M.~W.,  Pang P. T.~H.,  Bulla M.,  Heinzel J.,  Issa L.,
   Tews I.,   Antier S.,  2020

\bibitem[\protect\citeauthoryear{Doctor, Farr, Holz  \& Pürrer}{Doctor
  et~al.}{2017}]{DoFa2017}
Doctor Z.,  Farr B.,  Holz D.~E.,   Pürrer M.,  2017, \mn@doi [Phys. Rev.]
  {10.1103/PhysRevD.96.123011}, D96, 123011

\bibitem[\protect\citeauthoryear{{Ducoin} et~al.}{{Ducoin}
  et~al.}{2019}]{gcn26558}
{Ducoin} J.-G.,  et~al., 2019, GRB Coordinates Network, 26558

\bibitem[\protect\citeauthoryear{{Duque} et~al.}{{Duque}
  et~al.}{2019}]{gcn26386}
{Duque} R.,  et~al., 2019, GRB Coordinates Network, 26386

\bibitem[\protect\citeauthoryear{{Eappachen} et~al.}{{Eappachen}
  et~al.}{2019}]{gcn26397}
{Eappachen} D.,  et~al., 2019, GRB Coordinates Network, 26397, 1

\bibitem[\protect\citeauthoryear{{Eichler}, {Livio}, {Piran}  \&
  {Schramm}}{{Eichler} et~al.}{1989}]{1989Natur.340..126E}
{Eichler} D.,  {Livio} M.,  {Piran} T.,   {Schramm} D.~N.,  1989, \mn@doi
  [\nat] {10.1038/340126a0}, \href
  {http://adsabs.harvard.edu/abs/1989Natur.340..126E} {340, 126}

\bibitem[\protect\citeauthoryear{Etienne, Liu, Shapiro  \& Baumgarte}{Etienne
  et~al.}{2009}]{Etienne:2008re}
Etienne Z.~B.,  Liu Y.~T.,  Shapiro S.~L.,   Baumgarte T.~W.,  2009, \mn@doi
  [Phys. Rev.] {10.1103/PhysRevD.79.044024}, D79, 044024

\bibitem[\protect\citeauthoryear{{Evans} et~al.}{{Evans}
  et~al.}{2020a}]{gcn26763}
{Evans} P.,  et~al., 2020a, GRB Coordinates Network, 26763

\bibitem[\protect\citeauthoryear{{Evans} et~al.}{{Evans}
  et~al.}{2020b}]{gcn26798}
{Evans} P.,  et~al., 2020b, GRB Coordinates Network, 26798

\bibitem[\protect\citeauthoryear{Ezquiaga \& Zumalacárregui}{Ezquiaga \&
  Zumalacárregui}{2017}]{EzMa17}
Ezquiaga J.~M.,  Zumalacárregui M.,  2017, \mn@doi [Phys. Rev. Lett.]
  {10.1103/PhysRevLett.119.251304}, 119, 251304

\bibitem[\protect\citeauthoryear{{Ferrigno} et~al.}{{Ferrigno}
  et~al.}{2020}]{gcn26766}
{Ferrigno} C.,  et~al., 2020, GRB Coordinates Network, 26766

\bibitem[\protect\citeauthoryear{Foucart}{Foucart}{2012}]{Foucart:2012nc}
Foucart F.,  2012, \mn@doi [Phys. Rev.] {10.1103/PhysRevD.86.124007}, D86,
  124007

\bibitem[\protect\citeauthoryear{Foucart, Hinderer  \& Nissanke}{Foucart
  et~al.}{2018}]{Foucart:2018rjc}
Foucart F.,  Hinderer T.,   Nissanke S.,  2018, \mn@doi [Phys. Rev.]
  {10.1103/PhysRevD.98.081501}, D98, 081501

\bibitem[\protect\citeauthoryear{Goldstein et~al.}{Goldstein
  et~al.}{2019}]{GoAn2019}
Goldstein D.~A.,  et~al., 2019, \mn@doi [Astrophys. J.]
  {10.3847/2041-8213/ab3046}, 881, L7

\bibitem[\protect\citeauthoryear{Gomez et~al.,}{Gomez et~al.}{2019}]{GoHo2019}
Gomez S.,  et~al., 2019, \mn@doi [The Astrophysical Journal]
  {10.3847/2041-8213/ab4ad5}, 884, L55

\bibitem[\protect\citeauthoryear{Gompertz et~al.,}{Gompertz
  et~al.}{2020}]{GoCu2020}
Gompertz B.~P.,  et~al., 2020, Searching for Electromagnetic Counterparts to
  Gravitational-wave Merger Events with the Prototype Gravitational-wave
  Optical Transient Observer (GOTO-4) (\mn@eprint {arXiv} {2004.00025})

\bibitem[\protect\citeauthoryear{{Grado} et~al.}{{Grado}
  et~al.}{2019a}]{gcn24484}
{Grado} A.,  et~al., 2019a, GRB Coordinates Network, 24484

\bibitem[\protect\citeauthoryear{{Grado} et~al.}{{Grado}
  et~al.}{2019b}]{gcn25371}
{Grado} A.,  et~al., 2019b, GRB Coordinates Network, 25371

\bibitem[\protect\citeauthoryear{Graham et~al.,}{Graham
  et~al.}{2020}]{GrFo2020}
Graham M.~J.,  et~al., 2020, \mn@doi [Phys. Rev. Lett.]
  {10.1103/PhysRevLett.124.251102}, 124, 251102

\bibitem[\protect\citeauthoryear{{Gregory}}{{Gregory}}{2020}]{gcn27067}
{Gregory} S.,  2020, GRB Coordinates Network, 27067

\bibitem[\protect\citeauthoryear{{Groot} et~al.}{{Groot}
  et~al.}{2019}]{gcn25340}
{Groot} P.,  et~al., 2019, GRB Coordinates Network, 25340

\bibitem[\protect\citeauthoryear{{Han} et~al.}{{Han} et~al.}{2020}]{gcn26786}
{Han} X.~H.,  et~al., 2020, GRB Coordinates Network, 26786

\bibitem[\protect\citeauthoryear{Hankins et~al.}{Hankins
  et~al.}{2019a}]{gcn24284}
Hankins M.,  et~al., 2019a, GRB Coordinates Network, 24284

\bibitem[\protect\citeauthoryear{Hankins et~al.}{Hankins
  et~al.}{2019b}]{gcn25358}
Hankins M.,  et~al., 2019b, GRB Coordinates Network, 25358

\bibitem[\protect\citeauthoryear{{Ho} et~al.}{{Ho} et~al.}{2020}]{gcn27074}
{Ho} A.,  et~al., 2020, GRB Coordinates Network, 27074

\bibitem[\protect\citeauthoryear{Hosseinzadeh et~al.,}{Hosseinzadeh
  et~al.}{2019}]{HoCo2019}
Hosseinzadeh G.,  et~al., 2019, \mn@doi [The Astrophysical Journal]
  {10.3847/2041-8213/ab271c}, 880, L4

\bibitem[\protect\citeauthoryear{Hotokezaka \& Nakar}{Hotokezaka \&
  Nakar}{2019}]{HoNa2019}
Hotokezaka K.,  Nakar E.,  2019, arXiv, 1909.02581

\bibitem[\protect\citeauthoryear{Hotokezaka, Kiuchi, Kyutoku, Okawa, Sekiguchi,
  Shibata  \& Taniguchi}{Hotokezaka et~al.}{2013}]{HoKi13}
Hotokezaka K.,  Kiuchi K.,  Kyutoku K.,  Okawa H.,  Sekiguchi Y.-i.,  Shibata
  M.,   Taniguchi K.,  2013, \mn@doi [Phys. Rev.] {10.1103/PhysRevD.87.024001},
  D87, 024001

\bibitem[\protect\citeauthoryear{Hotokezaka, Nakar, Gottlieb, Nissanke, Masuda,
  Hallinan, Mooley  \& Deller}{Hotokezaka et~al.}{2019}]{Hotokezaka:2018dfi}
Hotokezaka K.,  Nakar E.,  Gottlieb O.,  Nissanke S.,  Masuda K.,  Hallinan G.,
   Mooley K.~P.,   Deller A.,  2019, \mn@doi [Nature Astron.]
  {10.1038/s41550-019-0820-1}

\bibitem[\protect\citeauthoryear{{Hussain}}{{Hussain}}{2020}]{gcn27043}
{Hussain} R.,  2020, GRB Coordinates Network, 27043

\bibitem[\protect\citeauthoryear{{Hussain} et~al.}{{Hussain}
  et~al.}{2019}]{gcn26349}
{Hussain} R.,  et~al., 2019, GRB Coordinates Network, 26349

\bibitem[\protect\citeauthoryear{Im et~al.}{Im et~al.}{2019}]{gcn24466}
Im M.,  et~al., 2019, GRB Coordinates Network, 24466

\bibitem[\protect\citeauthoryear{Kapadia et~al.,}{Kapadia
  et~al.}{2020}]{KaCa2019}
Kapadia S.~J.,  et~al., 2020, \mn@doi [Classical and Quantum Gravity]
  {10.1088/1361-6382/ab5f2d}, 37, 045007

\bibitem[\protect\citeauthoryear{Kasen, Metzger, Barnes, Quataert  \&
  Ramirez-Ruiz}{Kasen et~al.}{2017}]{KaMe2017}
Kasen D.,  Metzger B.,  Barnes J.,  Quataert E.,   Ramirez-Ruiz E.,  2017,
  Nature, 551, 80 EP

\bibitem[\protect\citeauthoryear{{Kasliwal} et~al.}{{Kasliwal}
  et~al.}{2019a}]{gcn24191}
{Kasliwal} M.,  et~al., 2019a, GRB Coordinates Network, 24191

\bibitem[\protect\citeauthoryear{{Kasliwal} et~al.}{{Kasliwal}
  et~al.}{2019b}]{gcn24283}
{Kasliwal} M.~M.,  et~al., 2019b, GRB Coordinates Network, 24283

\bibitem[\protect\citeauthoryear{Kasliwal et~al.,}{Kasliwal
  et~al.}{2020a}]{Kasliwal2020}
Kasliwal M.~M.,  et~al., 2020a, 2006.11306

\bibitem[\protect\citeauthoryear{{Kasliwal} et~al.}{{Kasliwal}
  et~al.}{2020b}]{gcn27051}
{Kasliwal} M.,  et~al., 2020b, GRB Coordinates Network, 27051

\bibitem[\protect\citeauthoryear{Kawaguchi, Kyutoku, Shibata  \&
  Tanaka}{Kawaguchi et~al.}{2016}]{Kawaguchi:2016ana}
Kawaguchi K.,  Kyutoku K.,  Shibata M.,   Tanaka M.,  2016, \mn@doi [Astrophys.
  J.] {10.3847/0004-637X/825/1/52}, 825, 52

\bibitem[\protect\citeauthoryear{{Kawaguchi}, {Shibata}  \&
  {Tanaka}}{{Kawaguchi} et~al.}{2020}]{KaSh20}
{Kawaguchi} K.,  {Shibata} M.,   {Tanaka} M.,  2020, \mn@doi [\apj]
  {10.3847/1538-4357/ab61f6}, \href
  {https://ui.adsabs.harvard.edu/abs/2020ApJ...889..171K} {889, 171}

\bibitem[\protect\citeauthoryear{{Kilpatrick} et~al.}{{Kilpatrick}
  et~al.}{2019}]{gcn25350}
{Kilpatrick} C.,  et~al., 2019, GRB Coordinates Network, 25350

\bibitem[\protect\citeauthoryear{{Kim} et~al.}{{Kim} et~al.}{2019}]{gcn25342}
{Kim} J.,  et~al., 2019, GRB Coordinates Network, 25342

\bibitem[\protect\citeauthoryear{Kiuchi, Kyutoku, Shibata  \& Taniguchi}{Kiuchi
  et~al.}{2019}]{Kiuchi:2019lls}
Kiuchi K.,  Kyutoku K.,  Shibata M.,   Taniguchi K.,  2019, \mn@doi [Astrophys.
  J.] {10.3847/2041-8213/ab1e45}, 876, L31

\bibitem[\protect\citeauthoryear{{Kool} et~al.}{{Kool} et~al.}{2019}]{gcn25616}
{Kool} E.,  et~al., 2019, GRB Coordinates Network, 25616

\bibitem[\protect\citeauthoryear{K{\"o}ppel, Bovard  \& Rezzolla}{K{\"o}ppel
  et~al.}{2019}]{Koppel:2019pys}
K{\"o}ppel S.,  Bovard L.,   Rezzolla L.,  2019, \mn@doi [Astrophys. J.]
  {10.3847/2041-8213/ab0210}, 872, L16

\bibitem[\protect\citeauthoryear{Korobkin, Rosswog, Arcones  \&
  Winteler}{Korobkin et~al.}{2012}]{KoRo2012}
Korobkin O.,  Rosswog S.,  Arcones A.,   Winteler C.,  2012, \mn@doi [Monthly
  Notices of the Royal Astronomical Society]
  {10.1111/j.1365-2966.2012.21859.x}, 426, 1940

\bibitem[\protect\citeauthoryear{{Korobkin} et~al.,}{{Korobkin}
  et~al.}{2020}]{KoWo20}
{Korobkin} O.,  et~al., 2020, arXiv e-prints, \href
  {https://ui.adsabs.harvard.edu/abs/2020arXiv200400102K} {p. arXiv:2004.00102}

\bibitem[\protect\citeauthoryear{{Kostrzewa-Rutkowska}
  et~al.}{{Kostrzewa-Rutkowska} et~al.}{2020}]{gcn26686}
{Kostrzewa-Rutkowska} Z.,  et~al., 2020, GRB Coordinates Network, 26686

\bibitem[\protect\citeauthoryear{Krüger \& Foucart}{Krüger \&
  Foucart}{2020}]{KrFo2020}
Krüger C.~J.,  Foucart F.,  2020

\bibitem[\protect\citeauthoryear{Kyutoku, Ioka, Okawa, Shibata  \&
  Taniguchi}{Kyutoku et~al.}{2015}]{Kyutoku:2015gda}
Kyutoku K.,  Ioka K.,  Okawa H.,  Shibata M.,   Taniguchi K.,  2015, \mn@doi
  [Phys. Rev.] {10.1103/PhysRevD.92.044028}, D92, 044028

\bibitem[\protect\citeauthoryear{{LIGO Scientific Collaboration} \& {Virgo
  Collaboration}}{{LIGO Scientific Collaboration} \& {Virgo
  Collaboration}}{2019a}]{2019ApJ...882L..24A}
{LIGO Scientific Collaboration} {Virgo Collaboration} 2019a, \mn@doi [\apjl]
  {10.3847/2041-8213/ab3800}, \href
  {https://ui.adsabs.harvard.edu/abs/2019ApJ...882L..24A} {882, L24}

\bibitem[\protect\citeauthoryear{{LIGO Scientific Collaboration} \& {Virgo
  Collaboration}}{{LIGO Scientific Collaboration} \& {Virgo
  Collaboration}}{2019b}]{SiEA2019a}
{LIGO Scientific Collaboration} {Virgo Collaboration} 2019b, GRB Coordinates
  Network, 24168

\bibitem[\protect\citeauthoryear{{LIGO Scientific Collaboration} \& {Virgo
  Collaboration}}{{LIGO Scientific Collaboration} \& {Virgo
  Collaboration}}{2019c}]{SiEA2019b}
{LIGO Scientific Collaboration} {Virgo Collaboration} 2019c, GRB Coordinates
  Network, 24228

\bibitem[\protect\citeauthoryear{{LIGO Scientific Collaboration} \& {Virgo
  Collaboration}}{{LIGO Scientific Collaboration} \& {Virgo
  Collaboration}}{2019d}]{ChEA2019a}
{LIGO Scientific Collaboration} {Virgo Collaboration} 2019d, GRB Coordinates
  Network, 24237

\bibitem[\protect\citeauthoryear{{LIGO Scientific Collaboration} \& {Virgo
  Collaboration}}{{LIGO Scientific Collaboration} \& {Virgo
  Collaboration}}{2019e}]{gcn24489}
{LIGO Scientific Collaboration} {Virgo Collaboration} 2019e, GRB Coordinates
  Network, 24489

\bibitem[\protect\citeauthoryear{{LIGO Scientific Collaboration} \& {Virgo
  Collaboration}}{{LIGO Scientific Collaboration} \& {Virgo
  Collaboration}}{2019f}]{ChEA2019b}
{LIGO Scientific Collaboration} {Virgo Collaboration} 2019f, GRB Coordinates
  Network, 25549

\bibitem[\protect\citeauthoryear{{LIGO Scientific Collaboration} \& {Virgo
  Collaboration}}{{LIGO Scientific Collaboration} \& {Virgo
  Collaboration}}{2019g}]{gcn25606}
{LIGO Scientific Collaboration} {Virgo Collaboration} 2019g, GRB Coordinates
  Network, 25606

\bibitem[\protect\citeauthoryear{{LIGO Scientific Collaboration} \& {Virgo
  Collaboration}}{{LIGO Scientific Collaboration} \& {Virgo
  Collaboration}}{2019h}]{gcn25695}
{LIGO Scientific Collaboration} {Virgo Collaboration} 2019h, GRB Coordinates
  Network, 25695

\bibitem[\protect\citeauthoryear{{LIGO Scientific Collaboration} \& {Virgo
  Collaboration}}{{LIGO Scientific Collaboration} \& {Virgo
  Collaboration}}{2019i}]{gcn25707}
{LIGO Scientific Collaboration} {Virgo Collaboration} 2019i, GRB Coordinates
  Network, 25707

\bibitem[\protect\citeauthoryear{{LIGO Scientific Collaboration} \& {Virgo
  Collaboration}}{{LIGO Scientific Collaboration} \& {Virgo
  Collaboration}}{2019j}]{gcn25814}
{LIGO Scientific Collaboration} {Virgo Collaboration} 2019j, GRB Coordinates
  Network, 25814

\bibitem[\protect\citeauthoryear{{LIGO Scientific Collaboration} \& {Virgo
  Collaboration}}{{LIGO Scientific Collaboration} \& {Virgo
  Collaboration}}{2019k}]{gcn26350}
{LIGO Scientific Collaboration} {Virgo Collaboration} 2019k, GRB Coordinates
  Network, 26350

\bibitem[\protect\citeauthoryear{{LIGO Scientific Collaboration} \& {Virgo
  Collaboration}}{{LIGO Scientific Collaboration} \& {Virgo
  Collaboration}}{2019l}]{gcn26402}
{LIGO Scientific Collaboration} {Virgo Collaboration} 2019l, GRB Coordinates
  Network, 26402

\bibitem[\protect\citeauthoryear{{LIGO Scientific Collaboration} \& {Virgo
  Collaboration}}{{LIGO Scientific Collaboration} \& {Virgo
  Collaboration}}{2019m}]{2019GCN.26413....1L}
{LIGO Scientific Collaboration} {Virgo Collaboration} 2019m, GRB Coordinates
  Network, \href {https://ui.adsabs.harvard.edu/abs/2019GCN.26413....1L}
  {26413, 1}

\bibitem[\protect\citeauthoryear{{LIGO Scientific Collaboration} \& {Virgo
  Collaboration}}{{LIGO Scientific Collaboration} \& {Virgo
  Collaboration}}{2019n}]{gcn26417}
{LIGO Scientific Collaboration} {Virgo Collaboration} 2019n, GRB Coordinates
  Network, 26417, 1

\bibitem[\protect\citeauthoryear{{LIGO Scientific Collaboration} \& {Virgo
  Collaboration}}{{LIGO Scientific Collaboration} \& {Virgo
  Collaboration}}{2020a}]{gcn26640}
{LIGO Scientific Collaboration} {Virgo Collaboration} 2020a, GRB Coordinates
  Network, 26640

\bibitem[\protect\citeauthoryear{{LIGO Scientific Collaboration} \& {Virgo
  Collaboration}}{{LIGO Scientific Collaboration} \& {Virgo
  Collaboration}}{2020b}]{gcn26657}
{LIGO Scientific Collaboration} {Virgo Collaboration} 2020b, GRB Coordinates
  Network, 26657

\bibitem[\protect\citeauthoryear{{LIGO Scientific Collaboration} \& {Virgo
  Collaboration}}{{LIGO Scientific Collaboration} \& {Virgo
  Collaboration}}{2020c}]{2020GCN.26665....1L}
{LIGO Scientific Collaboration} {Virgo Collaboration} 2020c, GRB Coordinates
  Network, \href {https://ui.adsabs.harvard.edu/abs/2020GCN.26665....1L}
  {26665, 1}

\bibitem[\protect\citeauthoryear{{LIGO Scientific Collaboration} \& {Virgo
  Collaboration}}{{LIGO Scientific Collaboration} \& {Virgo
  Collaboration}}{2020d}]{gcn26688}
{LIGO Scientific Collaboration} {Virgo Collaboration} 2020d, GRB Coordinates
  Network, 26688

\bibitem[\protect\citeauthoryear{{LIGO Scientific Collaboration} \& {Virgo
  Collaboration}}{{LIGO Scientific Collaboration} \& {Virgo
  Collaboration}}{2020e}]{gcn26759}
{LIGO Scientific Collaboration} {Virgo Collaboration} 2020e, GRB Coordinates
  Network, Circular Service, No.~26759, \#1 (2020/Jan-0), 26759

\bibitem[\protect\citeauthoryear{{LIGO Scientific Collaboration} \& {Virgo
  Collaboration}}{{LIGO Scientific Collaboration} \& {Virgo
  Collaboration}}{2020f}]{gcn26807}
{LIGO Scientific Collaboration} {Virgo Collaboration} 2020f, GRB Coordinates
  Network, 26807, 1

\bibitem[\protect\citeauthoryear{{LIGO Scientific Collaboration} \& {Virgo
  Collaboration}}{{LIGO Scientific Collaboration} \& {Virgo
  Collaboration}}{2020g}]{gcn27042}
{LIGO Scientific Collaboration} {Virgo Collaboration} 2020g, GRB Coordinates
  Network, 27042

\bibitem[\protect\citeauthoryear{{LIGO Scientific Collaboration} \& {Virgo
  Collaboration}}{{LIGO Scientific Collaboration} \& {Virgo
  Collaboration}}{2020h}]{gcn27096}
{LIGO Scientific Collaboration} {Virgo Collaboration} 2020h, GRB Coordinates
  Network, 27096

\bibitem[\protect\citeauthoryear{{LIGO-Virgo collaboration}}{{LIGO-Virgo
  collaboration}}{2019}]{gcn25876}
{LIGO-Virgo collaboration} 2019, GRB Coordinates Network, 25876

\bibitem[\protect\citeauthoryear{Lattimer}{Lattimer}{2012}]{La12}
Lattimer J.~M.,  2012, \mn@doi [Ann. Rev. Nucl. Part. Sci.]
  {10.1146/annurev-nucl-102711-095018}, 62, 485

\bibitem[\protect\citeauthoryear{{Lee} \& {Ramirez-Ruiz}}{{Lee} \&
  {Ramirez-Ruiz}}{2007}]{LeeRR2007}
{Lee} W.~H.,  {Ramirez-Ruiz} E.,  2007, \mn@doi [New Journal of Physics]
  {10.1088/1367-2630/9/1/017}, \href
  {http://adsabs.harvard.edu/abs/2007NJPh....9...17L} {9, 17}

\bibitem[\protect\citeauthoryear{Li \& Paczynski}{Li \&
  Paczynski}{1998}]{LiPa1998}
Li L.-X.,  Paczynski B.,  1998, The Astrophysical Journal Letters, 507, L59

\bibitem[\protect\citeauthoryear{Li et~al.}{Li et~al.}{2019}]{gcn24465}
Li B.,  et~al., 2019, GRB Coordinates Network, 24465

\bibitem[\protect\citeauthoryear{{Lien} et~al.}{{Lien} et~al.}{2020}]{gcn27058}
{Lien} A.,  et~al., 2020, GRB Coordinates Network, 27058

\bibitem[\protect\citeauthoryear{{Lipunov} et~al.}{{Lipunov}
  et~al.}{2017}]{LiGo2017}
{Lipunov} V.,  et~al., 2017, The Astrophysical Journal Letters, 850, L1

\bibitem[\protect\citeauthoryear{{Lipunov} et~al.}{{Lipunov}
  et~al.}{2019a}]{gcn24167}
{Lipunov} V.,  et~al., 2019a, GRB Coordinates Network, 24167

\bibitem[\protect\citeauthoryear{{Lipunov} et~al.}{{Lipunov}
  et~al.}{2019b}]{gcn24236}
{Lipunov} V.,  et~al., 2019b, GRB Coordinates Network, 24236

\bibitem[\protect\citeauthoryear{{Lipunov} et~al.}{{Lipunov}
  et~al.}{2019c}]{gcn24436}
{Lipunov} V.,  et~al., 2019c, GRB Coordinates Network, 24436

\bibitem[\protect\citeauthoryear{{Lipunov} et~al.}{{Lipunov}
  et~al.}{2019d}]{gcn25322}
{Lipunov} V.,  et~al., 2019d, GRB Coordinates Network, 25322

\bibitem[\protect\citeauthoryear{{Lipunov} et~al.}{{Lipunov}
  et~al.}{2019e}]{gcn25609}
{Lipunov} V.,  et~al., 2019e, GRB Coordinates Network, 25609

\bibitem[\protect\citeauthoryear{{Lipunov} et~al.}{{Lipunov}
  et~al.}{2019f}]{gcn25694}
{Lipunov} V.,  et~al., 2019f, GRB Coordinates Network, 25694

\bibitem[\protect\citeauthoryear{{Lipunov} et~al.}{{Lipunov}
  et~al.}{2019g}]{gcn25712}
{Lipunov} V.,  et~al., 2019g, GRB Coordinates Network, 25712

\bibitem[\protect\citeauthoryear{{Lipunov} et~al.}{{Lipunov}
  et~al.}{2019h}]{gcn25812}
{Lipunov} V.,  et~al., 2019h, GRB Coordinates Network, 25812

\bibitem[\protect\citeauthoryear{{Lipunov} et~al.}{{Lipunov}
  et~al.}{2019i}]{gcn26353}
{Lipunov} V.,  et~al., 2019i, GRB Coordinates Network, 26353

\bibitem[\protect\citeauthoryear{{Lipunov} et~al.}{{Lipunov}
  et~al.}{2019j}]{gcn26400}
{Lipunov} V.,  et~al., 2019j, GRB Coordinates Network, 26400

\bibitem[\protect\citeauthoryear{{Lipunov} et~al.}{{Lipunov}
  et~al.}{2020a}]{gcn26646}
{Lipunov} V.,  et~al., 2020a, GRB Coordinates Network, 26646

\bibitem[\protect\citeauthoryear{{Lipunov} et~al.}{{Lipunov}
  et~al.}{2020b}]{gcn26755}
{Lipunov} V.,  et~al., 2020b, GRB Coordinates Network, 26755

\bibitem[\protect\citeauthoryear{{Lipunov} et~al.}{{Lipunov}
  et~al.}{2020c}]{gcn27041}
{Lipunov} V.,  et~al., 2020c, GRB Coordinates Network, 27041

\bibitem[\protect\citeauthoryear{{Lipunov} et~al.}{{Lipunov}
  et~al.}{2020d}]{gcn27077}
{Lipunov} V.,  et~al., 2020d, GRB Coordinates Network, 27077

\bibitem[\protect\citeauthoryear{{Longo} et~al.,}{{Longo}
  et~al.}{2020}]{2020GCN.27361....1L}
{Longo} F.,  et~al., 2020, GRB Coordinates Network, \href
  {https://ui.adsabs.harvard.edu/abs/2020GCN.27361....1L} {27361, 1}

\bibitem[\protect\citeauthoryear{{Lundquist} et~al.,}{{Lundquist}
  et~al.}{2019}]{2019arXiv190606345L}
{Lundquist} M.~J.,  et~al., 2019, \mn@doi [\apjl] {10.3847/2041-8213/ab32f2},
  \href {https://ui.adsabs.harvard.edu/abs/2019ApJ...881L..26L} {881, L26}

\bibitem[\protect\citeauthoryear{{Malacaria}, {Fermi-GBM Team}  \&
  {GBM-LIGO/Virgo Group}}{{Malacaria} et~al.}{2019}]{2019GCN.26342....1M}
{Malacaria} C.,  {Fermi-GBM Team}  {GBM-LIGO/Virgo Group} 2019, GRB Coordinates
  Network, \href {https://ui.adsabs.harvard.edu/abs/2019GCN.26342....1M}
  {26342, 1}

\bibitem[\protect\citeauthoryear{Margalit \& Metzger}{Margalit \&
  Metzger}{2017}]{MaMe2017}
Margalit B.,  Metzger B.~D.,  2017, \mn@doi [Astrophys. J.]
  {10.3847/2041-8213/aa991c}, 850, L19

\bibitem[\protect\citeauthoryear{Margalit \& Metzger}{Margalit \&
  Metzger}{2019}]{MaMe2019}
Margalit B.,  Metzger B.~D.,  2019, \mn@doi [The Astrophysical Journal]
  {10.3847/2041-8213/ab2ae2}, 880, L15

\bibitem[\protect\citeauthoryear{{Matsumoto} \& {Kimura}}{{Matsumoto} \&
  {Kimura}}{2018}]{2018ApJ...866L..16M}
{Matsumoto} T.,  {Kimura} S.~S.,  2018, \mn@doi [\apjl]
  {10.3847/2041-8213/aae51b}, \href
  {https://ui.adsabs.harvard.edu/abs/2018ApJ...866L..16M} {866, L16}

\bibitem[\protect\citeauthoryear{McBrien et~al.,}{McBrien
  et~al.}{2019a}]{gcn24197}
McBrien O.,  et~al., 2019a, GRB Coordinates Network, 24197

\bibitem[\protect\citeauthoryear{{McBrien} et~al.}{{McBrien}
  et~al.}{2019b}]{gcn26485}
{McBrien} O.,  et~al., 2019b, GRB Coordinates Network, 26485

\bibitem[\protect\citeauthoryear{{Metzger} et~al.,}{{Metzger}
  et~al.}{2010}]{MeMa2010}
{Metzger} B.~D.,  et~al., 2010, \mn@doi [Monthly Notices of the Royal
  Astronomical Society] {10.1111/j.1365-2966.2010.16864.x}, \href
  {http://adsabs.harvard.edu/abs/2010MNRAS.406.2650M} {406, 2650}

\bibitem[\protect\citeauthoryear{{Mochkovitch}, {Hernanz}, {Isern}  \&
  {Martin}}{{Mochkovitch} et~al.}{1993}]{MocHer1993}
{Mochkovitch} R.,  {Hernanz} M.,  {Isern} J.,   {Martin} X.,  1993, \mn@doi
  [\nat] {10.1038/361236a0}, \href
  {http://adsabs.harvard.edu/abs/1993Natur.361..236M} {361, 236}

\bibitem[\protect\citeauthoryear{Mooley et~al.,}{Mooley
  et~al.}{2017}]{MoNa2017}
Mooley K.~P.,  et~al., 2017, Nature, 554, 207 EP

\bibitem[\protect\citeauthoryear{{Mroz} et~al.}{{Mroz} et~al.}{2020}]{gcn27085}
{Mroz} P.,  et~al., 2020, 27085

\bibitem[\protect\citeauthoryear{Nakar}{Nakar}{2007}]{Nakar2007}
Nakar E.,  2007, \mn@doi [Phys. Rept.] {10.1016/j.physrep.2007.02.005}, 442,
  166

\bibitem[\protect\citeauthoryear{{Narayan}, {Paczynski}  \& {Piran}}{{Narayan}
  et~al.}{1992}]{1992ApJ...395L..83N}
{Narayan} R.,  {Paczynski} B.,   {Piran} T.,  1992, \mn@doi [\apjl]
  {10.1086/186493}, \href {http://adsabs.harvard.edu/abs/1992ApJ...395L..83N}
  {395, L83}

\bibitem[\protect\citeauthoryear{Niino et~al.}{Niino et~al.}{2019}]{gcn24299}
Niino Y.,  et~al., 2019, GRB Coordinates Network, 24299

\bibitem[\protect\citeauthoryear{{Oates} et~al.}{{Oates}
  et~al.}{2020}]{gcn26808}
{Oates} S.~R.,  et~al., 2020, GRB Coordinates Network, 26808

\bibitem[\protect\citeauthoryear{{Onozato} et~al.}{{Onozato}
  et~al.}{2019}]{gcn26477}
{Onozato} H.,  et~al., 2019, GRB Coordinates Network, 26477

\bibitem[\protect\citeauthoryear{{Onozato} et~al.}{{Onozato}
  et~al.}{2020}]{gcn27066}
{Onozato} H.,  et~al., 2020, GRB Coordinates Network, 27066

\bibitem[\protect\citeauthoryear{{Paczynski}}{{Paczynski}}{1991}]{Paczynski1991}
{Paczynski} B.,  1991, \actaa, \href
  {http://adsabs.harvard.edu/abs/1991AcA....41..257P} {41, 257}

\bibitem[\protect\citeauthoryear{Pannarale, Tonita  \& Rezzolla}{Pannarale
  et~al.}{2011}]{Pannarale:2010vs}
Pannarale F.,  Tonita A.,   Rezzolla L.,  2011, \mn@doi [Astrophys. J.]
  {10.1088/0004-637X/727/2/95}, 727, 95

\bibitem[\protect\citeauthoryear{{Paterson} et~al.}{{Paterson}
  et~al.}{2019}]{gcn26360}
{Paterson} K.,  et~al., 2019, GRB Coordinates Network, 26360

\bibitem[\protect\citeauthoryear{{Pereyra} et~al.}{{Pereyra}
  et~al.}{2019}]{gcn25737}
{Pereyra} E.,  et~al., 2019, GRB Coordinates Network, 25737

\bibitem[\protect\citeauthoryear{{Perley} \& {Copperwheat}}{{Perley} \&
  {Copperwheat}}{2019}]{gcn26426}
{Perley} D.,  {Copperwheat} C.,  2019, GRB Coordinates Network, 26426

\bibitem[\protect\citeauthoryear{{Pian} et~al.,}{{Pian}
  et~al.}{2017}]{2017Natur.551...67P}
{Pian} E.,  et~al., 2017, \mn@doi [\nat] {10.1038/nature24298}, \href
  {http://adsabs.harvard.edu/abs/2017Natur.551...67P} {551, 67}

\bibitem[\protect\citeauthoryear{Radice \& Dai}{Radice \& Dai}{2019}]{RaDa2018}
Radice D.,  Dai L.,  2019, \mn@doi [Eur. Phys. J.]
  {10.1140/epja/i2019-12716-4}, A55, 50

\bibitem[\protect\citeauthoryear{Radice, Perego, Zappa  \& Bernuzzi}{Radice
  et~al.}{2018}]{RaPe2018}
Radice D.,  Perego A.,  Zappa F.,   Bernuzzi S.,  2018, The Astrophysical
  Journal Letters, 852, L29

\bibitem[\protect\citeauthoryear{{Reusch} et~al.}{{Reusch}
  et~al.}{2020}]{gcn27068}
{Reusch} S.,  et~al., 2020, GRB Coordinates Network, 27068

\bibitem[\protect\citeauthoryear{Rezzolla, Most  \& Weih}{Rezzolla
  et~al.}{2018}]{ReMo2017}
Rezzolla L.,  Most E.~R.,   Weih L.~R.,  2018, \mn@doi [Astrophys. J.]
  {10.3847/2041-8213/aaa401}, 852, L25

\bibitem[\protect\citeauthoryear{Roberts, Kasen, Lee  \& Ramirez-Ruiz}{Roberts
  et~al.}{2011}]{RoKa2011}
Roberts L.~F.,  Kasen D.,  Lee W.~H.,   Ramirez-Ruiz E.,  2011, The
  Astrophysical Journal Letters, 736, L21

\bibitem[\protect\citeauthoryear{{Sari}, {Piran}  \& {Narayan}}{{Sari}
  et~al.}{1998}]{SaPiNa1998}
{Sari} R.,  {Piran} T.,   {Narayan} R.,  1998, \mn@doi [\apjl]
  {10.1086/311269}, \href {http://adsabs.harvard.edu/abs/1998ApJ...497L..17S}
  {497, L17}

\bibitem[\protect\citeauthoryear{{Savaglio} et~al.}{{Savaglio}
  et~al.}{2020}]{gcn26823}
{Savaglio} S.,  et~al., 2020, GRB Coordinates Network, 26823

\bibitem[\protect\citeauthoryear{Savchenko et~al.,}{Savchenko
  et~al.}{2017}]{SaFe2017}
Savchenko V.,  et~al., 2017, \mn@doi [The Astrophysical Journal]
  {10.3847/2041-8213/aa8f94}, 848, L15

\bibitem[\protect\citeauthoryear{{Schutz}}{{Schutz}}{1986}]{Sch1986}
{Schutz} B.~F.,  1986, \mn@doi [Nature] {10.1038/323310a0}, \href
  {http://adsabs.harvard.edu/abs/1986Natur.323..310S} {323, 310}

\bibitem[\protect\citeauthoryear{{Shappee} et~al.}{{Shappee}
  et~al.}{2019}]{gcn24309}
{Shappee} B.,  et~al., 2019, GRB Coordinates Network, 24309

\bibitem[\protect\citeauthoryear{Singer, Price, Farr  et~al.}{Singer
  et~al.}{2014}]{SiPr2014}
Singer L.~P.,  Price L.~R.,  Farr B.,   et~al., 2014, \mn@doi [Astrophys. J.]
  {10.1088/0004-637X/795/2/105}, 795, 105

\bibitem[\protect\citeauthoryear{{Singer et al.}}{{Singer et
  al.}}{2019}]{gcn25343}
{Singer et al.} 2019, GRB Coordinates Network, 25343

\bibitem[\protect\citeauthoryear{{Smartt} et~al.}{{Smartt}
  et~al.}{2019a}]{gcn24517}
{Smartt} S.,  et~al., 2019a, GRB Coordinates Network, 24517

\bibitem[\protect\citeauthoryear{{Smartt} et~al.}{{Smartt}
  et~al.}{2019b}]{gcn25922}
{Smartt} S.,  et~al., 2019b, GRB Coordinates Network, 25922

\bibitem[\protect\citeauthoryear{{Smith} et~al.,}{{Smith}
  et~al.}{2019}]{gcn24210}
{Smith} K.~W.,  et~al., 2019, GRB Coordinates Network, 24210

\bibitem[\protect\citeauthoryear{{Soares-Santos et al.}}{{Soares-Santos et
  al.}}{2017}]{SoHo2017}
{Soares-Santos et al.} 2017, The Astrophysical Journal Letters, 848, L16

\bibitem[\protect\citeauthoryear{{Soares-Santos} et~al.}{{Soares-Santos}
  et~al.}{2019}]{gcn25336}
{Soares-Santos} M.,  et~al., 2019, GRB Coordinates Network, 25336

\bibitem[\protect\citeauthoryear{{Srivastav} \& {Smartt}}{{Srivastav} \&
  {Smartt}}{2020}]{gcn26839}
{Srivastav} S.,  {Smartt} S.~J.,  2020, GRB Coordinates Network, \href
  {https://ui.adsabs.harvard.edu/abs/2020GCN.26839....1S} {26839, 1}

\bibitem[\protect\citeauthoryear{{Srivastav} et~al.}{{Srivastav}
  et~al.}{2019}]{gcn25375}
{Srivastav} S.,  et~al., 2019, GRB Coordinates Network, 25375

\bibitem[\protect\citeauthoryear{Steeghs et~al.}{Steeghs
  et~al.}{2019a}]{gcn24224}
Steeghs D.,  et~al., 2019a, GRB Coordinates Network, 24224

\bibitem[\protect\citeauthoryear{Steeghs et~al.}{Steeghs
  et~al.}{2019b}]{gcn24291}
Steeghs D.,  et~al., 2019b, GRB Coordinates Network, 24291

\bibitem[\protect\citeauthoryear{{Steeghs} et~al.}{{Steeghs}
  et~al.}{2020}]{gcn26794}
{Steeghs} D.,  et~al., 2020, GRB Coordinates Network, 26794

\bibitem[\protect\citeauthoryear{{Stein} et~al.}{{Stein}
  et~al.}{2019a}]{gcn25722}
{Stein} R.,  et~al., 2019a, GRB Coordinates Network, 25722

\bibitem[\protect\citeauthoryear{{Stein} et~al.}{{Stein}
  et~al.}{2019b}]{gcn25899}
{Stein} R.,  et~al., 2019b, GRB Coordinates Network, 25899

\bibitem[\protect\citeauthoryear{{Stein} et~al.}{{Stein}
  et~al.}{2019c}]{gcn26437}
{Stein} R.,  et~al., 2019c, GRB Coordinates Network, 26437

\bibitem[\protect\citeauthoryear{{Stein} et~al.}{{Stein}
  et~al.}{2020}]{gcn26673}
{Stein} R.,  et~al., 2020, GRB Coordinates Network, 26673

\bibitem[\protect\citeauthoryear{Tanvir, Levan, Fruchter, Hjorth, Hounsell,
  Wiersema  \& Tunnicliffe}{Tanvir et~al.}{2013}]{TaLe2013}
Tanvir N.~R.,  Levan A.~J.,  Fruchter A.~S.,  Hjorth J.,  Hounsell R.~A.,
  Wiersema K.,   Tunnicliffe R.~L.,  2013, Nature, 500, 547 EP

\bibitem[\protect\citeauthoryear{Troja et~al.,}{Troja et~al.}{2017}]{TrPi2017}
Troja E.,  et~al., 2017, Nature, 551, 71 EP

\bibitem[\protect\citeauthoryear{Troja et~al.,}{Troja
  et~al.}{2020}]{troja2020thousand}
Troja E.,  et~al., 2020, {A thousand days after the merger: continued X-ray
  emission from GW170817} (\mn@eprint {arXiv} {2006.01150})

\bibitem[\protect\citeauthoryear{{Valenti et al.}}{{Valenti et
  al.}}{2017}]{VaSa2017}
{Valenti et al.} 2017, The Astrophysical Journal Letters, 848, L24

\bibitem[\protect\citeauthoryear{{Vallely}}{{Vallely}}{2019}]{gcn26508}
{Vallely} P.,  2019, GRB Coordinates Network, \href
  {https://ui.adsabs.harvard.edu/abs/2019GCN.26508....1V} {26508, 1}

\bibitem[\protect\citeauthoryear{Veitch et~al.,}{Veitch
  et~al.}{2015}]{VeRa2015}
Veitch J.,  et~al., 2015, \mn@doi [Phys. Rev. D] {10.1103/PhysRevD.91.042003},
  91, 042003

\bibitem[\protect\citeauthoryear{{Veres} et~al.}{{Veres}
  et~al.}{2020}]{gcn27056}
{Veres} P.,  et~al., 2020, GRB Coordinates Network, 27056

\bibitem[\protect\citeauthoryear{Watson et~al.}{Watson et~al.}{2019a}]{WaHa19}
Watson D.,  et~al., 2019a, \mn@doi [Nature] {10.1038/s41586-019-1676-3}, 574,
  497

\bibitem[\protect\citeauthoryear{Watson et~al.}{Watson
  et~al.}{2019b}]{gcn24310}
Watson A.~M.,  et~al., 2019b, GRB Coordinates Network, 24310

\bibitem[\protect\citeauthoryear{{Wei} et~al.}{{Wei} et~al.}{2019}]{gcn25648}
{Wei} J.,  et~al., 2019, GRB Coordinates Network, 25648

\bibitem[\protect\citeauthoryear{{Wyatt}, {Tohuvavohu}, {Arcavi}, {Lundquist},
  {Howell}  \& {Sand}}{{Wyatt} et~al.}{2020}]{2020arXiv200100588W}
{Wyatt} S.~D.,  {Tohuvavohu} A.,  {Arcavi} I.,  {Lundquist} M.~J.,  {Howell}
  D.~A.,   {Sand} D.~J.,  2020, arXiv e-prints, \href
  {https://ui.adsabs.harvard.edu/abs/2020arXiv200100588W} {p. arXiv:2001.00588}

\bibitem[\protect\citeauthoryear{Xu et~al.}{Xu et~al.}{2019a}]{gcn24190}
Xu D.,  et~al., 2019a, GRB Coordinates Network, 24190

\bibitem[\protect\citeauthoryear{Xu et~al.}{Xu et~al.}{2019b}]{gcn24285}
Xu D.,  et~al., 2019b, GRB Coordinates Network, 24285

\bibitem[\protect\citeauthoryear{Xu et~al.}{Xu et~al.}{2019c}]{gcn24286}
Xu D.,  et~al., 2019c, GRB Coordinates Network, 24286

\bibitem[\protect\citeauthoryear{Xu et~al.}{Xu et~al.}{2019d}]{gcn24476}
Xu D.,  et~al., 2019d, GRB Coordinates Network, 24476

\bibitem[\protect\citeauthoryear{{Xu} et~al.}{{Xu} et~al.}{2020}]{gcn27070}
{Xu} D.,  et~al., 2020, GRB Coordinates Network, 27070

\bibitem[\protect\citeauthoryear{Yoshida et~al.}{Yoshida
  et~al.}{2019}]{gcn24450}
Yoshida M.,  et~al., 2019, GRB Coordinates Network, 24450

\bibitem[\protect\citeauthoryear{{Zheng} et~al.}{{Zheng}
  et~al.}{2020}]{gcn27064}
{Zheng} W.,  et~al., 2020, GRB Coordinates Network, 27064

\bibitem[\protect\citeauthoryear{Zhu et~al.}{Zhu et~al.}{2019}]{gcn24475}
Zhu Z.,  et~al., 2019, GRB Coordinates Network, 24475

\bibitem[\protect\citeauthoryear{Özel \& Freire}{Özel \&
  Freire}{2016}]{OzFr16}
Özel F.,  Freire P.,  2016, \mn@doi [Ann. Rev. Astron. Astrophys.]
  {10.1146/annurev-astro-081915-023322}, 54, 401

\makeatother
\end{thebibliography}

% i get 20.5,20.5,21.9,21.3,21.7 for the apparent mags
\appendix
% i get 20.5,20.5,21.9,21.3,21.7 for the apparent mags
\begin{deluxetable*}{cccccccc}
\tablenum{2}
\setlength{\tabcolsep}{0pt}
\tablecaption{Reports of the observations by various teams of the sky localization area of gravitational-wave alerts of possible BNS candidates S191213g and S200213t, and BHNS candidates S191205ah, S200105ae, and S200115j. For ease of comparison to limits, assuming an absolute magnitude of $-$16\,mag, the median distances correspond to apparent magnitudes of 20.5, 20.5, 21.9, 21.3, and 21.7\,mag respectively. Teams that employed ``galaxy targeting''during their follow-up  or with less than 1\% coverage of the sky localisation area are not mentioned here. In the case where numbers were not reported or provided upon request in order to calculate the total coverage based on the most updated sky localization area, we recomputed some of them; if this was not possible, we add $-$.}
\label{tab:Tableobs}
\tablewidth{0pt}
\tablehead{\begin{scriptsize}Telescope\end{scriptsize} & \begin{scriptsize}Filter\end{scriptsize} & \begin{scriptsize}Limit mag\end{scriptsize} & \begin{scriptsize}Delay aft. GW \end{scriptsize} & \begin{scriptsize}Duration\end{scriptsize} & \multicolumn{2}{c}{\begin{scriptsize}GW sky localization area\end{scriptsize}}  & \begin{scriptsize}reference\end{scriptsize} \\  &  &  & \begin{scriptsize}(h)\end{scriptsize} & \begin{scriptsize}(h)\end{scriptsize} & \begin{scriptsize}name\end{scriptsize} & \begin{scriptsize}coverage (\%)\end{scriptsize} & }
\startdata
\multicolumn{8}{c}{\textbf{S191205ah (BHNS)}}  \\
%\begin{scriptsize} GRANDMA-FRAM-A    \end{scriptsize}         & \begin{scriptsize} R-band   \end{scriptsize}& \begin{scriptsize}$18$ \end{scriptsize}  &\begin{scriptsize} $8.2$  \end{scriptsize} &\begin{scriptsize} $1.4$    \end{scriptsize}    &\begin{scriptsize}  bayestar ini \end{scriptsize}&\begin{scriptsize} $1$ \end{scriptsize} &\begin{scriptsize} \citet{GRANDMA2020}\end{scriptsize} \\
\begin{scriptsize} GRANDMA-TCA    \end{scriptsize}         & \begin{scriptsize} Clear   \end{scriptsize}& \begin{scriptsize}$18$ \end{scriptsize}  &\begin{scriptsize} $18.9$  \end{scriptsize} &\begin{scriptsize} $50$    \end{scriptsize}    &\begin{scriptsize}  bayestar ini \end{scriptsize}&\begin{scriptsize} $3$ \end{scriptsize} &\begin{scriptsize} \citet{GRANDMA2020}\end{scriptsize} \\
\begin{scriptsize} GRANDMA-TCH   \end{scriptsize}         & \begin{scriptsize} Clear   \end{scriptsize}& \begin{scriptsize}$18$ \end{scriptsize}  &\begin{scriptsize} $2.9$  \end{scriptsize} &\begin{scriptsize} $54$    \end{scriptsize}    &\begin{scriptsize}  bayestar ini \end{scriptsize}&\begin{scriptsize} $1$ \end{scriptsize} &\begin{scriptsize} \citet{GRANDMA2020}\end{scriptsize} \\
\begin{scriptsize} MASTER-network \end{scriptsize}     & \begin{scriptsize}Clear\end{scriptsize}    & \begin{scriptsize}$ \approx 19$\end{scriptsize}   & \begin{scriptsize}$ \approx 0.1$ \end{scriptsize}  & \begin{scriptsize}$144$    \end{scriptsize}    &\begin{scriptsize} bayestar ini \end{scriptsize}&\begin{scriptsize} $\approx56$  \end{scriptsize} &\begin{scriptsize} \citet{gcn26353}\end{scriptsize} \\
\begin{scriptsize} SAGUARO     \end{scriptsize}     & \begin{scriptsize}G-band\end{scriptsize}    & \begin{scriptsize}$21.3$\end{scriptsize}   & \begin{scriptsize} 4.4 \end{scriptsize}  & \begin{scriptsize}$0.5$    \end{scriptsize}    &\begin{scriptsize} bayestar ini \end{scriptsize}&\begin{scriptsize} $9$ \end{scriptsize} &\begin{scriptsize} \citet{gcn26360,2020arXiv200100588W}, this work\end{scriptsize} \\
\begin{scriptsize} SVOM-GWAC  \end{scriptsize}         & \begin{scriptsize} R-band   \end{scriptsize}& \begin{scriptsize}$16$ \end{scriptsize}  &\begin{scriptsize} $\approx0.1$  \end{scriptsize} &\begin{scriptsize} $23$    \end{scriptsize}    &\begin{scriptsize}  bayestar ini \end{scriptsize}&\begin{scriptsize} $28$ \end{scriptsize} &\begin{scriptsize} \citet{gcn26386}, this work\end{scriptsize} \\
\begin{scriptsize} Zwicky Transient Facility     \end{scriptsize}        & \begin{scriptsize}g/r-band    \end{scriptsize}& \begin{scriptsize}$17.9$ \end{scriptsize}  &\begin{scriptsize} $10.7$  \end{scriptsize} &\begin{scriptsize} $167$   \end{scriptsize}    &\begin{scriptsize} bayestar ini \end{scriptsize}&\begin{scriptsize} $6$ \end{scriptsize} &\begin{scriptsize} \citet{gcn26416,Kasliwal2020}\end{scriptsize} \\
& & & & & & & \\
\hline
\multicolumn{8}{c}{\textbf{S191213g (BNS)}} \\
%\begin{scriptsize} GRANDMA-FRAM-C    \end{scriptsize}         & \begin{scriptsize} R-band   \end{scriptsize}& \begin{scriptsize}$16$ \end{scriptsize}  &\begin{scriptsize} $0.9$  \end{scriptsize} &\begin{scriptsize} $0.3$    \end{scriptsize}    &\begin{scriptsize}  LALInference \end{scriptsize}&\begin{scriptsize} $<0.1$ \end{scriptsize} &\begin{scriptsize} \citet{GRANDMA2020}\end{scriptsize} \\
\begin{scriptsize} GRANDMA-TCA  \end{scriptsize}         & \begin{scriptsize} Clear   \end{scriptsize}& \begin{scriptsize}$18$ \end{scriptsize}  &\begin{scriptsize} $47.6$  \end{scriptsize} &\begin{scriptsize} $73$    \end{scriptsize}    &\begin{scriptsize} LALInference \end{scriptsize}&\begin{scriptsize} $1$ \end{scriptsize} &\begin{scriptsize} \citet{GRANDMA2020}\end{scriptsize} \\
\begin{scriptsize} MASTER-network  \end{scriptsize}& \begin{scriptsize} Clear\end{scriptsize} & \begin{scriptsize} $\approx 18.5$\end{scriptsize} & \begin{scriptsize} $0.4$  \end{scriptsize}  & \begin{scriptsize} $144$  \end{scriptsize} &\begin{scriptsize} LALInference  \end{scriptsize}&\begin{scriptsize} $41$ \end{scriptsize} &\begin{scriptsize} \citet{gcn26400}  \end{scriptsize}\\
%\begin{scriptsize} Pan-STARRS     \end{scriptsize} \sa{ask}        & \begin{scriptsize}    \end{scriptsize}& \begin{scriptsize}$-$ \end{scriptsize}  &\begin{scriptsize} $\approx-$  \end{scriptsize} &\begin{scriptsize} $-$    \end{scriptsize}    &\begin{scriptsize} bayestar ini \end{scriptsize}&\begin{scriptsize} $-$ \end{scriptsize} &\begin{scriptsize} \end{scriptsize} \\
%\begin{scriptsize} Swift-UVOT  \sa{ask}   \end{scriptsize}         & \begin{scriptsize}    \end{scriptsize}& \begin{scriptsize}$-$ \end{scriptsize}  &\begin{scriptsize} $\approx-$  \end{scriptsize} &\begin{scriptsize} $-$    \end{scriptsize}    &\begin{scriptsize} bayestar ini \end{scriptsize}&\begin{scriptsize} $-$ \end{scriptsize} &\begin{scriptsize} \end{scriptsize} \\
\begin{scriptsize} Zwicky  Transient  Facility  \end{scriptsize}& \begin{scriptsize} g/r-band\end{scriptsize} & \begin{scriptsize} $20.4$\end{scriptsize} & \begin{scriptsize} $0.01$  \end{scriptsize}  & \begin{scriptsize} $\approx27.8$  \end{scriptsize} &\begin{scriptsize} LALInference  \end{scriptsize}&\begin{scriptsize} $28$ \end{scriptsize} &\begin{scriptsize} \citet{gcn26424,Kasliwal2020}  \end{scriptsize}\\
& & & & & & & \\
%\begin{scriptsize} \end{scriptsize}& \begin{scriptsize} \end{scriptsize} & \begin{scriptsize} \end{scriptsize} & \begin{scriptsize} \end{scriptsize}  & \begin{scriptsize}  \end{scriptsize} &\begin{scriptsize} \end{scriptsize}&\begin{scriptsize} \end{scriptsize} &\begin{scriptsize} \citet{2020arXiv200100588W}, this work \end{scriptsize} \\
\hline
\multicolumn{8}{c}{\textbf{S200105ae (BHNS)}}  \\
%\begin{scriptsize} GRANDMA-FRAM-A     \end{scriptsize}         & \begin{scriptsize} R-band   \end{scriptsize}& \begin{scriptsize}$18$ \end{scriptsize}  &\begin{scriptsize} $60.0$  \end{scriptsize} &\begin{scriptsize} $1.7$    \end{scriptsize}    &\begin{scriptsize} LALInference  \end{scriptsize}&\begin{scriptsize} $0.6$ \end{scriptsize} &\begin{scriptsize} \citet{GRANDMA2020}\end{scriptsize} \\
%\begin{scriptsize} GRANDMA-FRAM-C    \end{scriptsize}         & \begin{scriptsize} R-band   \end{scriptsize}& \begin{scriptsize}$17$ \end{scriptsize}  &\begin{scriptsize} $28.1$  \end{scriptsize} &\begin{scriptsize} $1.5$    \end{scriptsize}    &\begin{scriptsize} LALInference \end{scriptsize}&\begin{scriptsize} $0.2$ \end{scriptsize} &\begin{scriptsize} \citet{GRANDMA2020}\end{scriptsize} \\
\begin{scriptsize} GRANDMA-TCA    \end{scriptsize}         & \begin{scriptsize} Clear   \end{scriptsize}& \begin{scriptsize}$18$ \end{scriptsize}  &\begin{scriptsize} $27.5$  \end{scriptsize} &\begin{scriptsize} $50.4$    \end{scriptsize}    &\begin{scriptsize} LALInference \end{scriptsize}&\begin{scriptsize} $3$ \end{scriptsize} &\begin{scriptsize} \citet{GRANDMA2020}\end{scriptsize} \\
\begin{scriptsize} GRANDMA-TCH  \end{scriptsize}         & \begin{scriptsize} Clear   \end{scriptsize}& \begin{scriptsize}$18$ \end{scriptsize}  &\begin{scriptsize} $59.0$  \end{scriptsize} &\begin{scriptsize} $53.8$    \end{scriptsize}    &\begin{scriptsize} LALInference \end{scriptsize}&\begin{scriptsize} $3$ \end{scriptsize} &\begin{scriptsize} \citet{GRANDMA2020}\end{scriptsize} \\
\begin{scriptsize} GRANDMA-TRE  \end{scriptsize}         & \begin{scriptsize} Clear   \end{scriptsize}& \begin{scriptsize}$17$ \end{scriptsize}  &\begin{scriptsize} $48.0$  \end{scriptsize} &\begin{scriptsize} $26.5$    \end{scriptsize}    &\begin{scriptsize} LALInference  \end{scriptsize}&\begin{scriptsize} $10$ \end{scriptsize} &\begin{scriptsize} \citet{GRANDMA2020}\end{scriptsize} \\
\begin{scriptsize} MASTER-network    \end{scriptsize}     & \begin{scriptsize}Clear\end{scriptsize}    & \begin{scriptsize}$\approx 19.5$\end{scriptsize}   & \begin{scriptsize}$\approx3.2$ \end{scriptsize}  & \begin{scriptsize}$144$    \end{scriptsize}    &\begin{scriptsize} LALInference \end{scriptsize}&\begin{scriptsize} $43$ \end{scriptsize} &\begin{scriptsize} \citet{gcn26646}\end{scriptsize} \\
%\begin{scriptsize} Zwicky Transient Facility \sa{nook}     \end{scriptsize}        & \begin{scriptsize}g/r-band    \end{scriptsize}& \begin{scriptsize}$\sa{19.7TBC}$ \end{scriptsize}  &\begin{scriptsize} $\sa{33.5TBC}$  \end{scriptsize} &\begin{scriptsize} $\sa{3.2TBC --> 33.9 ?}$    \end{scriptsize}    &\begin{scriptsize} bayestar ini \end{scriptsize}&\begin{scriptsize} $\sa{52TBC}$ \end{scriptsize} &\begin{scriptsize} \citet{gcn26673}\end{scriptsize} \\
\begin{scriptsize} Zwicky Transient Facility    \end{scriptsize}        & \begin{scriptsize}g/r-band    \end{scriptsize}& \begin{scriptsize}$20.2$ \end{scriptsize}  &\begin{scriptsize} $9.96$  \end{scriptsize} &\begin{scriptsize} $34.6$    \end{scriptsize}    &\begin{scriptsize} LALInference \end{scriptsize}&\begin{scriptsize} 52 \end{scriptsize} &\begin{scriptsize} \citet{gcn26673,2020arXiv200100588W,Kasliwal2020},  \end{scriptsize} \\
% \begin{scriptsize} \end{scriptsize}& \begin{scriptsize} \end{scriptsize} & \begin{scriptsize} \end{scriptsize} & \begin{scriptsize} \end{scriptsize}  & \begin{scriptsize}  \end{scriptsize} &\begin{scriptsize} \end{scriptsize}&\begin{scriptsize} \end{scriptsize} &\begin{scriptsize}  this work \end{scriptsize} \\
& & & & & & & \\
\hline
\multicolumn{8}{c}{\textbf{S200115j (BHNS)}}  \\
\begin{scriptsize} GOTO    \end{scriptsize}         & \begin{scriptsize} g-band   \end{scriptsize}& \begin{scriptsize}$19.5$ \end{scriptsize}  &\begin{scriptsize} $0.2$  \end{scriptsize} &\begin{scriptsize} $26.4$    \end{scriptsize}    &\begin{scriptsize}  bayestar ini \end{scriptsize}&\begin{scriptsize} $52$ \end{scriptsize} &\begin{scriptsize} \citet{gcn26794}\end{scriptsize} \\
\begin{scriptsize} GRANDMA-FRAM-A     \end{scriptsize}         & \begin{scriptsize} R-band   \end{scriptsize}& \begin{scriptsize}$18$ \end{scriptsize}  &\begin{scriptsize} $20.8$  \end{scriptsize} &\begin{scriptsize} $1.7$    \end{scriptsize}    &\begin{scriptsize} LALInference  \end{scriptsize}&\begin{scriptsize} $2$ \end{scriptsize} &\begin{scriptsize} \citet{GRANDMA2020}\end{scriptsize} \\
\begin{scriptsize} GRANDMA-TCA    \end{scriptsize}         & \begin{scriptsize} Clear   \end{scriptsize}& \begin{scriptsize}$18$ \end{scriptsize}  &\begin{scriptsize} $12.9$  \end{scriptsize} &\begin{scriptsize} $72.2$    \end{scriptsize}    &\begin{scriptsize} LALInference  \end{scriptsize}&\begin{scriptsize} $4$ \end{scriptsize} &\begin{scriptsize} \citet{GRANDMA2020}\end{scriptsize} \\
\begin{scriptsize} GRANDMA-TCH    \end{scriptsize}         & \begin{scriptsize} Clear   \end{scriptsize}& \begin{scriptsize}$18$ \end{scriptsize}  &\begin{scriptsize} $0.3$  \end{scriptsize} &\begin{scriptsize} $148.3$    \end{scriptsize}    &\begin{scriptsize} LALInference  \end{scriptsize}&\begin{scriptsize} $7$ \end{scriptsize} &\begin{scriptsize} \citet{GRANDMA2020}\end{scriptsize} \\
\begin{scriptsize} GRANDMA-TRE \end{scriptsize}         & \begin{scriptsize} Clear   \end{scriptsize}& \begin{scriptsize}$17$ \end{scriptsize}  &\begin{scriptsize} $11.7$  \end{scriptsize} &\begin{scriptsize} $28.7$    \end{scriptsize}    &\begin{scriptsize} LALInference  \end{scriptsize}&\begin{scriptsize} $7$ \end{scriptsize} &\begin{scriptsize} \citet{GRANDMA2020}\end{scriptsize} \\
\begin{scriptsize} MASTER-network     \end{scriptsize}      & \begin{scriptsize} P/Clear   \end{scriptsize}& \begin{scriptsize}$ \approx 15/19$ \end{scriptsize}  &\begin{scriptsize} $\approx0.1$  \end{scriptsize} &\begin{scriptsize} $144$    \end{scriptsize}    &\begin{scriptsize} LALInference \end{scriptsize}&\begin{scriptsize} $62$ \end{scriptsize} &\begin{scriptsize} \citet{gcn26755}\end{scriptsize} \\
\begin{scriptsize} Pan-STARRS    \end{scriptsize}         & \begin{scriptsize}   w-band \end{scriptsize}& \begin{scriptsize}$21$ \end{scriptsize}  &\begin{scriptsize} $\approx72$  \end{scriptsize} &\begin{scriptsize} $\approx 24$    \end{scriptsize}    &\begin{scriptsize} $-$ \end{scriptsize}&\begin{scriptsize} $-$ \end{scriptsize} &\begin{scriptsize} \citet{gcn26839} \end{scriptsize} \\
\begin{scriptsize} SVOM-GWAC   \end{scriptsize}         & \begin{scriptsize} R-band  \end{scriptsize}& \begin{scriptsize}$\approx16$ \end{scriptsize}  &\begin{scriptsize} $7.1$  \end{scriptsize} &\begin{scriptsize} $10.6$    \end{scriptsize}    &\begin{scriptsize}  LALInference \end{scriptsize}&\begin{scriptsize} $41$ \end{scriptsize} &\begin{scriptsize} \citet{gcn26786}, this work \end{scriptsize} \\
\begin{scriptsize} Swift-UVOT  \end{scriptsize}         & \begin{scriptsize} u-band   \end{scriptsize}& \begin{scriptsize} $19.6$ \end{scriptsize}  &\begin{scriptsize} $2.0$  \end{scriptsize} &\begin{scriptsize} $80.1$    \end{scriptsize}    &\begin{scriptsize}  LALInference \end{scriptsize}&\begin{scriptsize} $3$ \end{scriptsize} &\begin{scriptsize} \citet{gcn26798,2020arXiv200100588W}, this work \end{scriptsize} \\
\begin{scriptsize} Zwicky Transient Facility     \end{scriptsize}        & \begin{scriptsize}g/r-band    \end{scriptsize}& \begin{scriptsize}$20.8$ \end{scriptsize}  &\begin{scriptsize} $0.24$  \end{scriptsize} &\begin{scriptsize} $> 1$    \end{scriptsize}    &\begin{scriptsize} LALInference \end{scriptsize}&\begin{scriptsize} $22$ \end{scriptsize} &\begin{scriptsize} \citet{gcn26767, Kasliwal2020}\end{scriptsize} \\
& & & & & & & \\
\hline
\multicolumn{8}{c}{\textbf{S200213t (BNS)}}  \\
\begin{scriptsize} DDOTI/OAN     \end{scriptsize}        & \begin{scriptsize}w-filter    \end{scriptsize}& \begin{scriptsize}$\approx 19$ \end{scriptsize}  &\begin{scriptsize} $0.75$  \end{scriptsize} &\begin{scriptsize} $2.17$ \end{scriptsize}    &\begin{scriptsize} LALInference \end{scriptsize}&\begin{scriptsize} $\approx 41$ \end{scriptsize} &\begin{scriptsize} \citet{gcn27061}, this work \end{scriptsize} \\
\begin{scriptsize} GOTO   \end{scriptsize}        & \begin{scriptsize} G-band   \end{scriptsize}& \begin{scriptsize}$18.4$ \end{scriptsize}  &\begin{scriptsize} $\approx0$  \end{scriptsize} &\begin{scriptsize} $26.5$    \end{scriptsize}    &\begin{scriptsize} bayestar ini \end{scriptsize}&\begin{scriptsize} $54$ \end{scriptsize} &\begin{scriptsize} \citet{gcn27069}\end{scriptsize} \\
\begin{scriptsize} GRANDMA-FRAM-C    \end{scriptsize}         & \begin{scriptsize} R-band   \end{scriptsize}& \begin{scriptsize}$17$ \end{scriptsize}  &\begin{scriptsize} $15.3$  \end{scriptsize} &\begin{scriptsize} $1.5$    \end{scriptsize}    &\begin{scriptsize} LALInference  \end{scriptsize}&\begin{scriptsize} $4$ \end{scriptsize} &\begin{scriptsize} \citet{GRANDMA2020}\end{scriptsize} \\
\begin{scriptsize} GRANDMA-OAJ   \end{scriptsize}         & \begin{scriptsize} r-band  \end{scriptsize}& \begin{scriptsize}$21$ \end{scriptsize}  &\begin{scriptsize} $15$  \end{scriptsize} &\begin{scriptsize} $1.5$    \end{scriptsize}    &\begin{scriptsize} LALInference  \end{scriptsize}&\begin{scriptsize} $18$ \end{scriptsize} &\begin{scriptsize} \citet{GRANDMA2020}\end{scriptsize} \\
\begin{scriptsize} GRANDMA-TCA   \end{scriptsize}         & \begin{scriptsize} Clear   \end{scriptsize}& \begin{scriptsize}$18$ \end{scriptsize}  &\begin{scriptsize} $0.4$  \end{scriptsize} &\begin{scriptsize} $43.6$    \end{scriptsize}    &\begin{scriptsize} LALInference  \end{scriptsize}&\begin{scriptsize} $30$ \end{scriptsize} &\begin{scriptsize} \citet{GRANDMA2020}\end{scriptsize} \\
\begin{scriptsize} MASTER-network    \end{scriptsize}         & \begin{scriptsize} Clear    \end{scriptsize}& \begin{scriptsize}$\approx 18.5$ \end{scriptsize}  &\begin{scriptsize} $0.1$  \end{scriptsize} &\begin{scriptsize} $144$    \end{scriptsize}    &\begin{scriptsize} LALInference \end{scriptsize}&\begin{scriptsize} $87$ \end{scriptsize} &\begin{scriptsize} \citet{gcn27041}\end{scriptsize} \\
\begin{scriptsize} Zwicky Transient Facility    \end{scriptsize}        & \begin{scriptsize}g/r-band    \end{scriptsize}& \begin{scriptsize}$21.2$ \end{scriptsize}  &\begin{scriptsize} $0.4$  \end{scriptsize} &\begin{scriptsize} $<25.7$    \end{scriptsize}    &\begin{scriptsize} LALInference \end{scriptsize}&\begin{scriptsize} $72$ \end{scriptsize} &\begin{scriptsize} \citet{gcn27051,Kasliwal2020}\end{scriptsize} \\
& & & & & & & \\
% \begin{scriptsize}    \end{scriptsize}        & \begin{scriptsize}    \end{scriptsize}& \begin{scriptsize} \end{scriptsize}  &\begin{scriptsize} $1.1$  \end{scriptsize} &\begin{scriptsize} $22.0$    \end{scriptsize}    &\begin{scriptsize} LALInference \end{scriptsize}&\begin{scriptsize} $72$ \end{scriptsize} &\begin{scriptsize} \citet{Kasliwal2020} \end{scriptsize} \\
\hline
\hline
\enddata
\end{deluxetable*}

\end{document}